\documentclass[final,5p,times,twocolumn]{elsarticle}

\makeatletter
\def\@author#1{\g@addto@macro\elsauthors{\normalsize%
		\def\baselinestretch{1}%
		\upshape\authorsep#1\unskip\textsuperscript{%
			\ifx\@fnmark\@empty\else\unskip\sep\@fnmark\let\sep=,\fi
			\ifx\@corref\@empty\else\unskip\sep\@corref\let\sep=,\fi
		}%
		\def\authorsep{\unskip,\space}%
		\global\let\@fnmark\@empty
		\global\let\@corref\@empty  
		\global\let\sep\@empty}%
	\@eadauthor={#1}
}
\makeatother

\usepackage{etoolbox}
\usepackage{amssymb}
\usepackage[utf8]{inputenc}
\usepackage{microtype}
\usepackage{graphicx}
\usepackage{dblfloatfix}
\usepackage{filecontents}
\usepackage[hyphens]{url}
\usepackage{breakurl}
\usepackage[hidelinks, bookmarks=false, breaklinks=true]{hyperref}
\usepackage{flushend}
\usepackage{amsmath}
\usepackage{multirow}
\usepackage{todonotes}
\usepackage{enumitem}
\usepackage{subcaption}
\usepackage{csquotes}
\usepackage{mdframed}
\usepackage{supertabular}
\usepackage{colortbl}
\usepackage{varwidth}
\usepackage{hhline}
\usepackage{footnote}
\usepackage{fancyhdr}
\usepackage[misc,geometry]{ifsym}
\mdfsetup{skipabove=4pt,skipbelow=4pt}

\expandafter\def\expandafter\UrlBreaks\expandafter{\UrlBreaks
	\do\a\do\b\do\c\do\d\do\e\do\f\do\g\do\h\do\i\do\j%
	\do\k\do\l\do\m\do\n\do\o\do\p\do\q\do\r\do\s\do\t%
	\do\u\do\v\do\w\do\x\do\y\do\z\do\A\do\B\do\C\do\D%
	\do\E\do\F\do\G\do\H\do\I\do\J\do\K\do\L\do\M\do\N%
	\do\O\do\P\do\Q\do\R\do\S\do\T\do\U\do\V\do\W\do\X%
	\do\Y\do\Z\do\*\do\-\do\~\do\'\do\"\do\-}%

\journal{Journal of Systems and Software}

\newcommand*\circled[1]{\tikz[baseline=(char.base)]{
		\node[shape=circle,draw,inner sep=1pt] (char) {#1};}}

\makeatletter
\patchcmd{\cortextmark}
{*}
{+}
{}{}

\makeatother

\fancypagestyle{footer}{\fancyhf{}\lfoot{\textit{\footnotesize Preprint accepted for publication in Journal of Systems and Software.}\\ {\footnotesize \url{https://doi.org/10.1016/j.jss.2019.110479}}\\ \vspace{10pt} {\footnotesize © 2019. This manuscript is made available under the CC-BY-NC-ND 4.0 license.\\ \url{http://creativecommons.org/licenses/by-nc-nd/4.0/}}} \cfoot{} \rfoot{\textit{\footnotesize \today}}}

\begin{document}

\begin{frontmatter}




\title{Representing Software Project Vision by Means of Video:\\ A Quality Model for Vision Videos}

\author[germany]{Oliver Karras\corref{cor1}}
\ead{oliver.karras@inf.uni-hannover.de}
\author[germany]{Kurt Schneider}
\ead{kurt.schneider@inf.uni-hannover.de}
\author[switzerland,sweden]{Samuel A. Fricker}
\ead{samuel.fricker@\{fhnw.ch, bth.se\}}
\address[germany]{Software Engineering Group, Leibniz Universität Hannover, 
30167 Hannover, Germany}
\address[switzerland]{Institute for Interactive Technologies, Fachhochschule Nordwestschweiz, 5210 Windisch, Switzerland}
\address[sweden]{Software Engineering Research Laboratory (SERL-Sweden), 
Blekinge Institute of Technology, 371 79 Karlskrona, Sweden}
\cortext[cor1]{Corresponding author}

\begin{abstract}
Establishing a shared software project vision is a key challenge in Requirements Engineering (RE). Several approaches use videos to represent visions. However, these approaches omit how to produce a good video. This missing guidance is one crucial reason why videos are not established in RE. We propose a quality model for videos representing a vision, so-called vision videos. Based on two literature reviews, we elaborate ten quality characteristics of videos and five quality characteristics of visions which together form a quality model for vision videos that includes all $15$ quality characteristics. We provide two representations of the quality model: (a) A hierarchical decomposition of vision video quality into the quality characteristics and (b) A mapping of these characteristics to the video production and use process. While the hierarchical decomposition supports the evaluation of vision videos, the mapping provides guidance for video production. In an evaluation with $139$ students, we investigated whether the $15$ characteristics are related to the overall quality of vision videos perceived by the subjects from a developer's the point of view. Six characteristics (\textit{video length}, \textit{focus}, \textit{prior knowledge}, \textit{clarity}, \textit{pleasure}, and \textit{stability}) correlated significantly with the likelihood that the subjects perceived a vision video as good. These relationships substantiate a fundamental relevance of the proposed quality model. Therefore, we conclude that the quality model is a sound basis for future refinements and extensions.
\end{abstract}


\end{frontmatter}


\section{Introduction: Visions on Video}
\label{sec:introduction-visions-on-video}
\thispagestyle{footer}
A clear vision of the desired result can accelerate software development projects and increase the likelihood of developing a successful system \cite{Lynn.2001}. It provides the flag around which all involved parties can rally. Thus, a dialog can emerge to define the scope of the future system that contributes to accomplishing the vision. Clarity of a vision refers to having a vision that is well articulated, easy to understand, and represented in an accessible way to all product partners consisting of customers, business, and technology \cite{Gottesdiener.2014}. All of them need to share the same system vision to achieve a shared understanding guiding their project activities \cite{Glinz.2015}. \enquote{Only when they all [customer, supplier, and development team] share a common vision, scope, and desired outcome is the project likely to be successful} \cite[p. 1]{Bruegge.b}.

Creighton et al. \cite{Creighton.} as well as Antón and Potts \cite{Anton.1998} emphasize the lack of a clear vision as a key challenge in requirements engineering (RE). The establishment of a shared and clearly defined vision is a challenging task regardless of whether stakeholders and project members meet in person \cite{Ochodek.2018} or not \cite{Ambler.2002, Fricker.2010}. Therefore, successful communication of a vision depends on suitable documentation options qualified for conveying the stakeholders' needs comprehensibly between all involved parties.

One of the most widely used documentation options to convey stakeholders' needs is a written specification as suggested by standards such as ISO/IEC/IEEE 29148:2011 \cite{ISO29148.2011}. However, the use of written specifications for communication is cumbersome since textual representations including digital versions have the lowest communication richness and effectiveness \cite{Ambler.2002}. The supposedly simple handover of a written specification insufficiently supports the rich knowledge transfer that is necessary to develop an acceptable system \cite{Fricker.2010}. Abad et al. \cite{Abad.2016} found the need for better support of requirements communication that exceeds pictorial representations documented in written specifications. They proposed to invest more efforts in addressing interactive visualizations such as storytelling with videos \cite{Abad.2016}.

The application of videos in RE has been discussed in recent years and its contributions have interesting potential \cite{Carter.2009, Fricker.2016, Schneider.2017, Karras.2016b}. Thirty-five years ago, different researchers \cite{Feeney.1983, Brouse.1992, BrunCottan.1995} already proposed the application of videos to support knowledge transfer due to the communication richness and effectiveness of video \cite{Ambler.2002}. According to Carter and Karatsolis \cite{Carter.2009}, short videos of well-expressed key concepts, such as a vision, are an effective and persuasive tool which can produce significant value as a documentation option. Creighton et al. support this perspective by emphasizing that a video, as a timed medium, \enquote{needs to focus on the essentials of the visionary system} \cite[p. 9]{Creighton.}.
Therefore, videos seem to be adequate to visualize visions and their future impacts. Since $1992$, several approaches \cite{Creighton., Brouse.1992, Karras.2017a} focused on the use of videos to represent a software project vision or parts of it (problem, solution, and improvement \cite{Wiegers.2003, Kittlaus.2017, Moore.1991}). For example, Creighton et al. \cite{Creighton.} used videos to capture problems of as-is situations, illustrate envisioned solutions and thus highlight the improvement. A vision or parts of it are suitable contents for videos since a vision is a key concept that provides an overview of a project with its goals and total extent of the future system \cite{Kittlaus.2017, Pohl.2010, Karras.2016}. Karras \cite{Karras.2019a} defines the concept of videos that represent a vision or parts of it as under the term \textit{vision video} as follows:

\begin{mdframed}
	\begin{itemize}[leftmargin=-2.5mm]
		\item[] \textbf{Definition:}
		A vision video is a video that represents a vision or parts of it for achieving shared understanding among all parties involved by disclosing, discussing, and aligning their mental models of the future system.
	\end{itemize}
\end{mdframed}

The audience of vision videos includes the stakeholders for validating the vision and the project team which is supposed to implement the vision. Stakeholders are the initial source of requirements; they need to comply with the vision. The project team also needs to be aware of the vision to push the development in the right direction to implement a software that supports the vision.

\subsection{Guidance for Good Vision Videos}
\label{sec:guidance-for-good-vision-videos}
Although existing approaches use vision videos, the required \textit{video production is often considered a secondary task} \cite{Karras.2018}. For example, Brill et al. \cite{Brill.2010} demonstrated the benefits of using vision videos compared to textual use cases to clarify the requirements of a future system with stakeholders. Brill et al. clearly stated: \enquote{We give no guidance for creating good videos – this remains future work} \cite[p. 2]{Brill.2010}. Many other approaches also apply vision videos but omit the details on how to produce good videos that are suitable for their respective purpose \cite{Karras.2018, Xu.2012}. At first, this neglect does not seem to be a problem since viewers of a video are mainly interested in what a video shows and tells. It is unlikely that they are concerned about how the production was done unless they get bored or the technology becomes obtrusive and distracting \cite{Owens.2011}. Every video producer needs to avoid any defects to increase the quality of a video. Thus, they focus the viewers' attention on the conveyed content so that viewers can fulfill the goals of their individual underlying information needs. 

According to the findings of Karras \cite{Karras.2018b}, software professionals lack knowledge on how to produce good videos for visual communication. So far, little research encountered the challenge to encourage and enable software professionals to produce good videos on their own \cite{Karras.2018}. Karras and Schneider \enquote{assert that software professionals could enrich their communication and thus RE abilities if they knew what constitutes a good video} \cite[p. 4]{Karras.2018}. Thus, the missing guidance for producing good videos is one crucial reason why videos are still not an established documentation option in RE \cite{Karras.2018}.

Whether a video is good or not depends on its perceived quality. However, \textit{video quality} is a rather ill-defined concept due to numerous factors \cite{Winkler.2008}. On the one hand, there are technical factors such as video properties (size, resolution, brightness, etc.), record and display devices, or sound quality. On the other hand, there are subjective factors such as the individual interests, quality expectations, video experiences, and viewing conditions of viewers. The wide variety and subjectivity of these factors indicate the complexity of video quality. This complexity impedes the prediction of \textit{how different viewers assess the quality of a video} due to the several factors that affect their attitude \cite{Karras.2018}.

Inspired by the ISO/IEC FDIS $25010$:$2010$ \cite{ISO25010.2010} for system and software quality models, Karras and Schneider \cite{Karras.2018} propose as a plan of action to develop a quality model for videos. In consideration of the ISO/IEC FDIS $25010$:$2010$ \cite{ISO25010.2010}, the comprehensive specification and evaluation of the quality of a video requires the definition of the necessary and desired quality characteristics associated with the producers' and viewers' goals and objectives of a video. A quality model for videos following the ISO/IEC FDIS $25010$:$2010$ \cite{ISO25010.2010} can be used to identify relevant video characteristics that can be further used to establish requirements, their criteria for satisfaction, and corresponding measures for a particular video. Such a quality model can encourage and enable software professionals to produce good videos by guiding their video production and use process in terms of planning, recording, editing, and viewing a video. Thus, software professionals should be able to produce good videos on their own at affordable costs and efforts, yet sufficient quality.

\subsection{Developing a Quality Model for Vision Videos}
\label{sec:objective-approach-and-contribution}
In this article, we follow the line of thought of Karras and Schneider \cite{Karras.2018}. Our objective is to present a quality model for vision videos that can be used for (a) evaluating a given vision video and (b) guiding the video production by software professionals. Thus, we strive for a tailored approach that offers the essence of what constitutes a good video for the purposes associated with conveying a software project vision.

Quality models such as the ISO/IEC FDIS $25010$:$2010$ are structured as a hierarchical decomposition which provides a convenient breakdown of the quality of a product. The individual quality characteristics of such a model can also be mapped to the respective steps of the development and use process of a product. In particular, a quality model distinguishes between \textit{product quality} and \textit{quality in use}. Thus, a quality model for a product includes sensorial characteristics of its \textit{representation}, perceptual characteristics of its \textit{content} as well as emotional characteristics regarding its \textit{impact} on its target recipients.

In our case, the product is a vision video. Thereby, the representation format is a video and the content is a vision. For the emotional characteristics, we need to consider the impact of a vision video on its target audience consisting of stakeholders and project members. We consider both structuring options to accommodate our two goals. The hierarchical decomposition supports goal (a) by breaking down vision video quality into individual quality characteristics that need to be assessed to evaluate the overall quality of a given video. For goal (b), we need the mapping of the individual characteristics to the steps of the production and use process of a video. This structuring shows how one upstream quality characteristic affects downstream characteristics which helps to guide the video production by software professionals.

This article is structured as follows: Section \ref{sec:research-approach} provides an overview of our research process consisting of two literature reviews and an evaluation. Section \ref{sec:background} deals with the background and discusses related work. Section \ref{sec:quality-model-for-vision-videos} summarizes the idea of a quality model for vision videos. Section \ref{sec:literature-review-of-video-production-guidelines} and \ref{sec:literature-review-on-vision-characteristics} present the two literature reviews on video production guidelines and software project vision. Based on the results of the literature reviews, section \ref{sec:represent-vision-by-means-of-video-the-quality-model} propose the quality model for vision videos. Section \ref{sec:evaluation} reports the design, analysis, and results of the evaluation. Section \ref{sec:threats-to-validity} describes the threats to validity separately according to the literature reviews and the evaluation. While section \ref{sec:discussion} discusses our findings and future work, section \ref{sec:conclusion} concludes the article.

\section{Research Approach}
\label{sec:research-approach}
Below, we present our research approach to develop a quality model for vision videos. We show the details of our concrete research process and its contributions. 

\subsection{Research Process}
\begin{figure}[!b]
	\centering
	\includegraphics[width=\columnwidth]{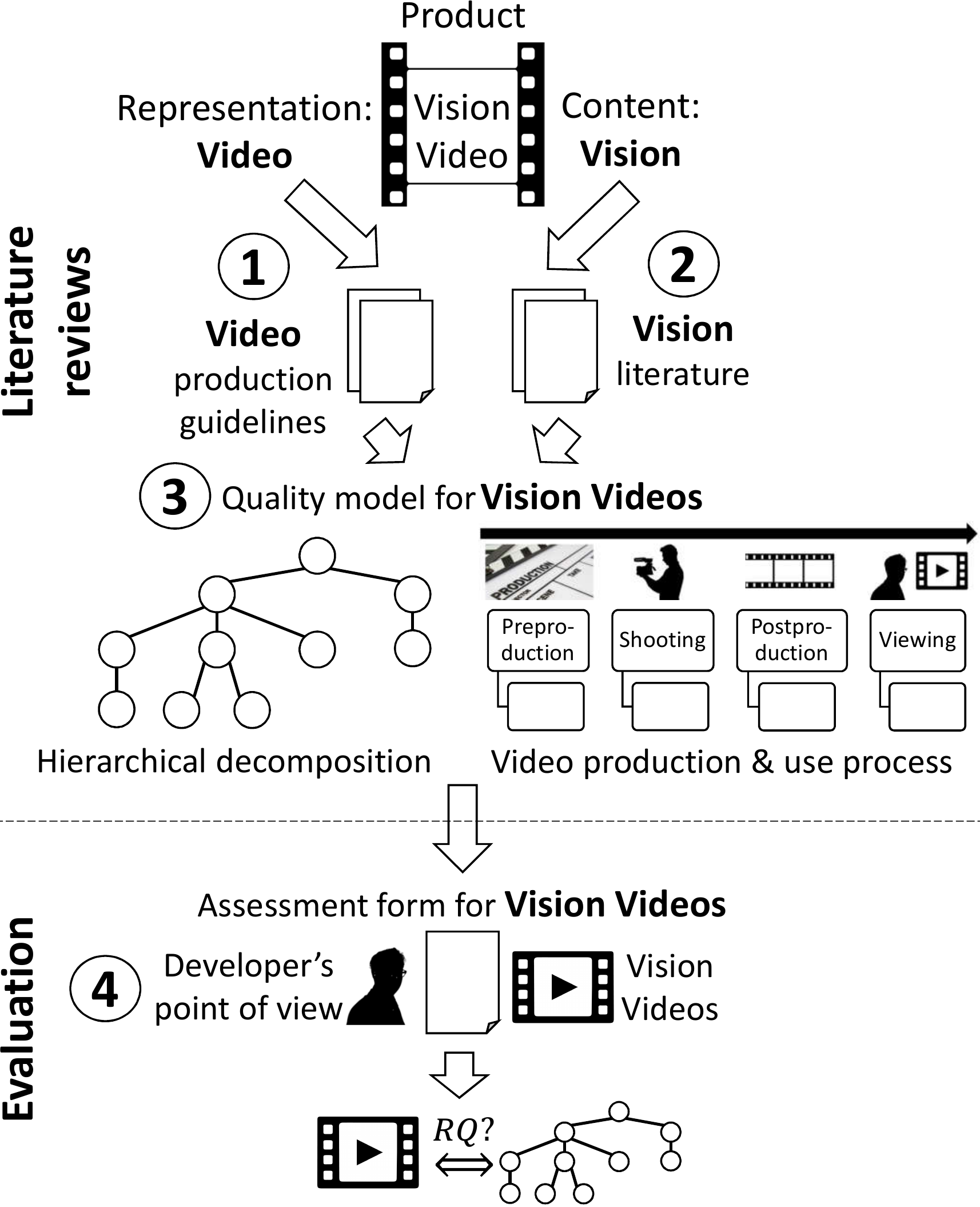}
	\caption{Research process}
	\label{fig:fig1}
\end{figure}

In \figurename{ \ref{fig:fig1}}, we illustrate our research process consisting of two phases: \textit{literature reviews} and \textit{evaluation}. 

\paragraph{Literature Reviews} We derived our quality model for vision videos from two literature reviews. According to Karras and Schneider \cite{Karras.2018}, we conducted a review of video production guidelines (see \figurename{ \ref{fig:fig1}}, \circled{\textbf{1}}). However, a video is only a representation format for arbitrary contents that a video producer wants to convey to diverse viewers. Therefore, we also conducted a review on software project vision to consider the content of a vision video (see \figurename{ \ref{fig:fig1}}, \circled{\textbf{2}}). Thus, we cover video and vision with their representations, contents, and intended purposes. The obtained characteristics form a quality model for vision videos which fulfills our two goals (see \figurename{ \ref{fig:fig1}}, \circled{\textbf{3}}). 

\paragraph{Evaluation}
The proposed quality model is only a theoretical consideration of the relevant characteristics of a vision video that needs to be evaluated \cite{Stachowiak.1973}. According to Gorschek et al. \cite{Gorschek.2006}, any proposed candidate solution, such as the quality model for vision videos, must be initially validated in academia before it is presented in the industry. We need to ensure that the quality model is of fundamental relevance and soundness to be suitable for presentation to industry experts. We investigate how the individual quality characteristics relate to the overall quality of vision videos from a developer's point of view (see \figurename{ \ref{fig:fig1}}, \circled{\textbf{4}}). We ask the following research question:

\begin{mdframed}
	\begin{itemize}[leftmargin=-2.5mm]
		\item[] \textbf{RQ:}
		How do the individual quality characteristics relate to the overall quality of vision videos from a developer's point of view?
	\end{itemize}
\end{mdframed}

According to Seshadrinathan and Bovik \cite{Seshadrinathan.2010} subjective judgment of video quality collected by asking humans for their opinion is considered as the ultimate standard and right way to assess video quality. We conducted a within-subjects experiment with $139$ undergraduate students who had the role of a developer and actively developed software in projects with real customers at the time of the experiment. Therefore, the subjects can be considered as developers due to their experience at the moment of the experiment. The undergraduate students subjectively assessed the overall quality and each quality characteristic of eight vision videos by completing an assessment form for each video (see \figurename{ \ref{fig:fig1}}, \circled{\textbf{4}}).
Based on the collected data \cite{Karras.2019b}, we investigated the relationship between the perceived overall quality of vision videos and the individual subjectively assessed quality characteristics. We examined whether the structures in the data match with our proposed quality model. Thus, we determined a set of individual quality characteristics that correlate with the overall quality of vision videos from the perspective of our subjects. Our findings substantiate the fundamental relevance of the proposed quality model and thus justify the next validation step in the industry which is part of our future work.

\begin{figure*}[!b]
	\captionsetup{justification=centering}
	\begin{subfigure}{.3\textwidth}
		\centering
		\includegraphics[width=\columnwidth]{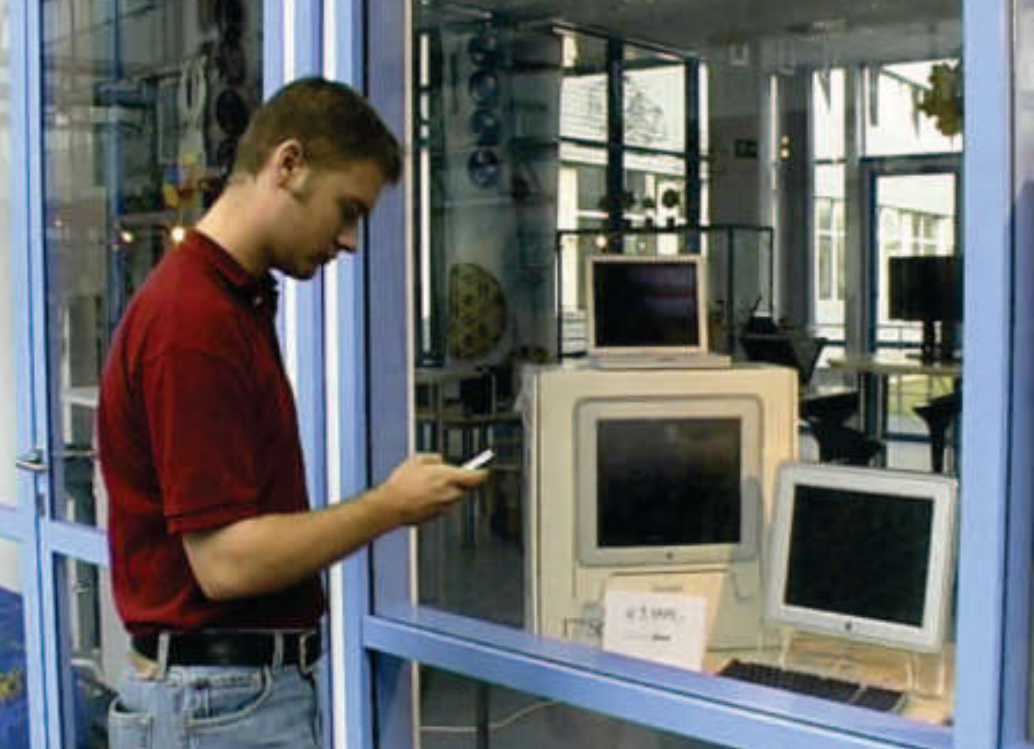}
		\caption{System for interactive window shopping\\ by Creighton et al. \cite{Creighton.}}
		\label{fig2:sfig1}
	\end{subfigure}
	\begin{subfigure}{.3\textwidth}
		\centering
		\includegraphics[width=\columnwidth]{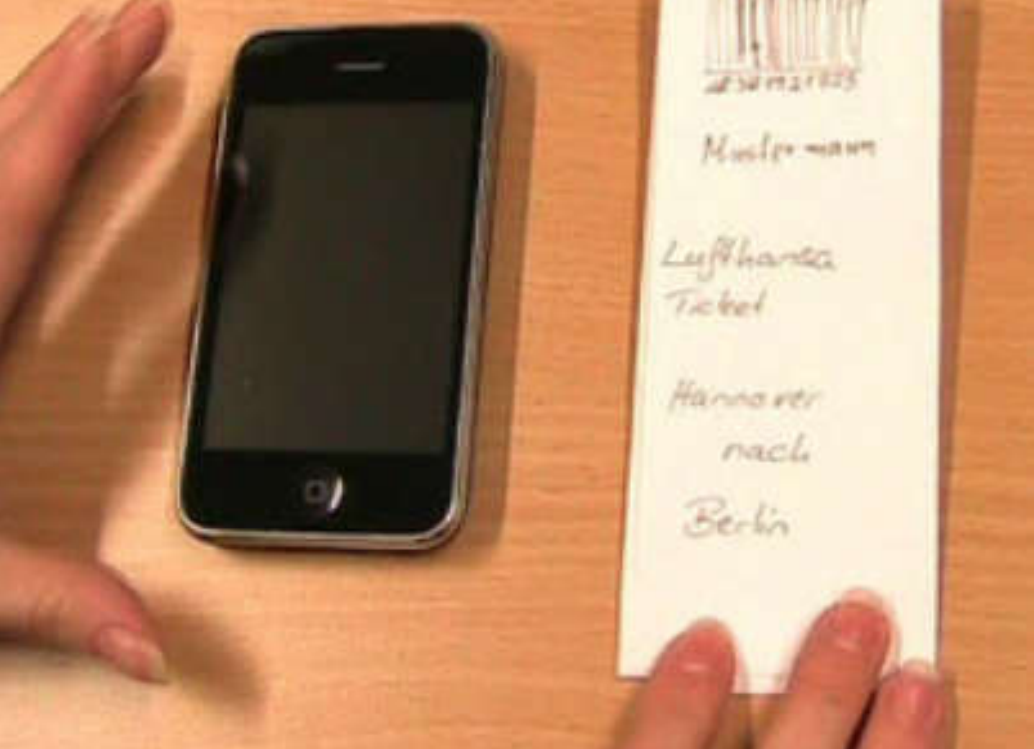}
		\caption{System for checking in at an airport\\ by Brill et al. \cite{Brill.2010}}
		\label{fig2:sfig2}
	\end{subfigure}
	\begin{subfigure}{.3\textwidth}
		\centering
		\includegraphics[width=\columnwidth]{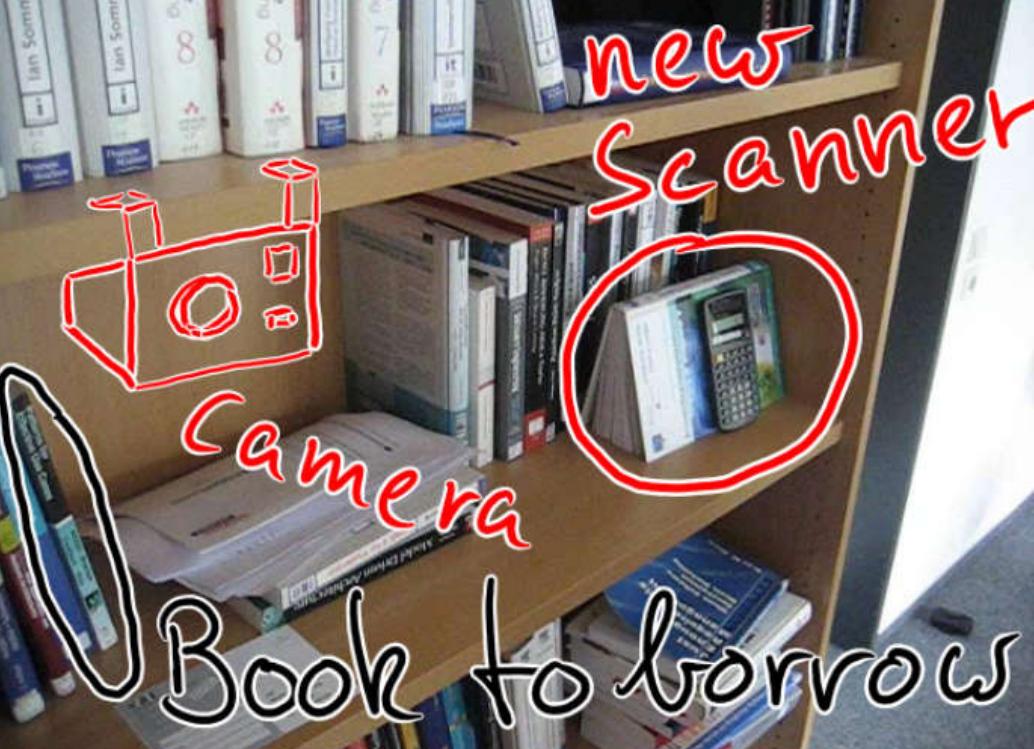}
		\caption{System for borrowing books at a library\\ by Pham et al. \cite{Pham.}}
		\label{fig2:sfig3}
	\end{subfigure}
	\begin{subfigure}{.3\textwidth}
		\centering
		\includegraphics[width=\columnwidth]{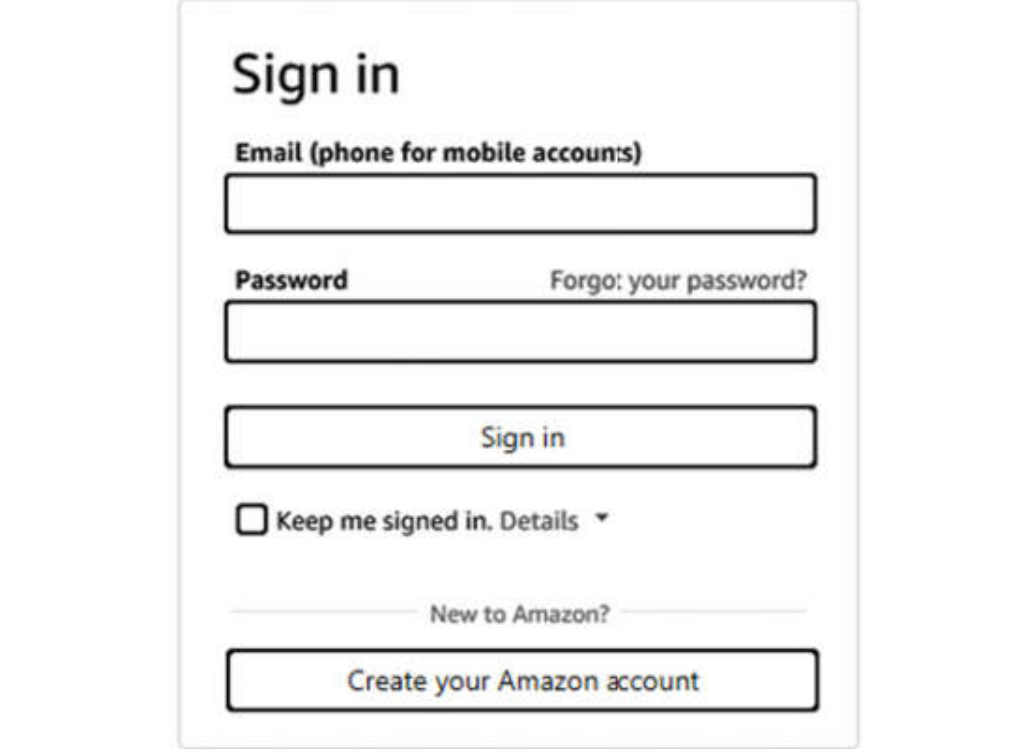}
		\caption{System for buying products online by Karras et al. \cite{Karras.2017a}}
		\label{fig2:sfig4}
	\end{subfigure}
	\begin{subfigure}{.3\textwidth}
		\centering
		\includegraphics[width=\columnwidth]{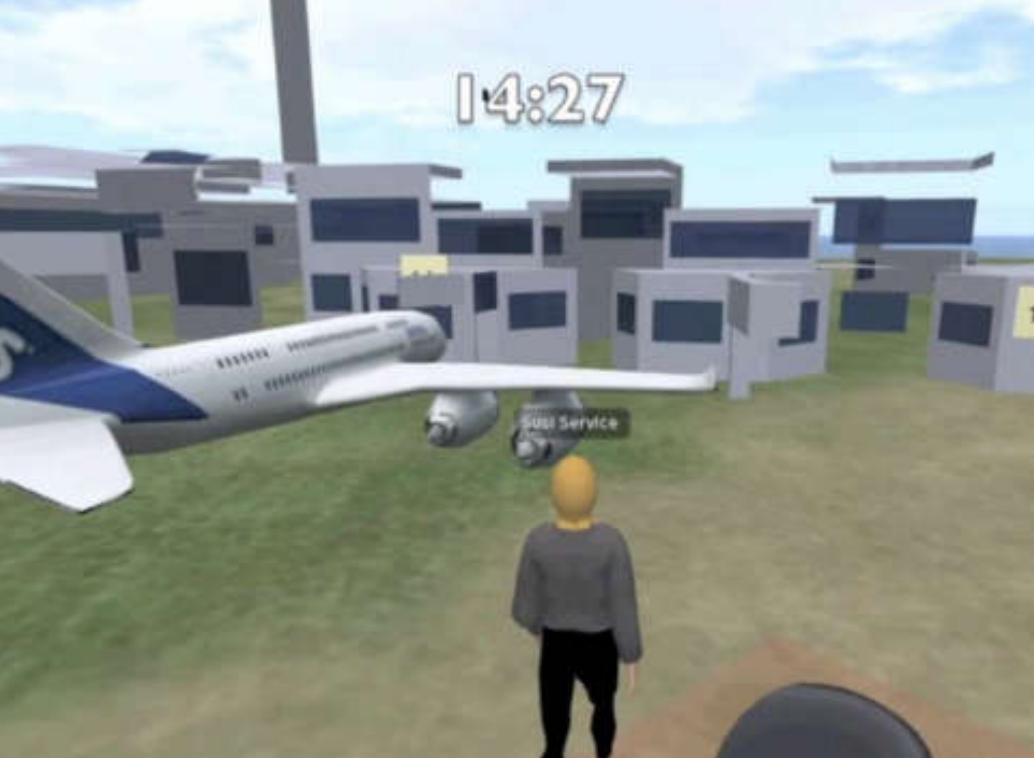}
		\caption{System for handling VIPs at an airport by Xu et al. \cite{Xu.2012}}
		\label{fig2:sfig5}
	\end{subfigure}
	\begin{subfigure}{.3\textwidth}
		\centering
		\includegraphics[width=\columnwidth]{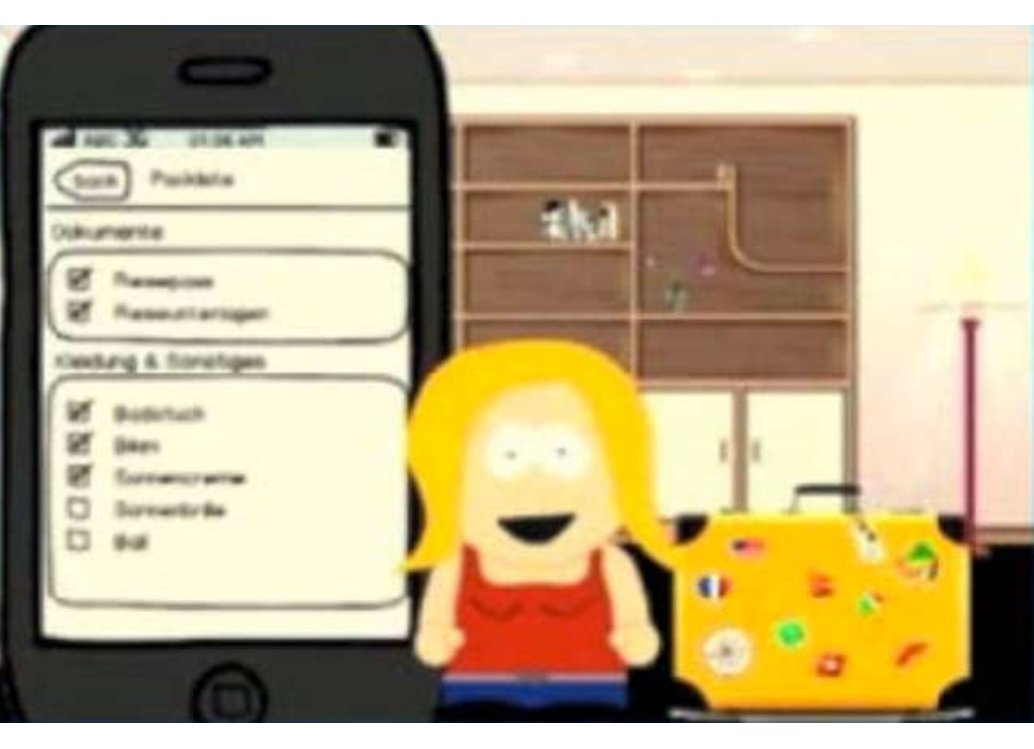}
		\caption{System for planing trips and vacations by Xu et al. \cite{Xu.2013}}
		\label{fig2:sfig6}
	\end{subfigure}
	\centering
	\caption{Examples of vision videos created according to different approaches}
	\label{fig:fig2}
\end{figure*}

\subsection{Contribution}
In comparison to unguided video production, the proposed quality model is supposed to simplify video production by software professionals and improve the overall quality of a produced vision video. In general, it is usually not practical to specify a whole quality model for a product in detail \cite{ISO25010.2010}. Therefore, the experimentally verified quality characteristics with relationships to the overall quality provide an additional benefit. Software professionals may spend more care and efforts on these quality characteristics since they affect the overall quality of a vision video according to the subjects who viewed and assessed the videos from a developer's point of view. However, the relative importance of quality characteristics depends on the high-level goals of the target audience and thus the intended use of the video. Hence, the findings should not be overgeneralized due to the experimental setting which excludes the perspective of specific stakeholders and video producers such as customers and requirements engineers. We contribute the following insights:

\begin{itemize}[leftmargin=0mm, topsep=0.5mm]
	\setlength\itemsep{-.25mm}
	\item[] ($1$) \textit{A Quality Model for Vision Videos}\\
	We offer a clearer definition of the hitherto ill-defined concept of video quality in the context of representing a software project vision. We illustrate the proposed quality model for vision videos in two representations: (a) The hierarchical decomposition provides a structured overview of the individual quality characteristics that constitute the overall quality of vision videos. (b) The mapping of the individual quality characteristics to the steps of video production and use process serves as a checklist to ensure the comprehensive treatment of the vision video quality over the entire process. While the first representation is suited to evaluate given vision videos, the second one is intended to guide video production by software professionals.
	
	\item[] ($2$) \textit{Relevance of the Quality Model for Vision Videos}\\
	Based on the first validation in academia, we found that there are specific relationships between six of the $15$ quality characteristics and the likelihood that the subjects perceive a vision video as good. These six quality characteristics of vision videos are \textit{video length}, \textit{focus}, \textit{prior knowledge}, \textit{clarity}, \textit{pleasure}, and \textit{stability}. The identified relationships, however, are limited in their validity due to the experimental design. Nevertheless, our key finding is the presence of relationships between the overall quality and individual quality characteristics of vision videos. These relationships substantiate a fundamental relevance of the proposed quality model for vision videos. Thus, we are confident that we developed a sound quality model that is suited for future refinements and extensions.
\end{itemize}

\section{Background}
\label{sec:background}
In the following, we provide a brief overview of related work on vision videos in RE. Furthermore, we present the fundamentals of video quality assessment to show the basic problem of video quality regarding its definition and measurement.

\subsection{Vision Videos in RE}
\label{sec:vision-videos-in-re}
Several approaches focus on the use of videos to represent a vision of parts of it. In \figurename{ \ref{fig:fig2}}, we illustrate the use of vision videos in RE with examples of different approaches.

Creighton et al. \cite{Creighton.} employed videos to describe as-is and visionary scenarios of a system for users, customers, and requirements engineers. While the as-is scenarios illustrate current problems in work practice, the visionary scenario videos show how the envisioned system may look, work, or be used (see \figurename{ \ref{fig2:sfig1}}). Creighton et al. \cite{Creighton.} introduced the role of video producer which can be fulfilled by either a member of the development team or an external video professional. This approach combines the created videos with UML diagrams to trace videos and requirements in later development phases.
Brill et al. \cite{Brill.2010} used low-effort, ad-hoc videos created by requirements engineers to represent use cases of a future system to elicit and clarify requirements with customers (see \figurename{ \ref{fig2:sfig2}}). Their experimental results yielded that such videos help to avoid misunderstandings and clarify requirements better than textual use cases.
Pham et al. \cite{Pham.} proposed an interactive storyboard to support requirements engineers to elicit, validate, and document requirements and visions of stakeholders. The interactive storyboard enables the creation of a special kind of videos which was enhanced by multimedia technologies (see \figurename{ \ref{fig2:sfig3}}).
Karras et al. \cite{Karras.2017a} proposed an approach to generate videos, which demonstrate interaction sequences on hand- or digitally drawn mockups, as additional support for textual scenarios. These videos are created as a by-product of digital prototyping by capturing and replaying interaction events of responsive controls without any implementation (see \figurename{ \ref{fig2:sfig4}}). They showed that such videos allow a faster understanding of textual scenarios by developers compared to static mockups.
Xu et al. \cite{Xu.2012} proposed 'Evolutionary Scenario-Based Design' which uses visionary scenario videos of unimplemented parts of a system for requirements elicitation and system demonstration purposes throughout the project lifecycle. These videos were created by members of the development team using virtual world technology (see \figurename{ \ref{fig2:sfig5}}). Xu et al. \cite{Xu.2012} report five short lessons learned about how to produce videos using virtual reality. Summarized, these lessons suggest the use of a storyboard to tell a story-driven video, the involvement of developers and customers to achieve a better understanding, and the full use of text or audio to express ideas. However, they stated explicitly that detailed guidance on the design and evaluation of video-based scenarios remains future work.
Xu et al. \cite{Xu.2013} extended their approach of video-based scenarios to represent user models. They described how they have employed videos to illustrate as-is, visionary, and demo scenarios for problem definition, elicitation, and validation (see \figurename{ \ref{fig2:sfig6}}). They provide five short lessons learned, similar to their previous ones \cite{Xu.2012}, to create videos of demo scenarios.

The presented examples illustrate how different vision videos of various approaches can be. While some vision videos show real-world scenes, others present animations of computer-generated contents. There are also clear differences in the representation of persons, systems, and interactions. As a consequence, vision videos of the individual approaches seem to have their specific characteristics. However, only two out of these six approaches provided some guidance on how to produce videos with the required characteristics so that the videos are suitable for their respective approach.
Besides the six previously presented approaches with concrete examples of vision videos, we found $14$ more approaches that also deal with the use of vision videos in RE but do not include examples of vision videos.

In \tablename{ \ref{tbl:t4}}, we summarized all $20$ related approaches to vision videos in RE regarding supported RE activity, focused part of a vision, video content, target audience, target video producer, and given guidance on video production.
All approaches used vision videos mainly to support the RE activities: problem definition, goal definition, elicitation, validation, and documentation. Thereby, the videos illustrated (1) problems, that need to be solved, (2) proposed solutions for a given problem, or (3) both problem and solution. Videos of problems presented environmental contexts and observations of users' work practice. Videos of proposed solutions showed visionary scenarios or use cases of the future system, prototypes, and software project visions. In the case of approaches presenting problem and solution, the produced videos could contain all previously mentioned contents and presentations of implemented parts of the future systems. The target audience consisted of stakeholders and members of the development team, i.e., decision-makers, users, managers, customers, domain experts, requirements engineers, developers, designers, and suppliers. The target video producers were mainly members of the development team, i.e., requirements engineers or arbitrary team members. Two approaches \cite{Rabiser.2006, Schneider.2011} focused on the use of videos created by users and four approaches \cite{Bruegge.b, Creighton.,Broll.2007, Bruegge.2008} introduced the role of a video producer which was fulfilled by either a member of the development team or an external video professional. Only five out of $20$ approaches \cite{Xu.2012, Xu.2013, Broll.2007, Mackay.1999, Jirotka.2006} provided a few brief tips, hints, or lessons learned for the production of videos for the respective approach. However, this guidance remains too abstract for a reader to understand how the specific videos need to be created to be of sufficient quality.

\subsection{Video Quality Assessment}\label{sec:video-quality-assessment}
Video quality can be assessed either subjectively or objectively. The subjective video quality assessment represents the most accurate method for obtaining quality ratings \cite{Winkler.2008}. In subjective experiments, the subjects (typically $15$ -- $30$) are asked to watch a set of videos and assess the quality of each video based on a defined $5$-point scale. The average assessment of all viewers for one video is defined as the Mean Opinion Score (MOS) \cite{Winkler.2008}. Due to the subjects' interests and expectations for a video, the variability in viewers' ratings cannot be excluded. There are several recommendations \cite{ITU.1998,ITU.2008,ITU.2012,ITU.2016} for conducting subjective assessments of video quality which provide precise instructions to mitigate this variability, such as standard viewing conditions, criteria for the selection of observers and test material, assessment procedures, and data analysis methods.

\begin{figure}[!b]
	\centering
	\includegraphics[width=\columnwidth]{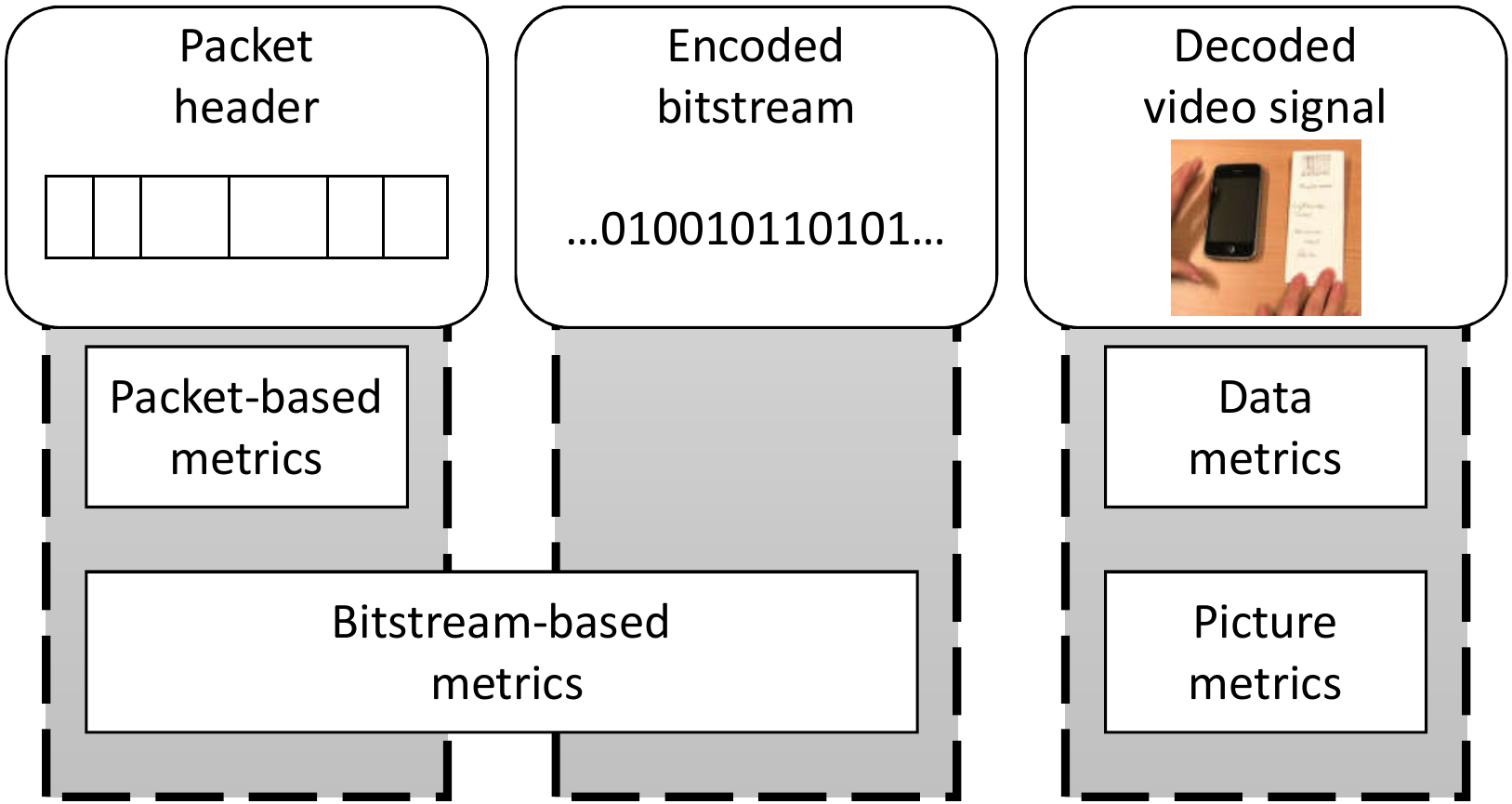}
	\caption{Objective quality metrics \cite{Winkler.2008}}
	\label{fig:fig3}
\end{figure}

Although subjective assessments are invaluable for evaluating video quality, their restricted applicability for a large number of viewers is their main disadvantage. This restriction limits the number of videos that can be assessed in a reasonable amount of time. Consequently, objective video quality assessments have been developed since the subjective assessment is neither intended nor practical for continuous monitoring of video quality \cite{Winkler.2008}.
Objective video quality assessments use algorithms which are designed to characterize video quality and predict the MOS of viewers. According to Winkler and Mohandas \cite{Winkler.2008}, there are three main types of objective metrics: \textit{Data metrics}, \textit{picture metrics}, and \textit{packet-} or \textit{bitstream-based metrics} (see \figurename{ \ref{fig:fig3}}).

The first two types are used for the analysis of a decoded video.
\textit{Data metrics} measure the fidelity of the signal based on byte-by-byte comparison without considering its content. No data metric is universally reliable since there is only an approximate relationship between data metrics and video quality perceived by viewers. Two of the most widely used data metrics are Mean Squared Error (MSE) and Peak Signal-Noise Ratio (PSNR) \cite{Winkler.2008}.

\textit{Picture metrics} treat the video data as the visual information by quantifying the effects of distortions and structural image contents on perceived quality. These metrics can be divided into \textit{vision modeling approaches} and \textit{engineering approaches}. 
Vision modeling approaches are based on modeling various components of the human visual system (HVS), such as color perception or contrast sensitivity. Some well-known HVS-based metrics are the Visual Differences Predictor (VDP) by Daly \cite{Daly.1992}, the Sarnoff JND (Just Noticeable Differences) metric by Lubin and Fibush \cite{Lubin.1997}, the Moving Picture Quality Metric (MPQM) by van den Branden Lambrecht and Verscheure \cite{VanDenBranden.1996}, and the Perceptual Distortion Metric (PDM) by Winkler \cite{Winkler.1999}. 
Engineering approaches are based on the analysis of image contents and distortions, e.g., contours or block artifacts. The Structural Similarity (SSIM) index by Wang et al. \cite{Wang.2004} and the Video Quality Metric (VQM) by Pinson and Wolf \cite{Pinson.2004} are two of the best-known metrics that belong to the engineering approaches.

The third type of metrics measures the impact of network losses on video quality. \textit{Packet-} or \textit{bitstream-based metrics} look at the packet header information and the encoded bitstream directly without fully decoding the video. Examples for such metrics are the V-Factor by Winkler and Mohandas \cite{Winkler.2008} and the approaches of Verscheure et al. \cite{Verscheure.1999} or Kanumuri et al. \cite{Kanumuri.2006,Kanumuri.2006a}.

\enquote{An important shortcoming of the existing metrics is that they measure image fidelity instead of perceived quality} \cite[p. 151]{Winkler.2007}. All these metrics mainly focus on measuring the visual fidelity of a video in terms of distortions introduced by various processing steps. Especially, the frequently applied metrics ignore the concrete content and emotional impact of a video. The consideration of all three dimensions of quality (\textit{representation}, \textit{content}, and \textit{impact}) for videos is an active research area \cite{Pereira.2005, Winkler.2009}.
In this context, the term video quality is currently being extended to the term \enquote{Quality of Experience} (QoE) \cite{Chikkerur.2011}. QoE describes quality from the perspective of the viewer by focusing on the perceived quality of the representation, the content, and the emotional impact of a video. 
Although several organizations, such as Video Quality Experts Group (VQEG\footnote{\url{www.vqeg.org}}) or International Telecommunication Union (ITU\footnote{\url{www.itu.int}}), are working on standards for video quality and significant progress has been made \cite{Winkler.2009}, \enquote{we are still a long way to video quality metrics that are widely applicable and universally recognized} \cite[p. 667]{Winkler.2008}.
 
Regardless of subjective or objective video quality assessment, the basic problem is that video quality is a rather ill-defined concept due to its numerous factors (see section \ref{sec:guidance-for-good-vision-videos}) \cite{Winkler.2008}. Therefore, it often remains unclear which individual quality characteristics are used to measure the overall video quality and how the quality can be influenced accordingly. Despite the sheer abundance of video quality metrics \cite{Winkler.2009}, to the best of our knowledge, there is no definition of video quality in form of its quality characteristics that need to be assessed to evaluate the overall quality of a given video, or in other words a quality model. 

Due to the lack of guidance for producing vision videos in RE and the ill-defined concept of video quality we decided to follow the line of thought by Karras and Schneider \cite{Karras.2018} by developing a \textit{quality model for vision videos}.

\section{A Quality Model for Vision Videos}
\label{sec:quality-model-for-vision-videos}
The standard ISO/IEC FDIS $25010$:$2010$ provides a suitable orientation to develop a quality model for vision videos. According to Moody \cite{Moody.2005}, the important features of a software quality model are:

\begin{itemize}[leftmargin=3.5mm, topsep=0.5mm]
	\setlength\itemsep{-.25mm}
	\item \textit{Hierarchical structure}: Definition of a hierarchy of quality (sub-) characteristics
	\item \textit{Familiar labels}: Use of single terms that are commonly understood in practice to identify each (sub-) characteristic
	\item \textit{Concise definitions}: Use of one single sentence to define each (sub-) characteristic
	\item \textit{Measurements}: Definition of metrics, measurement methods, and scales for all characteristics at the lowest model level
	\item \textit{Evaluation procedures}: Procedures that specify who should be involved in evaluations and how and when the evaluations should be conducted
\end{itemize}

This article focuses on the first three bullet points to propose a quality model for vision videos. The last two bullet points exceed the scope of this article and remain future work due to three reasons:
First, we need to define a quality model before we can establish corresponding measurements and evaluations. Second, the task of defining measurements and evaluation procedures is complex and extensive illustrated by the fact that three additional standards define the last two bullet points for software quality (ISO/IEC $25022$:$2016$ -- Measurements for Quality in Use \cite{ISO25022.2016}, ISO/IEC $25023$:$2016$ -- Measurements for System and Software Product Quality \cite{ISO25023.2016}, and ISO/IEC $25042$:$2012$ -- Evaluation Process \cite{ISO25040.2011}). Third, although there are several subjective and objective video quality assessments (see section \ref{sec:video-quality-assessment}), there is no established standard for video quality \cite{Winkler.2007}. Therefore, it is unclear which characteristics of a video are related to the defined video quality assessments and how they influence them. This impedes the mapping of existing assessments to the quality characteristics of vision videos. The decision of neglecting measurements and evaluation procedures affects the current usefulness and usability of the proposed quality model for vision videos by making it more difficult to carry out concrete evaluations. Based on their goals and questions, software professionals who want to use the quality model have to think about how they can make the quality characteristics measurable to assess them in appropriate evaluations. However, there are established approaches in software engineering, such as Goal Question Metric Paradigm \cite{Basili.1994, vanSolingen.1999}, to support this task. Furthermore, the recommendations for subjective video quality assessment \cite{ITU.1998,ITU.2008,ITU.2012,ITU.2016} are suitable sources that can be used for orientation when planning evaluations and corresponding measurements.

In consideration of ISO/IEC FDIS $25010$:$2010$ \cite{ISO25010.2010}, we define the quality of a vision video as the degree to which the vision video satisfies the stated and implied needs of its target audience consisting of all product partners to provide value. These needs can be represented by a quality model that categorizes the quality of a vision video into characteristics which in some cases can be further divided into sub-characteristics. A set of sub-characteristics associated with one characteristic needs to be representative of typical concerns regarding a vision video without necessarily being exhaustive. The benefit of such a quality model is its guidance for further specifying requirements, establishing measurements, and performing quality evaluations of vision videos. Software professionals can use the defined characteristics as a checklist for ensuring the comprehensive treatment of vision video quality requirements, thus providing a basis for estimating the consequent effort and activities that will be needed during the video production.

We want to illustrate the stated and implied needs of video producers and viewers with three exemplary situations. These situations are based on the \enquote{Major Software Quality Decision Points} of Boehm et al. \cite{Boehm.1976} which illustrated the needs for a quality model to describe and assess software quality.

\begin{itemize}[leftmargin=0mm, topsep=0.5mm]
	\setlength\itemsep{-.25mm}
	\item[] $(1)$ \textit{Describing what constitutes a good vision video}\\
	Formulating the required video duration, image, or sound quality is fairly easy nowadays. However, indicating the need for a particular plot or content in a vision video without a consistent terminology impedes the description.
	\item[] $(2)$ \textit{Producing a vision video for an existing RE approach}\\ 
	Producing a good vision video is cumbersome for unskilled software professionals. A novice may become discouraged due to the effort and risk of creating a potential improper video. These obstacles increase the threshold for producing vision videos.
	\item[] $(3)$ \textit{Making design trade-offs to reduce costs}\\
	Tight budgets and schedules may require trade-offs in video production. The decision-making can be supported by knowing what characteristics of a vision video are more important than others for a respective approach.
\end{itemize}

In the following, we present the details of the steps for developing a quality model for vision videos according to our research process (see \figurename{ \ref{fig:fig1}}). Section \ref{sec:literature-review-of-video-production-guidelines} and section \ref{sec:literature-review-on-vision-characteristics} present the individual literature reviews of video production guidelines (see \figurename{ \ref{fig:fig1}}, \circled{\textbf{1}}) and software project vision literature (see \figurename{ \ref{fig:fig1}}, \circled{\textbf{2}}). In section \ref{sec:represent-vision-by-means-of-video-the-quality-model}, we combine the results of the two literature reviews to propose the quality model for vision videos (see \figurename{ \ref{fig:fig1}}, \circled{\textbf{3}}). We have to emphasize that both literature reviews are not systematic literature reviews. We do not claim to provide systematic and comprehensive literature reviews that identify, analyze, and interpret all available evidence related to the topic of video and vision characteristics. Instead, these two literature reviews are intended to provide an initial overview based on a grounded and reflected body of knowledge to enable us to propose a first quality model for vision videos.

\section{Literature Review on Video Production Guidelines}
\label{sec:literature-review-of-video-production-guidelines}
According to our research process (see \figurename{ \ref{fig:fig1}}, \circled{\textbf{1}}), we started with a literature review on generic video production guidelines to deduce characteristics of videos. Although there are no universal rules for the production of high-quality videos, there are a lot of guiding recommendations \cite{Owens.2011}. These guidelines have been discovered through years of experience and thus represent best practices on how to produce a good video with specific characteristics \cite{Karras.2018, Owens.2011}. In particular, this literature review addresses the following research question:

\begin{mdframed}
	\begin{itemize}[leftmargin=-2.5mm]
		\item[] \textbf{RQ:}
		What characteristics of videos can be deduced from the recommendations of generic video production guidelines?
	\end{itemize}
\end{mdframed}

\subsection{Search Process}
We focused on generic video production guidelines that belong to gray literature which can provide valuable insights \cite{Mahood.2014, Kitchenham.2010}. According to Brings et al. \cite{Brings.2018}, gray literature is often not included in databases. Therefore, we decided to perform a web search using the two popular web search engines Google Scholar and Google as proposed by Mahood et al. \cite{Mahood.2014}. The search string was developed by using PICO (Population, Intervention, Comparison, Outcome) as suggested by Kitchenham and Charters \cite{Kitchenham.2007}. PICO helps to identify keywords to formulate a search string from the research question \cite{Petersen.2015}. PICO led to the following results and keywords (bold highlighting):

\begin{itemize}[leftmargin=3.5mm, topsep=0.5mm]
	\setlength\itemsep{-.25mm}
	\item \textit{Population}: We focus on the topic of \textbf{video production}.
	\item \textit{Intervention}: We investigate \textbf{guidelines}.
	\item \textit{Comparison}: We do not compare the intervention with anything else, but we analyze the content of the intervention.
	\item \textit{Outcome}: We expect to obtain a set of recommendations to produce generic videos.
\end{itemize}

In December $2017$, the first author of this article entered the resulting search string \enquote{video production guidelines} into both web search engines. He investigated the first $50$ results of each web search engine by applying the following exclusion and inclusion criteria (see section \ref{sec:exclusion-and-inclusion-criteria-lr1}) to each result. After the first $50$ results, the first author of this article scanned the further results but found none that met the exclusion and inclusion criteria. For this reason, the authors of this article decided to consider only the first $50$ results. The third author of this article reviewed the work of the first author.

\subsection{Exclusion and Inclusion Criteria}
\label{sec:exclusion-and-inclusion-criteria-lr1}
For each result, we applied the following criteria:\\

\noindent
\textit{Exclusion criteria}.
\begin{itemize}[leftmargin=8.5mm, topsep=0.5mm]
	\setlength\itemsep{-.25mm}
	\item[$EC_{1}$:] The result does not provide a downloadable document, e.g., a PDF file, representing a publication that was consciously created.
	\item[$EC_{2}$:] The document is not written in English.
	\item[$EC_{3}$:] The document is not provided by an official institution or one or more authors experienced in video production.
	\item[$EC_{4}$:] The document focuses only on a specific type of video, e.g., tutorial videos.
	\item[$EC_{5}$:] The document is only partially accessible.\\
\end{itemize}

\noindent
\textit{Inclusion criteria}.
\begin{itemize}[leftmargin=8.5mm, topsep=0.5mm]
	\setlength\itemsep{-.25mm}
	\item[$IC_{1}$:] The document contains an explicit statement that it is a guideline for video production.
	\item[$IC_{2}$:] The document contains an explicit list of recommendations for the production of videos.\\
\end{itemize}

\noindent
If none of the exclusion criteria $EC_{i}$, $1 \leq i \leq 5$ and both inclusion criteria $IC_{i}$, $i \in {1,2}$ were met, the result was selected: $$ \textrm{Result selected} \Leftrightarrow \neg (EC_{1} \lor EC_{2} \lor EC_{3} \lor EC_{4} \lor EC_{5}) \land (IC_{1} \land IC_{2})$$ 

\subsection{Data Collection and Analysis}
We analyzed the guidelines by performing manual coding according to Salda\~{n}a \cite{Saldana.2015}. Manual coding is a qualitative data analysis consisting of two coding cycles, each of which can be repeated iteratively. While the first coding cycle includes the initial coding of the data, the second cycle focuses on classifying, prioritizing, integrating, synthesizing, abstracting, and conceptualizing sub-categories as well as categories from the coded data. For each guideline, we extracted all coded passages into a spreadsheet to simplify the subsequent analysis.

\begin{figure}[htbp]
	\centering
	\includegraphics[width=.8\columnwidth]{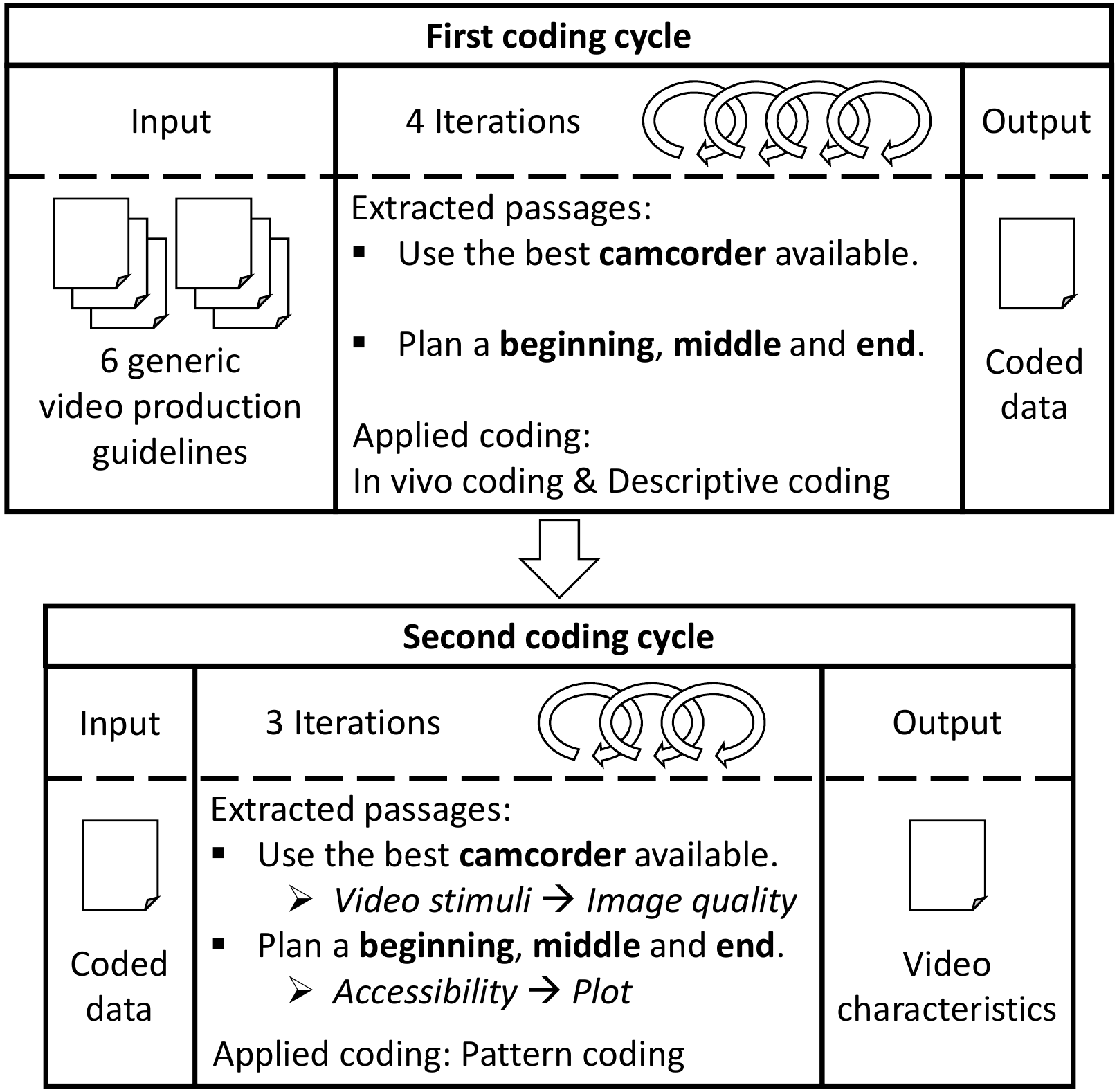}
	\caption{Video: Manual coding process}
	\label{fig:fig4}
\end{figure}

In \figurename{ \ref{fig:fig4}}, we show the manual coding process with two examples of extracted and coded passages. In the first coding cycle, we applied a combination of \textit{descriptive} and \textit{in vivo} coding. These two methods assign a word or phrase as a code to a passage (see \figurename{ \ref{fig:fig4}}, bold highlighting) in the qualitative data. While \textit{in vivo} codes are words or phrases found in the actual data, \textit{descriptive} codes are generated by the researcher. According to Salda\~{n}a \cite{Saldana.2015}, both methods are a good starting point for manual coding since they provide an essential groundwork for the second coding cycle. We focused on \textit{in vivo} coding to adhere to the video terminology in the guidelines. However, in the case of no recognizable terminology, we applied \textit{descriptive} coding. After four iterations of the first coding cycle, all three authors agreed on the extracted and coded passages. In the second coding cycle, we performed \textit{pattern} coding. This method groups the coded data into a smaller number of themes to develop sub-categories and categories from the data (see \figurename{ \ref{fig:fig4}}, italic highlighting). After three iterations of the second coding cycle, all three authors agreed on the identified categories and sub-categories which represent deduced video characteristics and sub-characteristics from the guidelines.

\subsection{Results}
\label{sec:results_video}
The search process resulted in six generic video production guidelines \cite{Owens.2011,MIT,Boise,Leeds,ARSC,Heath.2010} which we analyzed according to the previously described manual coding process. In the first coding cycle, we extracted $307$ passages and assigned a total of $586$ codes. In the second coding cycle, we grouped these codes into ten sub-categories which in turn led to four categories. Each identified sub-category is based on at least four out of the six guidelines to increase the validity of the results. \tablename{ \ref{tbl:coding_results}} shows the identified four categories/characteristics and ten sub-categories/sub-characteristics of videos with some exemplary codes, the respective coding frequencies, and the references from which the codes were extracted. 

We arranged the characteristics and sub-characteristics by the three dimensions of a quality model (\textit{representation}, \textit{content}, and \textit{impact}). In consideration of the important features of a quality model (see section \ref{sec:quality-model-for-vision-videos}), we selected \textit{familiar labels} and provided a \textit{concise definition} for each characteristic and the respective sub-characteristics (see \tablename{ \ref{tbl:dim_representation}}, \tablename{ \ref{tbl:dim_content1}}, \tablename{ \ref{tbl:dim_content2}}, and \tablename{ \ref{tbl:dim_impact}}). In the following, we briefly describe the individual characteristics and their respective sub-characteristics.

The first characteristic \textit{video stimuli} belongs to the \textit{representation} dimension (see \tablename{ \ref{tbl:dim_representation}}). This characteristic includes the sensorial stimuli of a video. The characteristic aggregates the three sub-characteristics \textit{image quality}, \textit{sound quality}, and \textit{video length} which a viewer perceives with his sensory organs.

\newcommand*{\MyIndent}{\hspace*{0.25cm}}%
\begin{table}[htbp]
	\captionsetup{justification=centering}
	\renewcommand{\arraystretch}{1.1}
	\centering
	\caption{Dimension: Representation -- Video stimuli}
	\label{tbl:dim_representation}
	\begin{tabular}{|l|l|l|l|l|}
		\hline
		\multicolumn{5}{|c|}{\textbf{Dimension: Representation}} \\
		\multicolumn{5}{|c|}{covers the sensorial characteristics of a video.} \\ \hline
		\multicolumn{5}{|p{8cm}|}{\raggedright \textbf{Characteristic:}\\ \textit{Video stimuli} considers the sensorial stimuli of a video.} \\ \hline
		\multicolumn{5}{|p{8cm}|}{\raggedright \textbf{Sub-characteristics:}\\ \textit{Image quality} considers the visual quality of the image of a video.\\ \textit{Sound quality} considers the auditory quality of the sound of a video.\\ \textit{Video length} considers the duration of a video.} \\ \hline
	\end{tabular}
\end{table}

The second characteristic is \textit{accessibility} which is part of the \textit{content} dimension (see \tablename{ \ref{tbl:dim_content1}}). \textit{Accessibility} focuses on the ease of access to the content of a video by containing the sub-characteristics \textit{plot} and \textit{prior knowledge}. A video is easier or harder to access depending on the structuring of the content and the presupposed prior knowledge.

\begin{table}[htbp]
	\captionsetup{justification=centering}
	\renewcommand{\arraystretch}{1.1}
	\centering
	\caption{Dimension: Content -- Accessibility}
	\label{tbl:dim_content1}
	\begin{tabular}{|l|l|l|l|l|}
		\hline
		\multicolumn{5}{|c|}{\textbf{Dimension: Content}} \\
		\multicolumn{5}{|c|}{covers the perceptual characteristics of a video.} \\ \hline
		\multicolumn{5}{|p{8cm}|}{\raggedright \textbf{Characteristic:}\\ \textit{Accessibility} considers the ease of access to the content of a video.} \\ \hline
		\multicolumn{5}{|p{8cm}|}{\raggedright \textbf{Sub-characteristics:}\\ \textit{Plot} considers the structured presentation of the content of a video.\\ \textit{Prior knowledge} considers the presupposed prior knowl- edge to understand the content of a video.} \\ \hline
	\end{tabular}
\end{table}

\textit{Relevance} is the third video characteristic and also belongs to the \textit{content} dimension (see \tablename{ \ref{tbl:dim_content2}}). This characteristic considers the presentation of valuable information in a video. The characteristic includes the sub-characteristics \textit{essence} and \textit{clutter} which distinguish between important core elements as well as disrupting and distracting elements. Important core elements, e.g., persons and locations, are planned content that a video producer wants to show to the target audience of the video. Disrupting and distracting elements, e.g., background actions and noises, are unintended contents that may distract the viewer from the important content of a video. While the important core elements are deliberately recorded since they are intended to be visible in a video, disrupting elements, which should not be included in a video, are inadvertently recorded.

\begin{table}[htbp]
	\captionsetup{justification=centering}
	\renewcommand{\arraystretch}{1.1}
	\centering
	\caption{Dimension: Content -- Relevance}
	\label{tbl:dim_content2}
	\begin{tabular}{|l|l|l|l|l|}
		\hline
		\multicolumn{5}{|c|}{\textbf{Dimension: Content}} \\
		\multicolumn{5}{|c|}{covers the perceptual characteristics of a video.} \\ \hline
		\multicolumn{5}{|p{8cm}|}{\raggedright \textbf{Characteristic:}\\ \textit{Relevance} considers the presentation of valuable content in a video.} \\ \hline
		\multicolumn{5}{|p{8cm}|}{\raggedright \textbf{Sub-characteristics:}\\ \textit{Essence} considers the important core elements, e.g., per- sons, locations, and entities, which are to be presented in a video.\\ \textit{Clutter} considers the disrupting and distracting elements, e.g., background actions and noises, that can be inadver- tently recorded in a video.} \\ \hline
	\end{tabular}
\end{table}

The fourth characteristic \textit{attitude} belongs to the \textit{impact} dimension (see \tablename{ \ref{tbl:dim_impact}}) since the included sub-characteristics \textit{pleasure}, \textit{sense of responsibility}, and \textit{intention} concern the emotional impact of a video. \textit{Pleasure} focuses on the enjoyment of watching a video. A video needs to be pleasant to watch to be interesting for its audience. In the context of video production and use, the \textit{sense of responsibility} is a crucial concern. The production and use need to comply with the legal regulations to ensure legal reliability for all involved parties who produce or use a video. A potential producer or viewer may reject a video as documentation option due to legal uncertainty. \textit{Intention} considers the intended purpose of a video. Especially, the intended purpose has a strong impact on a video and its content since it defines the reason why the video is necessary. In the following section \ref{sec:intentions-of-videos}, we outline the intents of videos in more detail.

\begin{table}[htbp]
	\captionsetup{justification=centering}
	\renewcommand{\arraystretch}{1.1}
	\centering
	\caption{Dimension: Impact -- Attitude}
	\label{tbl:dim_impact}
	\begin{tabular}{|l|l|l|l|l|}
		\hline
		\multicolumn{5}{|c|}{\textbf{Dimension: Impact}} \\
		\multicolumn{5}{|c|}{covers the emotional characteristics of a video.} \\ \hline
		\multicolumn{5}{|p{8cm}|}{\raggedright \textbf{Characteristic:}\\ \textit{Attitude} considers the humans' conception of a video.} \\ \hline
		\multicolumn{5}{|p{8cm}|}{\raggedright \textbf{Sub-characteristics:}\\ \textit{Pleasure} considers the enjoyment of watching a video.\\ \textit{Intention} considers the intended purpose of a video.\\ \textit{Sense of responsibility} considers the compliance of a video with the legal regulations.} \\ \hline
	\end{tabular}
\end{table}

\begin{table*}[!t]
	\captionsetup{justification=centering}
	\renewcommand{\arraystretch}{1.1}
	\centering
	\caption{Video: Manual coding results -- Categories/characteristics, sub-categories/sub-characteristics, and exemplary codes}
	\label{tbl:coding_results}
	\begin{tabular}{clll|c|c|}
		\hline
		\multicolumn{1}{|l}{\cellcolor{gray!45}} & \multicolumn{3}{l|}{\cellcolor{gray!45}\textit{Representation}} & \multicolumn{1}{c|}{\textbf{Coding frequency}} & \multicolumn{1}{c|}{\textbf{Extracted from}} \\ \hhline{|>{\arrayrulecolor{gray!45}}->{\arrayrulecolor{black}}|*{5}{-}|}
		\multicolumn{1}{|c|}{\cellcolor{gray!45}} & \cellcolor{gray!30} & \multicolumn{2}{l|}{\cellcolor{gray!30}\textit{Video stimuli}} & \multicolumn{1}{c|}{189} & \multicolumn{1}{c|}{\cite{Owens.2011, MIT, Boise, Leeds, ARSC, Heath.2010}} \\ \hhline{|>{\arrayrulecolor{gray!45}}->{\arrayrulecolor{black}}|>{\arrayrulecolor{gray!30}}->{\arrayrulecolor{black}}|*{4}{-}|}
		\multicolumn{1}{|c|}{\cellcolor{gray!45}} & \multicolumn{1}{l|}{\cellcolor{gray!30}} & \cellcolor{gray!15} & \textit{\cellcolor{gray!15}Image quality} & \multicolumn{1}{c|}{105} & \multicolumn{1}{c|}{\cite{Owens.2011, MIT, Boise, Leeds, ARSC, Heath.2010}} \\ \cline{4-6} 
		\multicolumn{1}{|c|}{\cellcolor{gray!45}} & \multicolumn{1}{l|}{\cellcolor{gray!30}} & \cellcolor{gray!15} & \multicolumn{3}{|p{15cm}|}{ \raggedright \footnotesize{ \textbf{Exemplary codes:} Wide shots, tight shots, long and short shots, camera, auto focus, white balance, automatic exposure controls, camcorder, point of view between shots, shooting height, lighting, graphic acquisition, H.264, legibility, resolution, compression, aspect ratio, high definition, format, mp4, Full HD, frame rate, zoom, equipment, framing, focus, stability, tripod, image, good camerawork}} \\ \hhline{|>{\arrayrulecolor{gray!45}}->{\arrayrulecolor{black}}|>{\arrayrulecolor{gray!30}}->{\arrayrulecolor{black}}|>{\arrayrulecolor{gray!15}}->{\arrayrulecolor{black}}*{3}{-}|}
		\multicolumn{1}{|c|}{\cellcolor{gray!45}} & \multicolumn{1}{l|}{\cellcolor{gray!30}} & \cellcolor{gray!15} & \cellcolor{gray!15}\textit{Sound quality} & \multicolumn{1}{c|}{54} & \multicolumn{1}{c|}{\cite{Owens.2011, MIT, Boise, Leeds, ARSC, Heath.2010}} \\ \cline{4-6} 
		\multicolumn{1}{|c|}{\cellcolor{gray!45}} & \multicolumn{1}{l|}{\cellcolor{gray!30}} & \cellcolor{gray!15} & \multicolumn{3}{|p{15cm}|}{ \raggedright \footnotesize{ \textbf{Exemplary codes:} Equipment, external microphone, sound level, stereo, shotgun microphone, audio, sound production, music, sound effects, ambient sound, check sound quality, avoid loud sounds, microphone distance, omni-directional, ambient sound, background music, environment noise, clear audible sound, loud speaking, matching of sound and image, AAC-LC, sound is essential, spoken word}} \\ \hhline{|>{\arrayrulecolor{gray!45}}->{\arrayrulecolor{black}}|>{\arrayrulecolor{gray!30}}->{\arrayrulecolor{black}}|>{\arrayrulecolor{gray!15}}->{\arrayrulecolor{black}}*{3}{-}|}
		\multicolumn{1}{|c|}{\cellcolor{gray!45}} & \multicolumn{1}{l|}{\cellcolor{gray!30}} & \cellcolor{gray!15} & \cellcolor{gray!15}\textit{Video length} & \multicolumn{1}{c|}{30} & \multicolumn{1}{c|}{\cite{Owens.2011, MIT, ARSC, Heath.2010}} \\ \cline{4-6} 
		\multicolumn{1}{|c|}{\multirow{-8}{*}{\rotatebox[origin=c]{90}{\cellcolor{gray!45}\textbf{Dimension}}}} & \multicolumn{1}{l|}{\multirow{-8}{*}{\rotatebox[origin=c]{90}{\cellcolor{gray!30}\textbf{Category}}}} & \multirow{-8}{*}{\rotatebox[origin=c]{90}{\cellcolor{gray!15}\textbf{Sub-category}}} & \multicolumn{3}{|p{15cm}|}{ \raggedright \footnotesize{ \textbf{Exemplary codes:} Number of new scenes/minutes, length of final tape, longer than 5 minutes, record additionally 5-10 seconds, short-format projects, longer-format projects (over 30 minutes), not exceed five minutes length, hold shot for some time, how long?}} \\ \hline \hline
		
		\multicolumn{1}{|l}{\cellcolor{gray!45}} & \multicolumn{3}{l|}{\cellcolor{gray!45}\textit{Content}} & \multicolumn{1}{c|}{\textbf{Coding frequency}} & \multicolumn{1}{c|}{\textbf{Extracted from}} \\ \hhline{|>{\arrayrulecolor{gray!45}}->{\arrayrulecolor{black}}|*{5}{-}|}
		\multicolumn{1}{|c|}{\cellcolor{gray!45}} & \cellcolor{gray!30} & \multicolumn{2}{l|}{\cellcolor{gray!30}\textit{Accessibility}} & \multicolumn{1}{c|}{79} & \multicolumn{1}{c|}{\cite{Owens.2011, MIT, Boise, Leeds, ARSC, Heath.2010}} \\ \hhline{|>{\arrayrulecolor{gray!45}}->{\arrayrulecolor{black}}|>{\arrayrulecolor{gray!30}}->{\arrayrulecolor{black}}|*{4}{-}|}
		\multicolumn{1}{|c|}{\cellcolor{gray!45}} & \multicolumn{1}{l|}{\cellcolor{gray!30}} & \cellcolor{gray!15} & \cellcolor{gray!15}\textit{Plot} & \multicolumn{1}{c|}{59} & \multicolumn{1}{c|}{\cite{Owens.2011, MIT, Boise, Leeds, ARSC}} \\ \cline{4-6} 
		\multicolumn{1}{|c|}{\cellcolor{gray!45}} & \multicolumn{1}{l|}{\cellcolor{gray!30}} & \cellcolor{gray!15} & \multicolumn{3}{|p{15cm}|}{ \raggedright \footnotesize{ \textbf{Exemplary codes:} Tell a story, composition, scheduling, script or event, storyline, timing, tempo, vision and goal for the production, main action of each scene, storyboard, guide audience's thought process, shooting plan, outline, action, introduction \& end, storytelling}} \\ \hhline{|>{\arrayrulecolor{gray!45}}->{\arrayrulecolor{black}}|>{\arrayrulecolor{gray!30}}->{\arrayrulecolor{black}}|>{\arrayrulecolor{gray!15}}->{\arrayrulecolor{black}}*{3}{-}|}
		\multicolumn{1}{|c|}{\cellcolor{gray!45}} & \multicolumn{1}{l|}{\cellcolor{gray!30}} & \cellcolor{gray!15} & \cellcolor{gray!15}\textit{Prior knowledge} & \multicolumn{1}{c|}{20} & \multicolumn{1}{c|}{\cite{Owens.2011, MIT, ARSC, Heath.2010}} \\ \cline{4-6} 
		\multicolumn{1}{|c|}{\cellcolor{gray!45}} & \multicolumn{1}{l|}{\cellcolor{gray!30}} & \multirow{-4}{*}{\rotatebox[origin=c]{90}{\cellcolor{gray!15}\textbf{Sub-category}}} & \multicolumn{3}{|p{15cm}|}{ \raggedright \footnotesize{ \textbf{Exemplary codes:} Target audience, appropriate to audience, accuracy and grammar, who is the program being made for?, familiarity, how the audience react, what is the program about, puzzling, do not know, specific background, level: basic or intermediate or advanced}} \\ \hhline{|>{\arrayrulecolor{gray!45}}->{\arrayrulecolor{black}}|>{\arrayrulecolor{gray!30}}->{\arrayrulecolor{black}}|*{4}{-}|}
		\multicolumn{1}{|c|}{\cellcolor{gray!45}} & \cellcolor{gray!30} & \multicolumn{2}{l|}{\cellcolor{gray!30}\textit{Relevance}} & \multicolumn{1}{c|}{145} & \multicolumn{1}{c|}{\cite{Owens.2011, MIT, Boise, Leeds, ARSC, Heath.2010}} \\ \hhline{|>{\arrayrulecolor{gray!45}}->{\arrayrulecolor{black}}|>{\arrayrulecolor{gray!30}}->{\arrayrulecolor{black}}|*{4}{-}|}
		\multicolumn{1}{|c|}{\cellcolor{gray!45}} & \multicolumn{1}{l|}{\cellcolor{gray!30}} & \cellcolor{gray!15} & \cellcolor{gray!15}\textit{Essence} & \multicolumn{1}{c|}{91} & \multicolumn{1}{c|}{\cite{Owens.2011, MIT, Boise, Leeds, ARSC, Heath.2010}} \\ \cline{4-6}
		\multicolumn{1}{|c|}{\cellcolor{gray!45}} & \multicolumn{1}{l|}{\cellcolor{gray!30}} & \cellcolor{gray!15} & \multicolumn{3}{|p{15cm}|}{ \raggedright \footnotesize{ \textbf{Exemplary codes:} Actions, location, primary themes, background, best illustrate your strategy, people, places, relevant shots, message, rolls, natural opportunities, subjects, ideas, information, atmosphere, real subject, broad concepts, person, object, key shots, what you want to cover, details, impression of reality, features of subject, topics, arguments, explanation, treatment, foreground, symbolism}} \\ \hhline{|>{\arrayrulecolor{gray!45}}->{\arrayrulecolor{black}}|>{\arrayrulecolor{gray!30}}->{\arrayrulecolor{black}}|>{\arrayrulecolor{gray!15}}->{\arrayrulecolor{black}}*{3}{-}|}
		\multicolumn{1}{|c|}{\cellcolor{gray!45}} & \multicolumn{1}{l|}{\cellcolor{gray!30}} & \cellcolor{gray!15} & \cellcolor{gray!15}\textit{Clutter} & \multicolumn{1}{c|}{54} & \multicolumn{1}{c|}{\cite{Owens.2011, MIT, Boise, Leeds, ARSC, Heath.2010}} \\ \cline{4-6}
		\multicolumn{1}{|c|}{\multirow{-12}{*}{\rotatebox[origin=c]{90}{\cellcolor{gray!45}\textbf{Dimension}}}} & \multicolumn{1}{c|}{\multirow{-12}{*}{\rotatebox[origin=c]{90}{\cellcolor{gray!30}\textbf{Category}}}} & \multirow{-4}{*}{\rotatebox[origin=c]{90}{\cellcolor{gray!15}\textbf{Sub-category}}} & \multicolumn{3}{|p{15cm}|}{ \raggedright \footnotesize{ \textbf{Exemplary codes:} Limitations, out of range, masking sounds, mistakes, too much information, reflections, windows, posters, flashing signs, distraction, attract attention, nearby noise, loud background noise, repetitious or boring parts, meaningless background, look at wrong things, simple background, fading, dead air, no ambient noise, messy background, cluttered room, ill-placed items, not cluttered}} \\ \hline \hline
		
		\multicolumn{1}{|l}{\cellcolor{gray!45}} & \multicolumn{3}{l|}{\cellcolor{gray!45}\textit{Impact}} & \multicolumn{1}{c|}{\textbf{Coding frequency}} & \multicolumn{1}{c|}{\textbf{Extracted from}} \\ \hhline{|>{\arrayrulecolor{gray!45}}->{\arrayrulecolor{black}}|*{5}{-}|}
		\multicolumn{1}{|c|}{\cellcolor{gray!45}} & \cellcolor{gray!30} & \multicolumn{2}{l|}{\cellcolor{gray!30}\textit{Attitude}} & \multicolumn{1}{c|}{173} & \multicolumn{1}{c|}{\cite{Owens.2011, MIT, Boise, Leeds, ARSC, Heath.2010}} \\ \hhline{|>{\arrayrulecolor{gray!45}}->{\arrayrulecolor{black}}|>{\arrayrulecolor{gray!30}}->{\arrayrulecolor{black}}|*{4}{-}|}
		\multicolumn{1}{|c|}{\cellcolor{gray!45}} & \multicolumn{1}{l|}{\cellcolor{gray!30}} & \cellcolor{gray!15} & \cellcolor{gray!15}\textit{Pleasure} & \multicolumn{1}{c|}{69} & \multicolumn{1}{c|}{\cite{Owens.2011, MIT, Boise, Leeds, ARSC, Heath.2010}} \\ \cline{4-6} 
		\multicolumn{1}{|c|}{\cellcolor{gray!45}} & \multicolumn{1}{l|}{\cellcolor{gray!30}} & \cellcolor{gray!15} & \multicolumn{3}{|p{15cm}|}{ \raggedright \footnotesize{ \textbf{Exemplary codes:} Capture interest, create special effects and mood, natural eye movement, hold viewer's interest, avoid motion sickness, impact, interesting and persuasive, convincingly, attention getting, audience impact, illusion of excitement, pictures for pleasure, enjoy, audience appeal, valid purpose, emotion and drama, imagination, overall impact, feeling of completeness, powerful}} \\ \hhline{|>{\arrayrulecolor{gray!45}}->{\arrayrulecolor{black}}|>{\arrayrulecolor{gray!30}}->{\arrayrulecolor{black}}|>{\arrayrulecolor{gray!15}}->{\arrayrulecolor{black}}*{3}{-}|}
		\multicolumn{1}{|c|}{\cellcolor{gray!45}} & \multicolumn{1}{l|}{\cellcolor{gray!30}} & \cellcolor{gray!15} & \cellcolor{gray!15}\textit{Intention} & \multicolumn{1}{c|}{21} & \multicolumn{1}{c|}{\cite{Owens.2011, MIT, Boise, Leeds, ARSC, Heath.2010}} \\ \cline{4-6}
		\multicolumn{1}{|c|}{\cellcolor{gray!45}} & \multicolumn{1}{l|}{\cellcolor{gray!30}} & \cellcolor{gray!15} & \multicolumn{3}{|p{15cm}|}{ \raggedright \footnotesize{ \textbf{Exemplary codes:} Introduce, demonstrate, fun, informational, documentation, training, entertainment, education, genre, promote, convey message, instructions, purpose of program, advertising, main purpose, production's impact, chief purpose, target audience}} \\ \hhline{|>{\arrayrulecolor{gray!45}}->{\arrayrulecolor{black}}|>{\arrayrulecolor{gray!30}}->{\arrayrulecolor{black}}|>{\arrayrulecolor{gray!15}}->{\arrayrulecolor{black}}*{3}{-}|}
		\multicolumn{1}{|c|}{\cellcolor{gray!45}} & \multicolumn{1}{l|}{\cellcolor{gray!30}} & \cellcolor{gray!15} & \cellcolor{gray!15}\textit{Sense of responsibility} & \multicolumn{1}{c|}{83} & \multicolumn{1}{c|}{\cite{Owens.2011, MIT, Boise, Leeds, ARSC, Heath.2010}} \\ \cline{4-6} 
		\multicolumn{1}{|c|}{\multirow{-7}{*}{\rotatebox[origin=c]{90}{\cellcolor{gray!45}\textbf{Dimension}}}} & \multicolumn{1}{l|}{\multirow{-7}{*}{\rotatebox[origin=c]{90}{\cellcolor{gray!30}\textbf{Category}}}} & \multirow{-7}{*}{\rotatebox[origin=c]{90}{\cellcolor{gray!15}\textbf{Sub-category}}} & \multicolumn{3}{|p{15cm}|}{ \raggedright \footnotesize{ \textbf{Exemplary codes:} Distribution, copyright, credits, protected by copyright, fonts, management guidelines, health and safety, permissions, consent form, recording rights, publishing rights, performer rights, product names and labels, parental permission, brand, approval process, privacy, copyright of video, comply to permissions, legalities, private property, danger factor, insurance, anonymity}} \\ \hline
	\end{tabular}
\end{table*}

\begin{figure}[htbp]
	\centering
	\includegraphics[width=\columnwidth]{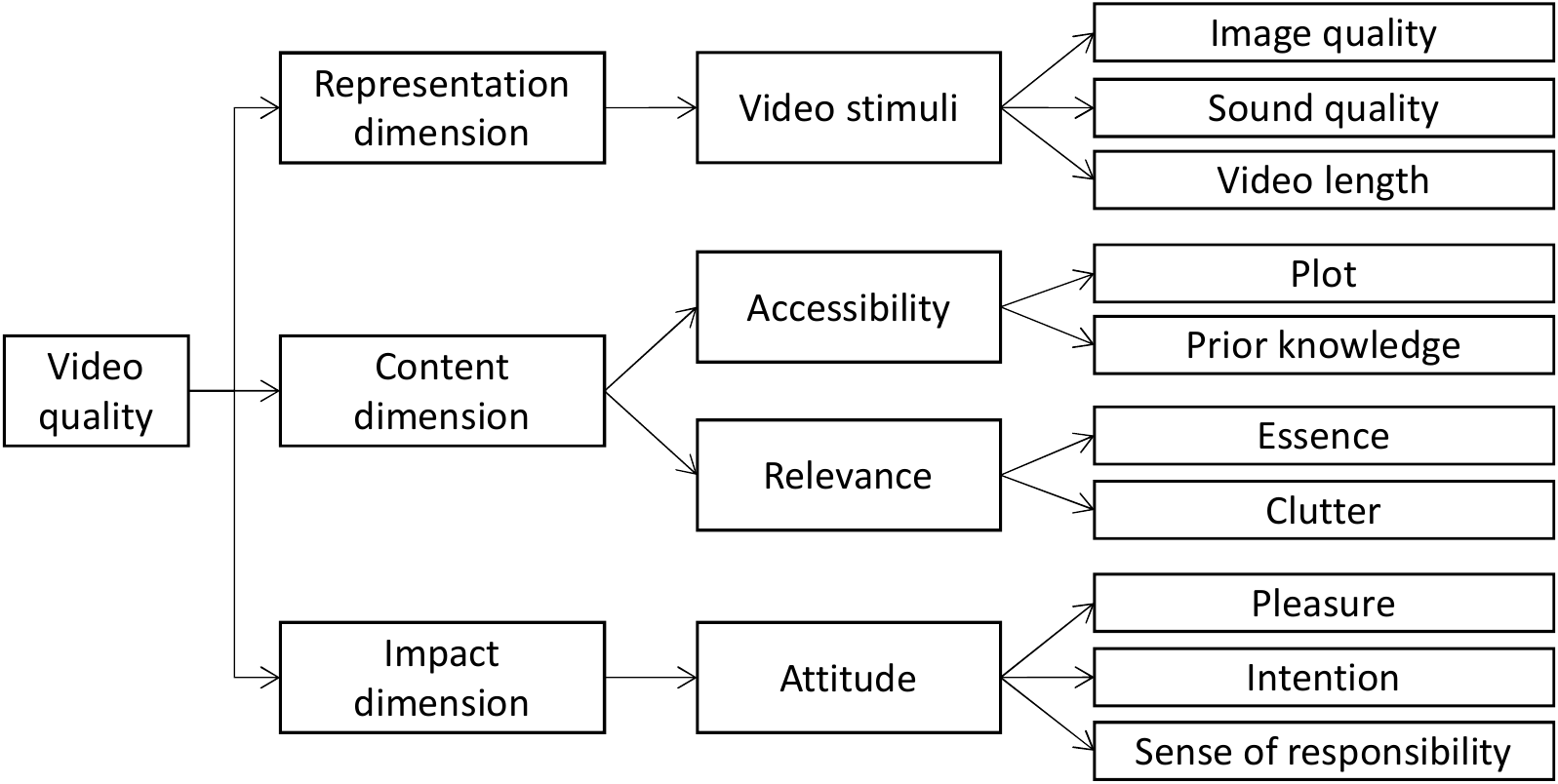}
	\caption{Hierarchical decomposition of video quality}
	\label{fig:fig6}
\end{figure}

We verified the identified video sub-characteristics by asking two raters to assign the identified sub-categories to the extracted passages on their own. The two raters are computer science researchers who have both been working in the software engineering group at Leibniz Universität Hannover for more than three years. Both researchers are familiar with RE but do not have much experience in video production. In this way, we ensured that the raters had a comparable perspective as the initial coders when assigning the sub-categories to the extracted passages. We evaluated the reliability of the raters' classification by using \textit{Cohen's kappa} \cite{Cohen.1960}. \textit{Cohen's kappa} is a robust measure of agreement between two raters since it takes into account the possibility of agreement occurring by chance. The calculated \textit{Cohen's kappa} value was $0.81$ which shows an almost perfect raters' agreement according to Landis and Koch \cite{Landis.1977}. Thus, we are confident that the identified sub-categories represent major themes in terms of video sub-characteristics in the analyzed guidelines. In \figurename{ \ref{fig:fig6}}, we present the hierarchical decomposition of the determined video (sub-) characteristics. As an answer to the research question, we can summarize:

\begin{mdframed}
	\begin{itemize}[leftmargin=-2.5mm]
		\item[] \textbf{Answer:}
		According to our analysis, we deduced the four video characteristics \textit{video stimuli}, \textit{accessibility}, \textit{relevance}, and \textit{attitude} and the ten sub-characteristics \textit{image quality}, \textit{sound quality}, \textit{video length}, \textit{plot}, \textit{prior knowledge}, \textit{essence}, \textit{clutter}, \textit{pleasure}, \textit{intention}, and \textit{sense of responsibility} from the recommendations of generic video production guidelines. 
	\end{itemize}
\end{mdframed}

\subsection{Intentions of Videos}
\label{sec:intentions-of-videos}
According to Femmer and Vogelsang \cite{Femmer.2018}, the quality of RE artifacts is essentially influenced by their specific purposes, all of which are intended to support a variety of stakeholders in their diverse project activities. For this reason, it is important to take a closer look at the intentions of videos to understand for which potential purposes videos can be used in RE.

Hanjalic et al. \cite{Hanjalic.2012} explained that it is important to understand \textit{why} a video is necessary to satisfy the viewers' information needs. This \textit{why} is understood as the reason, purpose, or immediate goal behind the application of a video and thus represents the underlying intention \cite{Hanjalic.2012}. Hanjalic et al. \cite{Hanjalic.2012} found the following five high-level intended purposes of videos:

\begin{itemize}[leftmargin=0mm, topsep=0.5mm]
	\setlength\itemsep{-.25mm}
	\item[] $(1)$ \textit{Information: Convey or obtain knowledge and/or new information (declarative knowledge)}.\\
	This intent covers the cases which have the goal of conveying or obtaining declarative knowledge, i.e., `knowing that'.
	\item[] $(2)$ \textit{Experience Learning: Convey or obtain skills or something practically by experience (procedural knowledge)}.\\	
	This intent covers knowledge acquisition regarding skills or procedural knowledge, i.e., `knowing how'.
	\item[] $(3)$ \textit{Experience Exposure: Convey or obtain particular experiences. The video serves as a replacement of an actual person, place, entity, or event}.\\
	This intent covers the cases which have the goal of conveying or obtaining exposure to real-life or another type of experience.
	\item[] $(4)$ \textit{Affect: Convey or obtain a mood or affective state. The video serves for relaxation or entertainment purpose}.\\
	This intent covers the cases which have the goal of conveying or obtaining a mood, affective, or physical state including relaxation and in general other effects of entertainment.
	\item[] $(5)$ \textit{Object: Convey or obtain content in form of a video to serve a particular purpose in a real-life situation}.\\
	This intent covers the cases which have the goal of conveying or obtaining a video suitable for a particular real-world situation.
\end{itemize}

The first four types of intended purposes have two generalizations in common \cite{Hanjalic.2012}. First, there is probably a large convergence between the producer's and viewer's main intended purpose of a video. Hanjalic et al. illustrate this statement with a series of examples: \enquote{News shows are made mainly to inform people; tutorials are made mainly to teach people skills; events are captured to expose people to experiences and films are produced to entertain} \cite[p. 4]{Hanjalic.2012}. Second, the goals of these four intents can also be accomplished without a video. Hanjalic et al. substantiate this statement with the following examples: \enquote{Information can be acquired from books and newspapers; skills can be acquired by taking lessons from a teacher; exposure can be gained by experiencing real-life persons, places, and things; and, finally users can change their affective state with live entertainment such as a concert or a play} \cite[p. 4]{Hanjalic.2012}. The difference between these four intents becomes more obvious by asking which real-world activity could help to achieve the same goal.

The fifth intent differs from the other ones in both generalizations \cite{Hanjalic.2012}. First, the producer's and viewer's intent may diverge radically from each other. This divergence is caused by the difference in the second generalization. The fifth intent covers the goals that necessarily require a video to be achieved. Thus, these goals cannot be accomplished by other means. Therefore, the viewer is looking for a video that fulfills his needs and intention which do not have to match with the producer's ones. Such a mismatch is possible since the five high-level intents are not mutually exclusive but overlap \cite{Hanjalic.2012}.

\section{Literature Review on Software Project Vision}
\label{sec:literature-review-on-vision-characteristics}
The previous elaboration of characteristics and intentions is based on the consideration of video as a representation format. Since we develop a specific quality model for vision videos, it is necessary to look more closely at the content, i.e., a vision, to substantiate the quality model that is generic so far. Therefore, we performed a second literature review on software project vision according to our research process (see \figurename{ \ref{fig:fig1}}, \circled{\textbf{2}}). This literature review addresses the following research question:

\begin{mdframed}
	\begin{itemize}[leftmargin=-2.5mm]
		\item[] \textbf{RQ:}
		What are the characteristics of a software project vision according to literature?
	\end{itemize}
\end{mdframed}

\subsection{Search Process}
The search process was a manual search in the internal library of the software engineering group at Leibniz Universität Hannover. We decided to perform a manual search instead of electronic search since different researchers \cite{Petersen.2015, Kitchenham.2013, Imtiaz.2013} strongly advocated the use of manual search due to its benefits of being more effective in identifying relevant literature. In December $2017$, the first author of this article checked all $428$ books in the internal library of the software engineering group by applying the following exclusion and inclusion criteria (see section \ref{sec:exclusion-and-inclusion-criteria-lr2}). The third author of this article reviewed the work of the first author. The third author also extended the identified literature by suggesting additional journal articles and books on software project vision which he knew due to his work on the book \enquote{Software Product Management: The ISPMA-Compliant Study Guide and Handbook} \cite{Kittlaus.2017} in which he has contributed to a chapter on software project vision.

\subsection{Exclusion and Inclusion Criteria}
\label{sec:exclusion-and-inclusion-criteria-lr2}
For each publication of the manual search and those proposed by the third author, we applied the following criteria:\\

\noindent
\textit{Exclusion criteria}.
\begin{itemize}[leftmargin=8.5mm, topsep=0.5mm]
	\setlength\itemsep{-.25mm}
	\item[$EC_{1}$:] The publication is neither a book nor a journal article.
	\item[$EC_{2}$:] The publication is not written in English.
	\item[$EC_{3}$:] The publication is only partially accessible.\\
\end{itemize}

\noindent
\textit{Inclusion criteria}.
\begin{itemize}[leftmargin=8.5mm, topsep=0.5mm]
	\setlength\itemsep{-.25mm}
	\item[$IC_{1}$:] The publication contains an individual section on software project vision.
	\item[$IC_{2}$:] The publication addresses the topic of vision and its influence, e.g., on a (software) project.\\
\end{itemize}

\noindent
If none of the exclusion criteria $EC_{i}$, $1 \leq i \leq 3$ and at least one of the two inclusion criteria $IC_{i}$, $i \in {1,2}$ were met, the publication was selected: $$ \textrm{Publication selected} \Leftrightarrow \neg (EC_{1} \lor EC_{2} \lor EC_{3}) \land (IC_{1} \lor IC_{2})$$ 

\subsection{Data Collection and Analysis}
As in the first literature review, we analyzed the selected publications by performing manual coding \cite{Saldana.2015}. For each document, we extracted all coded passages into a spreadsheet to simplify the subsequent analysis.

\begin{figure}[htbp]
	\centering
	\includegraphics[width=.8\columnwidth]{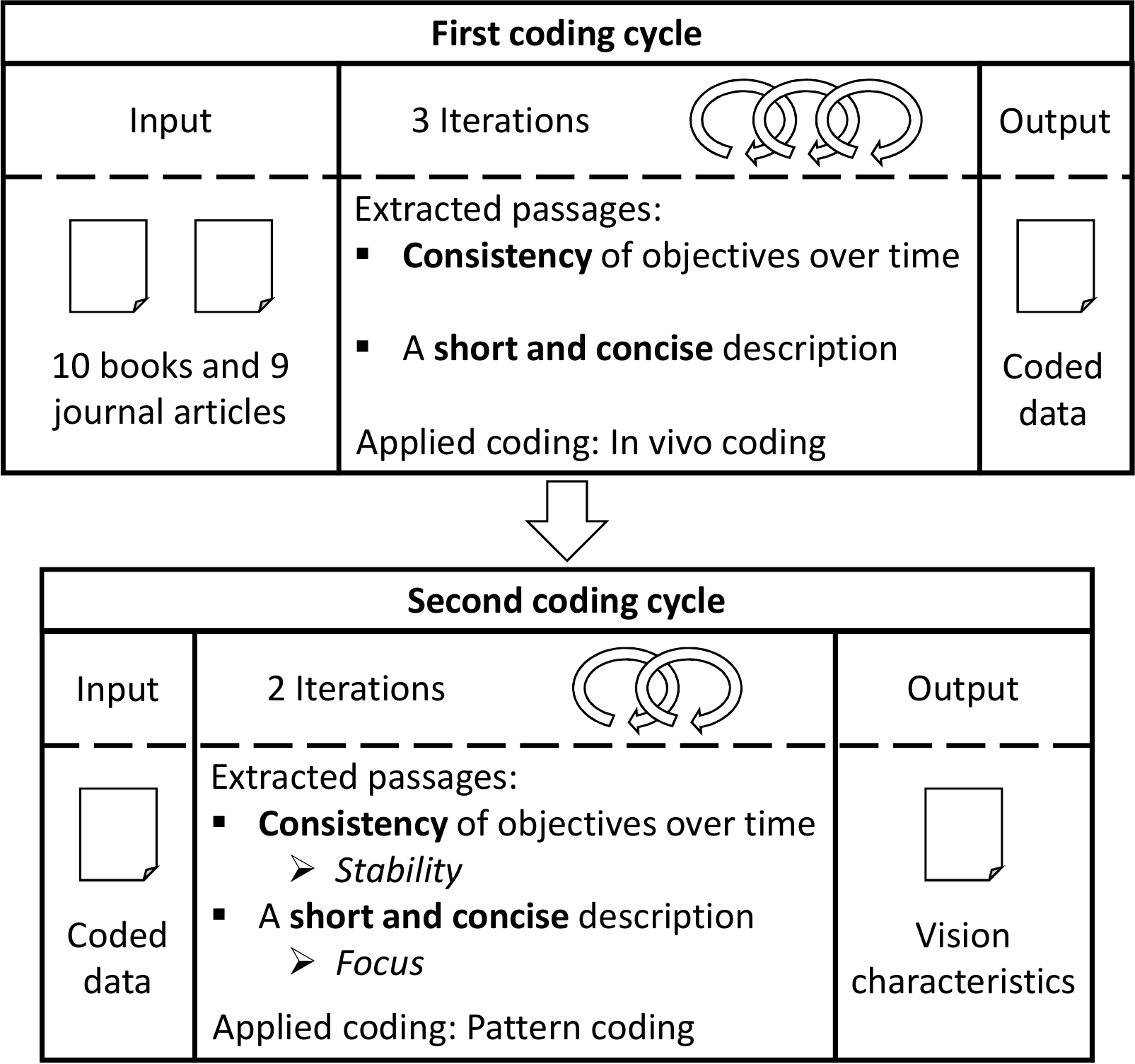}
	\caption{Vision: Manual coding process}
	\label{fig:coding_lr2}
\end{figure}

In \figurename{ \ref{fig:coding_lr2}}, we illustrate the manual coding process with two examples of extracted and coded passages. In the first coding cycle, we applied \textit{in vivo} coding to adhere to the terminology in the literature (see \figurename{ \ref{fig:coding_lr2}}, bold highlighting). After three iterations of the first coding cycle, all three authors agreed on the extracted and coded passages. In the second coding cycle, we used \textit{pattern} coding to group the data into a smaller number of themes to develop categories (see \figurename{ \ref{fig:coding_lr2}}, italic highlighting). After two iterations, all three authors agreed on the identified categories which represent deduced vision characteristics from the literature.

\subsection{Results}
\label{sec:results_vision}
The search process resulted in four books \cite{Wiegers.2003, Pohl.2010, Alexander.2005, Robertson.2012} from the internal library of the software engineering group as well as nine journal articles \cite{Lynn.2001, Brown.1995, Tessarolo.2007, Dyer.1999, Kessler.1996, Lynn.1999, Song.1998, Hamel.2005, Slater.1995} and six books \cite{Kittlaus.2017, Moore.1991, McGrath.2001, Crawford.2008, Clark.2010, Aaker.2009} proposed by the third author. In total, we identified $19$ relevant publications. In the first coding cycle, we extracted $46$ passages and assigned a total of $55$ codes. In the second coding cycle, we grouped these codes into five categories. Each category is based on at least five different publications to increase the validity of the results. In \tablename{ \ref{tbl:coding_results_lr2}}, we show the identified five categories/characteristics of a vision with some exemplary codes, the respective coding frequencies, and the references from which the codes were extracted.

The identified characteristics were arranged by the three dimensions of a quality model (\textit{representation}, \textit{content}, and \textit{impact}). As for the video characteristics, we selected \textit{familiar labels} and provided a \textit{concise definition} for each vision characteristic (see \tablename{ \ref{tbl:dim_representation_lr2}, \tablename{ \ref{tbl:dim_content1_lr2}}, \tablename{ \ref{tbl:dim_content2_lr2}}, \tablename{ \ref{tbl:dim_impact1_lr2}}, and \tablename{ \ref{tbl:dim_impact2_lr2}}). Below, we elaborate on the characteristics of a vision in detail.

The first vision characteristic is \textit{focus} which belongs to the \textit{representation} dimension (see \tablename{ \ref{tbl:dim_representation_lr2}}). According to McGrath \cite{McGrath.2001}, a vision needs to describe its essence compactly. A vision statement is documented in a few sentences or as a short text, no more than one page \cite{Kittlaus.2017}. Pohl \cite{Pohl.2010} also stated the need for a condensed and short description of the essence of an aspired change. The vision statement is a concise summary of the long-term goal of a new product \cite{Wiegers.2003}. However, a vision does not need to be presented as text. Alexander and Maiden \cite{Alexander.2005} suggested expressing a vision as a picture that describes the essence of a high-level goal. Such a model-based view of a new work practice synthesizes changes in the work structure and technology. No detailed requirements are included in a vision to keep the vision compact \cite{Wiegers.2003}. The definition of such information requires the consultation of customers, users, domain experts, documents, and existing systems \cite{Pohl.2010}. Thus, a vision shows a conceptual image of the future product as a \enquote{big picture} \cite{Kittlaus.2017, Alexander.2005}.

\begin{table}[htbp]
	\captionsetup{justification=centering}
	\renewcommand{\arraystretch}{1.1}
	\centering
	\caption{Dimension: Representation -- Focus}
	\label{tbl:dim_representation_lr2}
	\begin{tabular}{|l|l|l|l|l|}
		\hline
		\multicolumn{5}{|c|}{\textbf{Dimension: Representation}} \\
		\multicolumn{5}{|c|}{covers the sensorial characteristics of a vision.} \\ \hline
		\multicolumn{5}{|p{8cm}|}{\raggedright \textbf{Characteristic:}\\ \textit{Focus} considers the compact representation of a vision.} \\ \hline
	\end{tabular}
\end{table}

The second characteristic is \textit{completeness} which is part of the \textit{content} dimension (see \tablename{ \ref{tbl:dim_content1_lr2}}).
A vision consists of the addressed \textit{problem}, the key idea of the \textit{solution}, and how the solution \textit{improves} the state-of-the-art. Several researchers \cite{Wiegers.2003, Kittlaus.2017, Moore.1991} proposed different templates to define a vision, all of them covering these three content aspects.

A vision describes a business or user problem \cite{Robertson.2012}. The problem is described in a solution-neutral manner and explains the pain-points addressed by the solution \cite{Kittlaus.2017}. On the one hand, the problem represents the situation or background of a project \cite{Robertson.2012}. On the other hand, it justifies the efforts that an organization wants to address \cite{Robertson.2012}. The problem has to be anchored in the user needs and characterizes the work a user wants to do with the product \cite{Wiegers.2003, Robertson.2012}. These needs offer the reason for solving the problem \cite{Robertson.2012}.

A vision describes the key idea of the solution to the addressed problem \cite{McGrath.2001}. Thus, a vision provides product concepts for meeting the strategy of an organization and the corresponding market needs \cite{Kittlaus.2017, Alexander.2005, Brown.1995}. Robertson and Robertson suggest describing \enquote{how the product will solve [\dots] [the business] problem} \cite[p. 140]{Robertson.2012}. A vision needs to tell how manual practices, human interactions, and other tools come together within the planned system or product to better support the whole practice \cite{Alexander.2005}. An important part of the solution is clear objectives \cite{Tessarolo.2007}. These goals characterize business advantages such as increased market share, reduced operative costs, or improved customer service \cite{Robertson.2012}. Based on the goals, the success of a product will be determined. Therefore, a vision provides criteria for evaluating product success from the customer perspective \cite{Kittlaus.2017}. Whether the presence of a solution is necessary, however, remains a debate. Pohl stated that \enquote{the vision sets a goal, but does not define how this goal will be specifically achieved} \cite[p. 37]{Pohl.2010}. In contrast, Alexander and Maiden \cite{Alexander.2005} explain a vision as a high-level story of the \enquote{new world} of the customer population. \enquote{The vision synthesizes the findings and implications of customer data into a productive business response} \cite[p. 192]{Alexander.2005}.

A vision describes how the solution improves the state-of-the-art. A proposed solution needs to be different from the status quo, i.e., the essence of an aspired change \cite{Pohl.2010}. Representing a change offers the option to present an argumentation that allows a person to decide whether the product proposed by the vision is worthwhile \cite{Robertson.2012}. For the customer, a vision has to include a value proposition clarifying why the product is needed and cannot be replaced by an alternative \cite{Kittlaus.2017}. For the vendor, the business value and the reasons why the product will be successful are the required information of a vision \cite{Kittlaus.2017}.

\begin{table}[htbp]
	\captionsetup{justification=centering}
	\renewcommand{\arraystretch}{1.1}
	\centering
	\caption{Dimension: Content -- Completeness}
	\label{tbl:dim_content1_lr2}
	\begin{tabular}{|l|l|l|l|l|}
		\hline
		\multicolumn{5}{|c|}{\textbf{Dimension: Content}} \\
		\multicolumn{5}{|c|}{covers the perceptual characteristics of a vision.} \\ \hline
		\multicolumn{5}{|p{8cm}|}{\raggedright \textbf{Characteristic:}\\ \textit{Completeness} considers the coverage of the three contents of a vision, i.e., \textit{problem}, \textit{solution}, and \textit{improvement}.} \\ \hline
	\end{tabular}
\end{table}

\textit{Clarity} is the third vision characteristic which also belongs to the \textit{content} dimension (see \tablename{ \ref{tbl:dim_content2_lr2}}). An understandable vision statement depends on the ability of a company to define clear objectives \cite{Tessarolo.2007}. The inability to clearly define objectives greatly delays product development \cite{Dyer.1999}. Ambiguous concepts of a product allow speculations and conflicts about what should be produced \cite{Kessler.1996}. Instead of aspiring an unattainable goal, a clear and concise vision is associated with well-defined and verifiable goals that guide with explicit directions \cite{Lynn.2001, Pohl.2010, Lynn.1999}. Specific goals are important since they make the advantages of a vision measurable and thus allow to determine whether a product meets its vision \cite{Robertson.2012}. Although several researchers \cite{Kittlaus.2017, Pohl.2010, Robertson.2012, Tessarolo.2007, Crawford.2008} argue that clear and easy-to-understand goals are important for all stakeholders inside and outside of a company, Kessler and Chakrabarti \cite{Kessler.1996} found only limited empirical support that the clarity of a goal is a direct antecedent of fast development.

\begin{table}[htbp]
	\captionsetup{justification=centering}
	\renewcommand{\arraystretch}{1.1}
	\centering
	\caption{Dimension: Content -- Clarity}
	\label{tbl:dim_content2_lr2}
	\begin{tabular}{|l|l|l|l|l|}
		\hline
		\multicolumn{5}{|c|}{\textbf{Dimension: Content}} \\
		\multicolumn{5}{|c|}{covers the perceptual characteristics of a vision.} \\ \hline
		\multicolumn{5}{|p{8cm}|}{\raggedright \textbf{Characteristic:}\\ \textit{Clarity} considers the intelligibility of the aspired goals of a vision by all parties involved.} \\ \hline
	\end{tabular}
\end{table}

The fourth characteristic is \textit{support} belonging to the \textit{impact} dimension (see \tablename{ \ref{tbl:dim_impact1_lr2}}). A vision statement needs support in the development team \cite{Lynn.2001} and has to reflect a balanced view that satisfies the needs of diverse stakeholders \cite{Wiegers.2003}. All involved parties need to share and accept the same vision as their motivation and guidance of their actions and activities \cite{Kittlaus.2017, Tessarolo.2007}. A vision focuses on the future work of users by telling a futuristic story that can be idealistic \cite{Wiegers.2003, Alexander.2005}. However, the presented desirable and ambitious future needs to be achievable to be acceptable \cite{Kittlaus.2017}. Therefore, a vision has to be grounded in the realities of existing or anticipated markets, enterprise architectures, corporate strategies, and resource limitations \cite{Wiegers.2003}. The importance of a shared vision is debated. According to Song et al. \cite{Song.1998}, a team works efficiently when its members share a common perception of objectives, strategies, and the need to collaborate. In contrast, Lynn and Akgün \cite{Lynn.1999} did not find any significant relationships between a shared vision and development speed of a team.\\

\begin{table}[htbp]
	\captionsetup{justification=centering}
	\renewcommand{\arraystretch}{1.1}
	\centering
	\caption{Dimension: Impact -- Support}
	\label{tbl:dim_impact1_lr2}
	\begin{tabular}{|l|l|l|l|l|}
		\hline
		\multicolumn{5}{|c|}{\textbf{Dimension: Impact}} \\
		\multicolumn{5}{|c|}{covers the emotional characteristics of a vision.} \\ \hline
		\multicolumn{5}{|p{8cm}|}{\raggedright \textbf{Characteristic:}\\ \textit{Support} considers the level of acceptance of a vision, i.e., whether all parties involved share the vision.} \\ \hline
	\end{tabular}
\end{table}

\textit{Stability} is the fifth characteristic that is also part of the \textit{impact} dimension (see \tablename{ \ref{tbl:dim_impact2_lr2}}). A vision needs to be stable with consistent objectives over time \cite{Lynn.2001}. Vision stability is important \enquote{to lengthen the organization's attention span} \cite[p. 151]{Hamel.2005}. A stable vision provides consistency to short-term actions while leaving room for reinterpretation as new opportunities emerge. Thus, a vision helps to align a project by defining what needs to be done \cite{Lynn.2001}. Lynn and Akgün \cite{Lynn.2001} found that unsuccessful projects had a noticeably unstable vision. Unstable visions confuse and frustrate team members due to their frequent changes. Clark and Wheelwright \cite{Clark.2010} stated similar results by emphasizing that frequent changes of a vision confuse a team during a project. Vision stability enables an organization to learn and adapt to finish a project successfully \cite{Slater.1995}. In the case of dynamic goals, a stable vision helps to cope with a variety of uncertainties \cite{Aaker.2009}.

\begin{table}[htbp]
	\captionsetup{justification=centering}
	\renewcommand{\arraystretch}{1.1}
	\centering
	\caption{Dimension: Impact -- Stability}
	\label{tbl:dim_impact2_lr2}
	\begin{tabular}{|l|l|l|l|l|}
		\hline
		\multicolumn{5}{|c|}{\textbf{Dimension: Impact}} \\
		\multicolumn{5}{|c|}{covers the emotional characteristics of a vision.} \\ \hline
		\multicolumn{5}{|p{8cm}|}{\raggedright \textbf{Characteristic:}\\ \textit{Stability} considers the consistency of a vision over time.} \\ \hline
	\end{tabular}
\end{table}

\begin{table*}[!t]
	\captionsetup{justification=centering}
	\renewcommand{\arraystretch}{1.1}
	\centering
	\caption{Vision: Manual coding results -- Categories/characteristics and exemplary codes}
	\label{tbl:coding_results_lr2}
	\begin{tabular}{clll|c|c|}
		\hline
		\multicolumn{1}{|l}{\cellcolor{gray!45}} & \multicolumn{3}{l|}{\cellcolor{gray!45}\textit{Representation}} & \multicolumn{1}{c|}{\textbf{Coding frequency}} & \multicolumn{1}{c|}{\textbf{Extracted from}} \\ \hhline{|>{\arrayrulecolor{gray!45}}->{\arrayrulecolor{black}}|*{5}{-}|}
		\multicolumn{1}{|c|}{\cellcolor{gray!45}} & \cellcolor{gray!30} & \multicolumn{2}{l|}{\cellcolor{gray!30}\textit{Focus}} & \multicolumn{1}{c|}{10} & \multicolumn{1}{c|}{\cite{Wiegers.2003, Kittlaus.2017, Pohl.2010, Alexander.2005, McGrath.2001}} \\ \hhline{|>{\arrayrulecolor{gray!45}}->{\arrayrulecolor{black}}|>{\arrayrulecolor{gray!30}}->{\arrayrulecolor{black}}|*{4}{-}|}
		\multicolumn{1}{|c|}{\multirow{-2}{*}{\rotatebox[origin=c]{90}{\cellcolor{gray!45}\textbf{Dimension}}}} & \multicolumn{1}{l|}{\multirow{-1}{*}{\rotatebox[origin=c]{90}{\cellcolor{gray!30}\textbf{Category}}}} & \multicolumn{4}{|p{16cm}|}{ \raggedright \footnotesize{ \textbf{Exemplary codes:} Described its essence in a compact way, few sentences or a relatively short text, no more than one page of paper, short and concise description of the essence, concise summary,high-level story, model-based view of new work practice [\dots] into one high-level picture, be compact, big picture story of the future, condensed form, conceptual image of what the future product will be, not sufficient to elicit and elaborate detailed requirements}} \\ \hline \hline
		
		\multicolumn{1}{|l}{\cellcolor{gray!45}} & \multicolumn{3}{l|}{\cellcolor{gray!45}\textit{Content}} & \multicolumn{1}{c|}{\textbf{Coding frequency}} & \multicolumn{1}{c|}{\textbf{Extracted from}} \\ \hhline{|>{\arrayrulecolor{gray!45}}->{\arrayrulecolor{black}}|*{5}{-}|}
		\multicolumn{1}{|c|}{\cellcolor{gray!45}} & \cellcolor{gray!30} & \multicolumn{2}{l|}{\cellcolor{gray!30}\textit{Completeness}} & \multicolumn{1}{c|}{17} & \multicolumn{1}{c|}{\cite{Wiegers.2003, Kittlaus.2017, Moore.1991, Pohl.2010, Alexander.2005, Robertson.2012, McGrath.2001, Brown.1995, Tessarolo.2007}} \\ \hhline{|>{\arrayrulecolor{gray!45}}->{\arrayrulecolor{black}}|>{\arrayrulecolor{gray!30}}->{\arrayrulecolor{black}}|*{4}{-}|}
		\multicolumn{1}{|c|}{\cellcolor{gray!45}} & \multicolumn{1}{l|}{\cellcolor{gray!30}} & \multicolumn{4}{|p{16cm}|}{ \raggedright \footnotesize{ \textbf{Exemplary codes:} What the future product will be, why it is needed, and why it will be successful, problem, key idea, and improves state-of-art, aspired change, customer value proposition and business value, statement of need and key benefit unlike alternative, problem and solution, justifying the project based on seriousness of problem and reasons for solving the problem, description of the work the user wants to do}} \\ 		\hhline{|>{\arrayrulecolor{gray!45}}->{\arrayrulecolor{black}}|>{\arrayrulecolor{gray!30}}->{\arrayrulecolor{black}}|*{4}{-}|}
		\multicolumn{1}{|c|}{\cellcolor{gray!45}} & \cellcolor{gray!30} & \multicolumn{2}{l|}{\cellcolor{gray!30}\textit{Clarity}} & \multicolumn{1}{c|}{11} & \multicolumn{1}{c|}{\cite{Lynn.2001, Kittlaus.2017, Pohl.2010, Robertson.2012, Tessarolo.2007, Dyer.1999, Kessler.1996, Lynn.1999, Crawford.2008}} \\ \hhline{|>{\arrayrulecolor{gray!45}}->{\arrayrulecolor{black}}|>{\arrayrulecolor{gray!30}}->{\arrayrulecolor{black}}|*{4}{-}|} 
		\multicolumn{1}{|c|}{\multirow{-5}{*}{\rotatebox[origin=c]{90}{\cellcolor{gray!45}\textbf{Dimension}}}} & \multicolumn{1}{l|}{\multirow{-3}{*}{\rotatebox[origin=c]{90}{\cellcolor{gray!30}\textbf{Category}}}} & \multicolumn{4}{|p{16cm}|}{ \raggedright \footnotesize{ \textbf{Exemplary codes:} Firm's ability to define clear objectives, need for objective measures to determine whether the product is worthwhile and meets the goal, clearly define product and market objectives, ambiguous project concepts allow for greater speculation and conflict, unsuccessful projects lacked a clear vision, clarity, well-defined and verifiable goal, clearly signals [\dots] development goals, clear and easy-to-understand goal}} \\ \hline \hline
		
		\multicolumn{1}{|l}{\cellcolor{gray!45}} & \multicolumn{3}{l|}{\cellcolor{gray!45}\textit{Impact}} & \multicolumn{1}{c|}{\textbf{Coding frequency}} & \multicolumn{1}{c|}{\textbf{Extracted from}} \\ \hhline{|>{\arrayrulecolor{gray!45}}->{\arrayrulecolor{black}}|*{5}{-}|}
		\multicolumn{1}{|c|}{\cellcolor{gray!45}} & \cellcolor{gray!30} & \multicolumn{2}{l|}{\cellcolor{gray!30}\textit{Support}} & \multicolumn{1}{c|}{10} & \multicolumn{1}{c|}{\cite{Lynn.2001, Wiegers.2003, Kittlaus.2017, Alexander.2005, Tessarolo.2007, Song.1998, Lynn.1999}} \\ \hhline{|>{\arrayrulecolor{gray!45}}->{\arrayrulecolor{black}}|>{\arrayrulecolor{gray!30}}->{\arrayrulecolor{black}}|*{4}{-}|}
		\multicolumn{1}{|c|}{\cellcolor{gray!45}} & \multicolumn{1}{l|}{\cellcolor{gray!30}} & \multicolumn{4}{|p{16cm}|}{ \raggedright \footnotesize{ \textbf{Exemplary codes:} Balanced view that satisfy the needs of diverse stakeholders, to share these objectives and strategy with all those involved, motivating effect on stakeholders inside and outside of the company, support, guiding star, grounded in the realities of existing or anticipated markets, team tells story, desirable and ambitious but achievable future, share a common perception of objectives and strategy}} \\ 		\hhline{|>{\arrayrulecolor{gray!45}}->{\arrayrulecolor{black}}|>{\arrayrulecolor{gray!30}}->{\arrayrulecolor{black}}|*{4}{-}|}
		\multicolumn{1}{|c|}{\cellcolor{gray!45}} & \cellcolor{gray!30} & \multicolumn{2}{l|}{\cellcolor{gray!30}\textit{Stability}} & \multicolumn{1}{c|}{7} & \multicolumn{1}{c|}{\cite{Lynn.2001, Hamel.2005, Clark.2010, Slater.1995, Aaker.2009}} \\ \hhline{|>{\arrayrulecolor{gray!45}}->{\arrayrulecolor{black}}|>{\arrayrulecolor{gray!30}}->{\arrayrulecolor{black}}|*{4}{-}|} 
		\multicolumn{1}{|c|}{\multirow{-5}{*}{\rotatebox[origin=c]{90}{\cellcolor{gray!45}\textbf{Dimension}}}} & \multicolumn{1}{l|}{\multirow{-3}{*}{\rotatebox[origin=c]{90}{\cellcolor{gray!30}\textbf{Category}}}} & \multicolumn{4}{|p{16cm}|}{ \raggedright \footnotesize{ \textbf{Exemplary codes:} Stability, consistency of objectives over time, strategic intent is stable over time, consistency to short-term action while leaving room for reinterpretation, vision stability means that a company's vision remains consistent over time, having a stable vision reduces confusion, successful projects [\dots] had a stable vision, on the unsuccessful projects the visions were noticeably unstable, stable vision}} \\ \hline
	\end{tabular}
\end{table*}

As in the first literature review, we asked the same two raters to assign the identified categories to the extracted passages on their own to verify the vision characteristics. The reliability of the raters' classification was evaluated by using \textit{Cohen's kappa} \cite{Cohen.1960}. The calculated \textit{Cohen's kappa} value was $0.84$ which shows an almost perfect raters' agreement \cite{Landis.1977}. This result indicates that the identified categories represent major themes in terms of vision characteristics in the analyzed publications. In \figurename{ \ref{fig:fig7}}, we present the hierarchical decomposition of vision quality. As an answer to the research question, we can summarize:

\begin{mdframed}
	\begin{itemize}[leftmargin=-2.5mm]
		\item[] \textbf{Answer:}
		According to literature, the characteristics of a vision are \textit{focus}, \textit{completeness}, \textit{clarity}, \textit{support}, and \textit{stability}. 
	\end{itemize}
\end{mdframed}

\begin{figure}[htbp]
	\centering
	\includegraphics[width=\columnwidth]{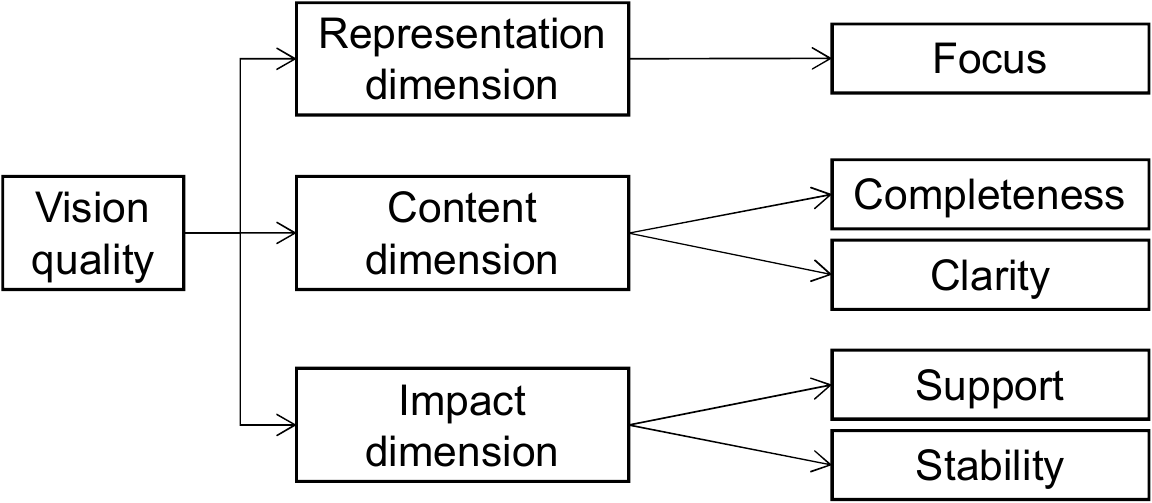}
	\caption{Hierarchical decomposition of vision quality}
	\label{fig:fig7}
\end{figure}

\subsection{Intentions of Visions}
Although a video as a representation format has high-level intentions (see section \ref{sec:intentions-of-videos}), its content, i.e., the vision, primarily influences the purpose of a vision video. The success of a vision depends on the fulfillment of its associated intention which needs to be identified early in software projects \cite{Wiegers.2003}. Thus, we also elaborated the intentions of visions by considering literature and referring these purpose to the high-level intentions of videos. In the analyzed literature, we identified the following three types of intended purposes of visions:

\begin{itemize}[leftmargin=0mm, topsep=0.5mm]
	\setlength\itemsep{-.25mm}
	\item[] $(1)$ \textit{Share an integrated view within a heterogeneous group of stakeholders to align their actions and views}.\\
	A vision synthesizes the contributions, findings, and implications solicited from employees and stakeholders \cite{Alexander.2005, Gorschek.2010, Chesbrough.2006}. The vision serves as a guiding principle for all stakeholders to align their actions \cite{Pohl.2010, Robertson.2012}.
	
	\item[] $(2)$ \textit{Share an integrated view of a project with the development team that will implement the vision}.\\
	A vision is a \enquote{guiding star} for the project members to develop a productive business response for the stakeholders \cite{Alexander.2005} since it provides the context for decision-making \cite{Wiegers.2003, Robertson.2012}. This support is especially important in the initial phase of a project when a team conceives and develops the first version of a product \cite{Kittlaus.2017}.
	
	\item[] $(3)$ \textit{Convey an integrated view of a future system and its use for validating this view and for eliciting new or diverging aspects}.\\
	According to Pohl \cite{Pohl.2010}, the RE process needs to establish a vision in the relevant system context. Therefore, an integrated view of the potential future system and its use is necessary to validate whether different stakeholders share this view \cite{Robertson.2012, Tessarolo.2007}. In the case of inconsistencies or ambiguities, all involved parties can negotiate to make joint decisions to conceive and develop a satisfying system \cite{Kittlaus.2017, Alexander.2005}.
\end{itemize}

The first two intended purposes of visions are related to the intended purpose of videos: \textit{Information}. In both cases, declarative knowledge is shared, whereby only the respective target audience is different. The third intention corresponds to the intended purpose of videos: \textit{Experience Exposure}. In this case, the video serves as a replacement for the future system and its use so that the viewers can experience the envisioned product.

\section{Represent Vision by Means of Video: The Quality Model}
\label{sec:represent-vision-by-means-of-video-the-quality-model}
According to our objective (see section \ref{sec:objective-approach-and-contribution}), we need both structuring options of a quality model due to our two goals. In the following, we first present the hierarchical decomposition of vision video quality. This representation provides a convenient breakdown of the quality into the individual quality characteristics that need to be assessed to evaluate the overall quality of a vision video (see section \ref{sec:hierarchical-decomposition-of-vision-video-quality}).
We also present a mapping of the individual quality characteristics to the steps of the production and use process of a video. This representation helps to guide video production by software professionals by highlighting which quality characteristics can be affected by which process step (see section \ref{sec:vision-video-quality-along-the-production-and-use-steps}).

\subsection{Hierarchical Decomposition of Vision Video Quality}
\label{sec:hierarchical-decomposition-of-vision-video-quality}
A quality model is typically structured as a hierarchical decomposition of its quality characteristics and sub-characteristics. This hierarchy shows how the overall quality of a product, i.e., a vision video, is composed of the individual quality characteristics which can be further divided into sub-characteristics.

\subsubsection{Procedure}
Based on the previous results, we obtained the quality model for vision videos by combining the two hierarchical decompositions of video (see \figurename{ \ref{fig:fig6}}) and vision (see \figurename{ \ref{fig:fig7}}). The combination of the two individual hierarchies is based on the simple merge of the three dimensions of a quality model which cover the sensorial (\textit{representation}), perceptual (\textit{content}), and emotional (\textit{impact}) characteristics of a product. 

\subsubsection{Result}
In \figurename{ \ref{fig:fig8}}, we present the quality model for vision videos structured as hierarchical decomposition. All quality characteristics and sub-characteristics are placed as before in the single hierarchical decompositions of video and vision. The labels used in \figurename{ \ref{fig:fig8}} follow the previous explanations and definitions (see section \ref{sec:results_video} and section \ref{sec:results_vision}).

\begin{figure}[htbp]
	\centering
	\includegraphics[width=\columnwidth]{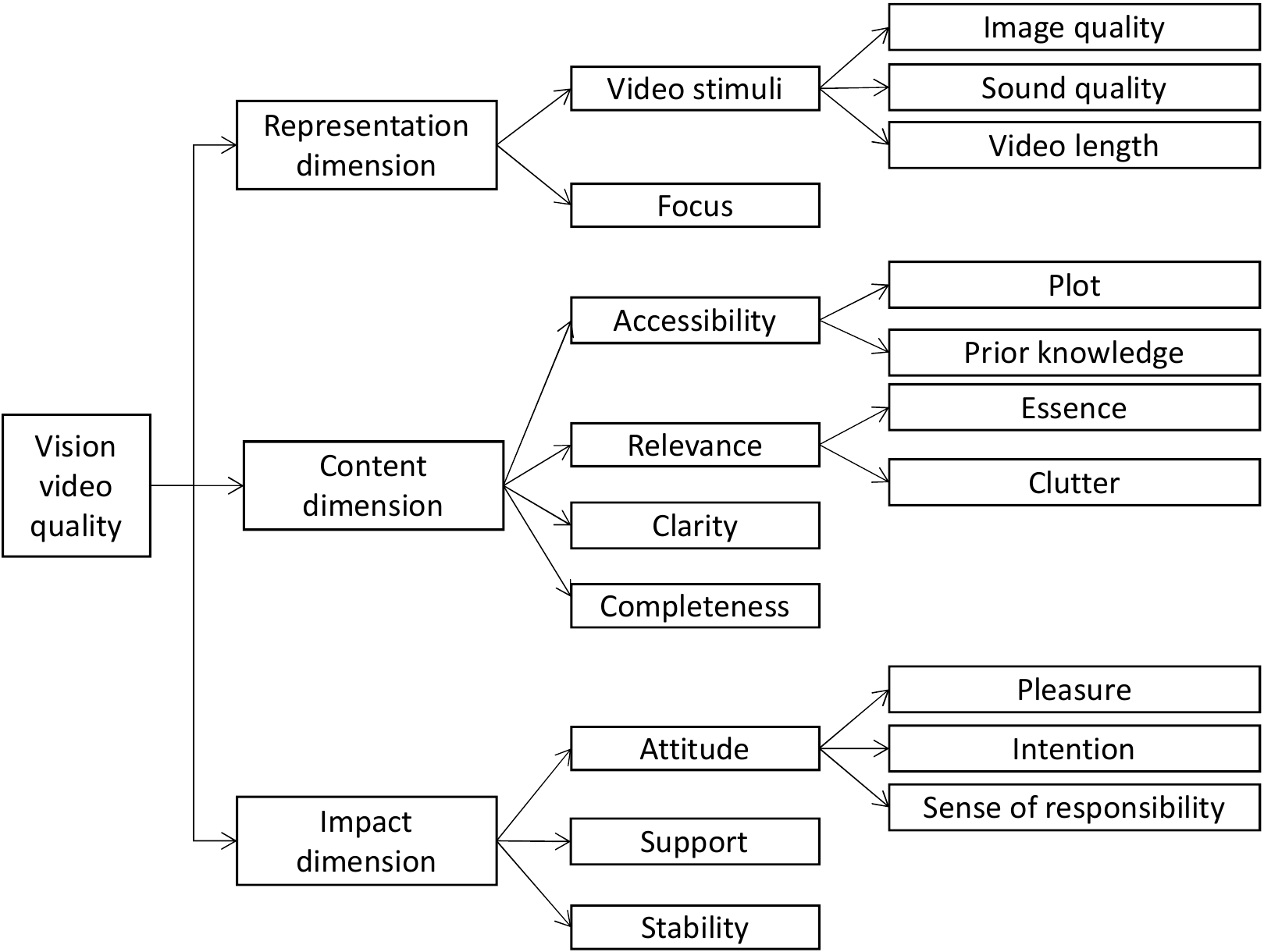}
	\caption{Hierarchical decomposition of vision video quality}
	\label{fig:fig8}
\end{figure}

\subsection{Vision Video Quality along the Video Production and Use}
\label{sec:vision-video-quality-along-the-production-and-use-steps}
The typical video production process consists of three main steps: \textit{preproduction}, \textit{shooting}, and \textit{postproduction} \cite{Owens.2011, Creighton.2006}. We extended the video production process by the step \textit{viewing} since the production of a video generally does not include its viewing which is the typical use of a video \cite{Owens.2011}. Each of these steps is supposed to add value to the video. However, poor performance and violation of quality characteristics in any of the steps can diminish the final value of a video. The dynamic perspective on video production and use highlights which quality characteristics can be affected in which steps. Poor quality in an early step will often constrain the quality that can be achieved in later steps. In a way, this phenomenon resembles the V-model of a software development lifecycle. Poor performance and misunderstandings in the requirements elicitation cannot be compensated by good design. Instead, any early flaw decreases and limits possible value.

\subsubsection{Procedure}
The mapping of the quality characteristics and process steps resulted from the following procedure.

For the video characteristics, we assigned one or more potential process steps to each of the $586$ assigned codes of the $307$ extracted passages from the video production guidelines. Some guidelines \cite{MIT, ARSC, Owens.2011} arranged at least partially their recommendations by the process steps which simplified the assignment. If no step was stated by the guidelines, we analyzed the extracted passage for keywords, such as plan, record, edit, view, or similar words, that indicate one or more potential process steps. As a result, we obtained the frequencies for each video characteristic and process step pair (see \figurename{ \ref{fig:fig9}}). Based on the mean frequency of all pairs ($M = 15$), we developed a $3$-point scale that ranges from \textit{strong} (frequency $\geq 15$) to \textit{medium} ($7 <$ frequency $< 15$) to \textit{weak} (frequency $\leq 7$). This scale is a rough indicator of how much a particular video characteristic can be affected in a particular process step according to the analyzed guidelines.

For the vision characteristics, we were not able to apply a similar procedure since all analyzed books and journal articles did not provide any detailed information about the process of creating a vision. Therefore, the assignment of the vision characteristics to the production steps is only based on the following considerations of the three authors of this article. At first, the vision is the key content of a vision video. Hence, it represents the starting point for the video production. Furthermore, any poor quality in an early step constrains the quality in later steps. Therefore, we conclude that the vision characteristics should be addressed as early as possible and thus in the \textit{preproduction}. Nevertheless, we assume that the vision characteristics can and should be considered in later steps. In particular, we suggest addressing \textit{focus}, \textit{completeness}, and \textit{clarity} in the \textit{postproduction} to ensure that the final vision video fulfills these quality characteristics after editing and digital postprocessing. \textit{Support} and \textit{stability} are important to the \textit{viewing} step since they mainly affect the target audience of a vision video. The stakeholders and development team need to support the vision by sharing and accepting the vision as their motivation and guidance. Thereby, it is important to clarify how stable the vision is. A vision video of a stable vision is suitable to be shared with the stakeholders and the development team as guidance. In contrast, a vision video of a less stable vision is unsuitable as guidance but beneficial for elicitation and validation.

\begin{figure*}[!t]
	\centering
	\includegraphics[width=\textwidth]{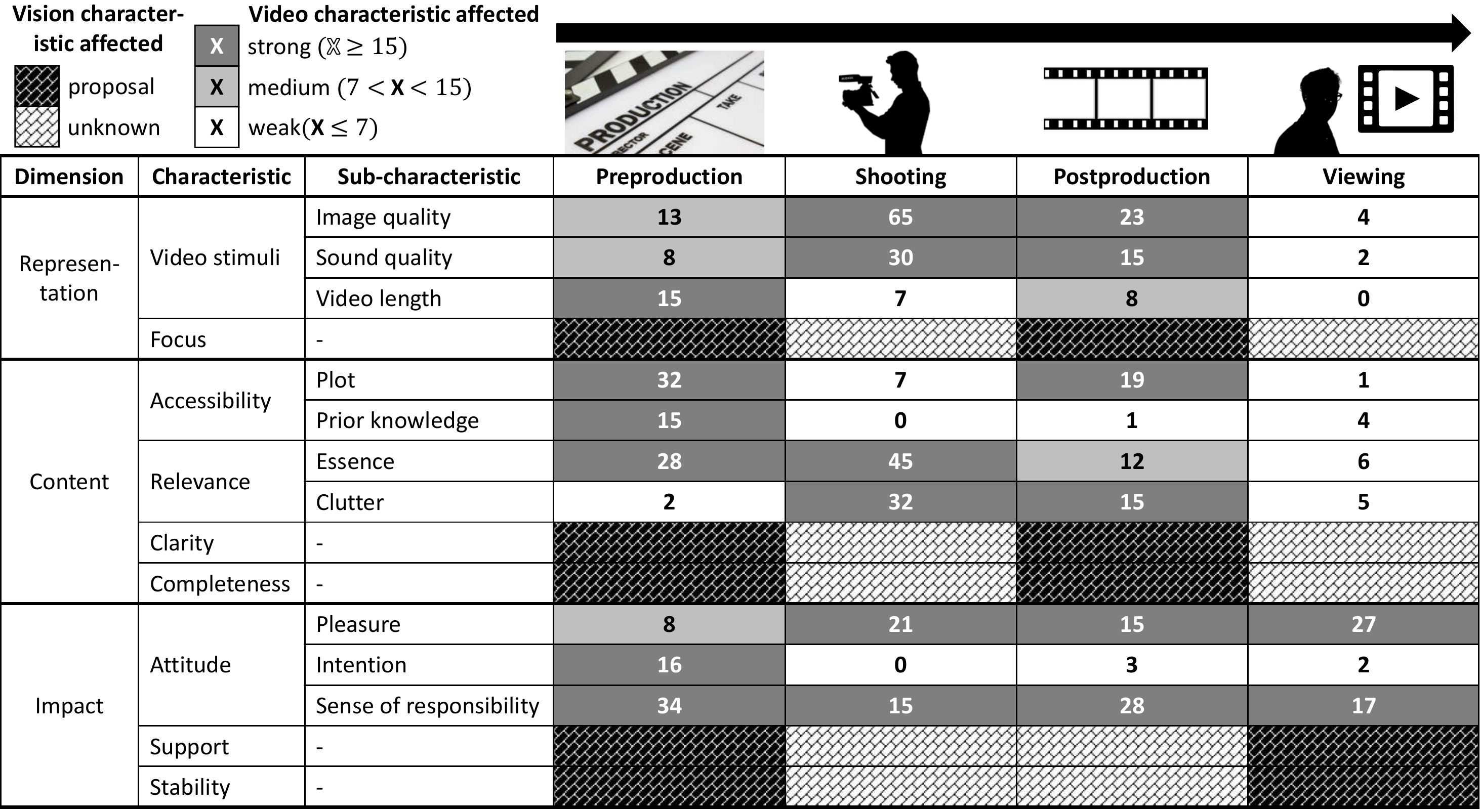}
	\caption{Vision video quality along the video production and use process (The numbers represent the frequencies of each video characteristic and process step pair.)}
	\label{fig:fig9}
\end{figure*}

\subsubsection{Result}
In \figurename{ \ref{fig:fig9}}, we present the mapping of the individual quality characteristics to the steps of the video production and use process. 
We want to emphasize that this mapping is discussible since the frequencies of the video characteristic and process step pairs show that several video characteristics can be affected in several process steps. However, the later a quality characteristic is addressed, the higher is the risk of diminishing the final value of a vision video. Therefore, any quality characteristic should be addressed as early as possible. This conclusion is supported by Owens and Millerson who stated that \enquote{ninety percent of the work on a [video] production usually goes into the planning and preparation phase} \cite[p. 37]{Owens.2011}.
For low-cost production of a vision video, this perspective provides orientation and guidance by answering the question: What are crucial quality characteristics to consider in a given step? Below, we briefly explain the single steps and why specific characteristics are strongly affected in the respective step.

\paragraph{Preproduction} In this step, the planning of a vision video takes place. Preliminaries, preparations, and the organization need to be done before the shooting begins. In this phase, the purpose of a vision video (\textit{intention}), its story and single scenes (\textit{plot}), the aspired duration (\textit{video length}), the necessary \textit{prior knowledge}, and the relevant contents (\textit{essence}) are defined so that the shooting can be done quickly and easily. It is important to specify clear goals of a vision (\textit{clarity}) to define its essential parts (\textit{completeness}) that need to be presented compactly (\textit{focus}) in the vision video. The more stable a vision, the easier the \textit{support} may be achieved since \textit{stability} can contribute to acceptance.

\paragraph{Shooting} In this step, the vision video is recorded. A short video clip is recorded for each scene of a story. These single video clips are later combined into one video. In this phase, the relevant contents (\textit{essence}) of a vision video need to be captured pleasantly (\textit{pleasure}) by avoiding any disrupting or distracting contents (\textit{clutter}) such as background noise or actions. The video stimuli (\textit{image quality} and \textit{sound quality}) are mainly affected by the recording of the single video clips.

\paragraph{Postproduction} In this step, the whole vision video is created by editing and digitally postprocessing the image and sound of the video. Besides the video stimuli (\textit{image quality} and \textit{sound quality}), the \textit{focus} and the \textit{plot} of the vision video are mainly affected since the single video clips are combined and possibly rearranged to convey the entire planned vision. Especially, one important task of the postproduction is to remove \textit{clutter} from the video, such as background noise or unnecessary parts in video clips. During the entire postprocessing, it must be ensured that the final video clearly presents the complete vision (\textit{clarity} and \textit{completeness}) in a pleasant and interesting way for the target audience (\textit{pleasure}).

\paragraph{Viewing} In this step, the vision video exists and is viewed as a whole. Although specific quality characteristics of the representation and content dimension may be more or less important for the audience, the most affected quality characteristics come from the impact dimension. A vision video must be pleasant (\textit{pleasure}) for the audience to focus their attention on the conveyed content. Thereby, the \textit{stability} of a vision is crucial. A vision used for validation or elicitation might be less stable to adapt to changes and new insights. However, a vision used to inform stakeholders and developers needs to be stable to align their actions and to avoid frustration and confusion. Similar to \textit{stability}, it is important to consider \textit{support}. This characteristic does not need to be fulfilled for the validation and elicitation purpose. However, all stakeholders and developers should share and accept the defined vision when it is conveyed to align their activities and actions.\\

The quality characteristic \textit{sense of responsibility} needs to be considered separately. The entire production and use process of a vision video requires to ensure legal reliability. In the preproduction, it has to be identified which permissions must be obtained, such as consent forms to record persons or places. During the shooting, compliance with the permissions must be ensured. In the postproduction, copyrights have to be fulfilled in case of used music or images by third parties. The postproduction also needs to ensure that the contents of the video are not falsified due to the editing. For viewing, it is important to ensure the right to share and distribute the video.

\section{Evaluation}
\label{sec:evaluation}
In the evaluation (see \figurename{ \ref{fig:fig1}}, \circled{\textbf{4}}), we investigated whether the identified quality characteristics characterize the overall impression of a vision video. In case of relationships between the quality characteristics and the perceived overall quality of vision videos, we assume that the proposed quality model is of fundamental relevance. We ensured that the experimental design is well-defined by following the recommendations for experimentation in software engineering by Wohlin et al. \cite{Wohlin.2012}.

\begin{mdframed}
	\begin{itemize}[leftmargin=-2.5mm]
		\item[] \textbf{Goal definition:}
		We \textit{analyze} the $15$ characteristics of the proposed quality model \textit{for the purpose of} evaluating the relationship between each characteristic and the perceived overall quality of vision videos \textit{from the point of view of} students who have the role of a developer and actively develop software \textit{in the context of} a software project course.
	\end{itemize}
\end{mdframed}

\subsection{Experimental Design}

\subsubsection{Hypotheses}
\label{sec:hypotheses}
We specified the criterion for measuring fundamental relevance as follows. The proposed quality model for vision videos is of fundamental relevance if:

\begin{itemize}[topsep=0.5mm]
	\setlength\itemsep{-.25mm}
	\item[($1$)] All three dimensions of the quality model are related to the overall quality of a vision video.
	
	
	\item[($2$)] The related quality characteristics include both vision and video characteristics.
\end{itemize}

\noindent
According to our goal definition and the criterion for measuring fundamental relevance, we formulated the following global null and alternative hypothesis:

\begin{itemize}[topsep=0.5mm, leftmargin=10mm]
	\setlength\itemsep{-.25mm}
	\item[$gH_{0}$:] None of the characteristics of the proposed quality model for vision videos affects the likelihood that the subjects perceive the overall quality of a vision video as good from a developer's point of view.
	
	\item[$gH_{1}$:] There are characteristics, covering both vision and video characteristics as well as all three dimensions of the proposed quality model for vision videos, that affect the likelihood that the subjects perceive the overall quality of a vision video as good from a developer's point of view.
\end{itemize}

\noindent
In \tablename{ \ref{tbl:hypotheses}}, we formulated a specific alternative hypothesis for each of the $15$ quality characteristics to concretize the global alternative hypothesis $gH_{1}$ . Each specific null hypothesis $Hi_{0}$, $i \in \{1, \dots, 15\}$ considers that there is no relationship between the respective quality characteristic and the likelihood that the subjects perceive the overall quality of a vision video as good from a developer's point of view.

\subsubsection{Dependent and Independent Variables}
The dependent variable was the overall quality of a vision video, perceived by the subjects, with two levels: \textit{bad} and \textit{good}.
The independent variables were the $15$ characteristics at the lowest level of the proposed quality model for vision videos. While \textit{video length} was measured in seconds, the subjects assessed all other $14$ characteristics on a 5-point Likert scale (cf. \cite{Karras.2019b}) ranging from $-2$ (most negative answer option, e.g., very unclear) to $2$ (most positive answer option, e.g., very clear).

\subsubsection{Context and Material}
\label{sec:context-and-material}
In a yearly course called \textit{Software Project}, we conduct multiple real software projects with real customers. The project course consists of a $4$-week requirements analysis phase followed by the development phase divided into two $3$-week iterations and one $2$-week polish phase. Each project team creates a vision video at the end of the second week of the requirements analysis phase. They illustrate the given problem, their proposed solution, and the improvement of the problem due to the solution. These videos are intended to convey an integrated view of the future system and its use to validate the overall product goals with the customers. The teams produce their videos with simple equipment, i.e., smartphones and open-source software.

For the evaluation, we used eight vision videos of eight different projects from $2017$. Each project covered a different domain with real customers, e.g., the Central Crime Service of the Police Administration Hannover who needed an investigation software for personal data in publicly available sources on the internet. The project teams who created the videos consisted of ten undergraduate students of computer science which were at least in their $5$th academic semester. Apart from a maximum duration of $3$ minutes, there were no further restrictions on the video production. On average, all eight vision videos had a duration of $103.4$ seconds. The minimum and maximum duration of all videos were $69$ respectively $155$ seconds. We are not allowed to distribute these videos since we have to follow the guidelines of the central ethics committee of our university to secure good scientific practice\footnote{\url{https://www.uni-hannover.de/en/universitaet/profil/ziele/gute-wissenschaftliche-praxis/}}. This committee regulates subjects' information and rights. We do not have the explicit consent of the actors to distribute the vision videos. For this reason, we have to archive the vision videos internally for future reference since recognizable persons shall not be visible on distributed videos without their explicit consent.

\begin{table*}[!t]
	\captionsetup{justification=centering}
	\renewcommand{\arraystretch}{1.1}
	\centering
	\caption{Specific alternative hypotheses of the $15$ quality characteristics}
	\label{tbl:hypotheses}
	\begin{tabular}{|c|p{16cm}|}
		\hline
		\multicolumn{1}{|c|}{\textbf{ID}} & \multicolumn{1}{c|}{\textbf{Specific alternative hypothesis}} \\ \hline
		\multirow{2}{*}{$H1_{1}$} & The higher the value for \textit{image quality}, the higher the likelihood that the subjects perceive the overall quality of a vision video as good from a developer's point of view. \\ \hline
		\multirow{2}{*}{$H2_{1}$} & The higher the value for \textit{sound quality}, the higher the likelihood that the subjects perceive the overall quality of a vision video as good from a developer's point of view. \\ \hline
		\multirow{2}{*}{$H3_{1}$} & The closer the \textit{video length} is to the given maximum duration of $3$ minutes (cf. section \ref{sec:context-and-material}), the higher the likelihood that the subjects perceive the overall quality of a vision video as good from a developer's point of view. \\ \hline
		\multirow{2}{*}{$H4_{1}$} & The higher the value for \textit{focus}, the higher the likelihood that the subjects perceive the overall quality of a vision video as good from a developer's point of view. \\ \hline
		\multirow{2}{*}{$H5_{1}$} & The higher the value for \textit{plot}, the higher the likelihood that the subjects perceive the overall quality of a vision video as good from a developer's point of view. \\ \hline
		\multirow{2}{*}{$H6_{1}$} & The lower the value for \textit{prior knowledge}, the higher the likelihood that the subjects perceive the overall quality of a vision video as good from a developer's point of view. \\ \hline
		\multirow{2}{*}{$H7_{1}$} & The higher the value for \textit{clarity}, the higher the likelihood that the subjects perceive the overall quality of a vision video as good from a developer's point of view. \\ \hline
		\multirow{2}{*}{$H8_{1}$} & The higher the value for \textit{essence}, the higher the likelihood that the subjects perceive the overall quality of a vision video as good from a developer's point of view. \\ \hline
		\multirow{2}{*}{$H9_{1}$} & The lower the value for \textit{clutter}, the higher the likelihood that the subjects perceive the overall quality of a vision video as good from a developer's point of view. \\ \hline
		\multirow{2}{*}{$H10_{1}$} & The higher the value for \textit{completeness}, the higher the likelihood that the subjects perceive the overall quality of a vision video as good from a developer's point of view. \\ \hline
		\multirow{2}{*}{$H11_{1}$} & The higher the value for \textit{pleasure}, the higher the likelihood that the subjects perceive the overall quality of a vision video as good from a developer's point of view. \\ \hline
		\multirow{2}{*}{$H12_{1}$} & The higher the value for \textit{intention}, the higher the likelihood that the subjects perceive the overall quality of a vision video as good from a developer's point of view. \\ \hline
		\multirow{2}{*}{$H13_{1}$} & The higher the value for \textit{sense of responsibility}, the higher the likelihood that the subjects perceive the overall quality of a vision video as good from a developer's point of view. \\ \hline
		\multirow{2}{*}{$H14_{1}$} & The higher the value for \textit{support}, the higher the likelihood that the subjects perceive the overall quality of a vision video as good from a developer's point of view. \\ \hline
		\multirow{2}{*}{$H15_{1}$} & The higher the value for \textit{stability}, the higher the likelihood that the subjects perceive the overall quality of a vision video as good from a developer's point of view. \\ \hline
	\end{tabular}
\end{table*}

\subsubsection{Subject Selection}
Although experiments with students are often associated with a lack of realism \cite{Sjoberg.2002}, we consciously decided to select students as subjects for the first evaluation due to the following reasons. According to Höst et al. \cite{Hoest.2000}, there are only minor differences between students, which are close to their graduation, and software professionals concerning their ability to perform relatively small tasks of assessment. Therefore, students are suitable subjects for this evaluation since the task to be performed requires to subjectively assess vision videos. Due to this subjective assessment, we needed a large number of subjects to draw grounded conclusions. Furthermore, the proposed quality model for vision videos must first be validated in academia to ensure its soundness before being presented to industry experts \cite{Gorschek.2006}.

\begin{table}[htbp]
	\captionsetup{justification=centering}
	\renewcommand{\arraystretch}{1.1}
	\centering
	\caption{Distribution of subjects in terms of years of experience as a developer}
	\label{tbl:yoe}
	\begin{tabular}{c|c|c|c|}
		\cline{2-4}
		\multicolumn{1}{l|}{} & \begin{tabular}[c]{@{}c@{}}\textbf{Years of}\\ \textbf{experience}\end{tabular} & \begin{tabular}[c]{@{}c@{}}\textbf{Number of}\\ \textbf{subjects}\end{tabular} & \textbf{Proportion} \\ \hline
		\multicolumn{1}{|c|}{\begin{tabular}[c]{@{}c@{}}\textbf{Inexperienced}\\\textbf{subjects}\end{tabular}} & 0 & 39 & 32.77\% \\ \hline
		\multicolumn{1}{|c|}{\multirow{12}{*}{\begin{tabular}[c]{@{}c@{}}\textbf{Experienced}\\\textbf{subjects}\end{tabular}}} & 1 & 10 & \multirow{12}{*}{67.23\%} \\ \cline{2-3}
		\multicolumn{1}{|c|}{} & 2 & 27 &  \\ \cline{2-3}
		\multicolumn{1}{|c|}{} & 3 & 13 &  \\ \cline{2-3}
		\multicolumn{1}{|c|}{} & 4 & 8 &  \\ \cline{2-3}
		\multicolumn{1}{|c|}{} & 5 & 7 &  \\ \cline{2-3}
		\multicolumn{1}{|c|}{} & 6 & 6 &  \\ \cline{2-3}
		\multicolumn{1}{|c|}{} & 7 & 2 &  \\ \cline{2-3}
		\multicolumn{1}{|c|}{} & 8 & 3 &  \\ \cline{2-3}
		\multicolumn{1}{|c|}{} & 9 & 1 &  \\ \cline{2-3}
		\multicolumn{1}{|c|}{} & 10 & 1 &  \\ \cline{2-3}
		\multicolumn{1}{|c|}{} & 12 & 1 &  \\ \cline{2-3}
		\multicolumn{1}{|c|}{} & 15 & 1 &  \\ \hline
	\end{tabular}
\end{table}

Due to these reasons, we conducted the experiment during the \textit{Software-Project} course with $139$ undergraduate students of computer science in $2018$. The students participated in the experiment voluntarily. There was no financial reward and the experiment did not influence the success of passing the course. All subjects had the role of a developer and were actively developing software at the time of the experiment. The subjects were at least in their $5$th academic semester and close to their graduation. Thus, the subjects basically formed a fairly homogeneous group. \tablename{ \ref{tbl:yoe}} shows the distribution of the subjects in terms of their years of experience as a developer. While about one-third of the subjects had less than one year of experience as a developer, two-third had at least one year of experience as a developer. On average, the subjects had $2.4$ years of experience as a developer with a minimum of $0$ and a maximum of $15$ years of experience. The difference in the experience of up to $15$ years increased the heterogeneity in the sample. This heterogeneity is representative of an industrial context since variations among software professionals are generally expected to be even greater than variations among students due to a more varied educational background and working experience \cite{Sjoberg.2002}. In total, we expected the subjects to be suitable to assess the vision videos from a developer's point of view.


\subsubsection{Setting and Procedure}
\label{sec:setting-and-procedure}
We had the permission of the lecturer of \textit{Software Project} course to conduct the evaluation at the sixth of twelve weekly one-hour appointments that were binding for all participants of the course. We conducted the experiment in a lecture hall with all $139$ subjects at the same time. The vision videos were presented via a projector with a resolution of $1920 \times 1080$ px and the sound was played through the sound system of the room. We used a random number generator to ensure a random presentation order of the videos. The whole experiment lasted $60$ minutes. While all videos together had a duration of $13$:$47$ minutes, the assessment of a single video took $5$ to $6$ minutes. 

Before the experiment, we explained to all course participants that their participation in the subsequent experiment is entirely voluntary, has no influence on passing the course, and that they can leave the lecture hall if they do not want to participate. We communicated to all remaining persons that they give their explicit and informed consent to participate in the experiment if they stay in the lecture hall and complete the assessment form.

Subsequently, all subjects got an introduction in which we briefly explained the experimental procedure with the task of assessing the presented vision videos and the provided assessment form to collect the data. When explaining the assessment form, we also presented and explained the individual quality characteristics. We asked all subjects to ask questions at any time if they needed clarification. All subjects were put in the situation that they join an ongoing project in their familiar role as a developer. In this context, we later showed each of the eight vision videos always with the intent to share the vision of the particular project with the subjects. Afterward, the subjects completed a pre-questionnaire. This questionnaire collected each subject's demographic data such as academic semester and years of experience as a developer.

After the introduction, we repeated the following two steps for each video. First, we played the vision video once for all subjects together. Second, each subject completed an assessment form by himself to rate the perceived overall quality and the perceived level of each characteristic of the video. The assessment form (cf. \cite{Karras.2019b}) contained a $2$-point scale to rate the overall quality as either \textit{good} or \textit{bad} and a $5$-point Likert scale for each characteristic. Thereby, we provided an incomplete statement for each characteristic which the subjects completed by selecting one item of the $5$-point Likert scale.

\subsubsection{Data Set}
\label{sec:data-set-description}
After data cleaning, the data set contains $952$ vision video quality assessments of $119$ subjects for the eight vision videos. Each entry consists of a rating for the overall quality, the ratings for $14$ characteristics of vision videos, and the duration of the particular video in seconds. While $281$ $(29.5\%)$ of these assessments rated the overall quality as bad, $671$ $(70.5\%)$ rated the overall quality as good. We present all the details of the data set in the tables \ref{tbl:t5}, \ref{tbl:t6}, and \ref{tbl:t7}. We also published all collected data and the assessment form online to increase the transparency of our results \cite{Karras.2019b}.

\subsection{Analysis: Impact of Experience on the Assessments}
First, we examined whether the years of experience as a developer had an impact on the assessments of the individual variables due to the increased heterogeneity in our sample. In particular, we investigated the relationships between the years of experience and each assessed variable. For the analysis, we used the \textit{Spearman's} rank correlation since all variables were assessed based on Likert scales, i.e., ordinal scales. 

\tablename{ \ref{tbl:correlation_coefficients}} shows the \textit{Spearman's} rank correlation coefficient and $p$-values for each of the $15$ pairs consisting of years of experience and assessed variable. Only one of the $15$ analyzes yielded a significant result. However, the resulting correlation coefficient ($\rho = 0.067 \approx 0$) indicates no correlation between years of experience and the assessments of the independent variable \textit{stability} (see \tablename{ \ref{tbl:correlation_coefficients}}, light gray cells). Based on these results, there are no relationships between the years of experience and the assessments of the individual variables. Therefore, we conclude that the years of experience and thus the increased heterogeneity of our sample had no major impact on the assessments.

\begin{table}[htbp]
	\captionsetup{justification=centering}
	\renewcommand{\arraystretch}{1.1}
	\centering
	\caption{Spearman's rank correlation coefficients between years of experience and each variable assessed by the subjects\\ (The light gray cells show the statistically significant result.)}
	\label{tbl:correlation_coefficients}
	\begin{tabular}{|l|p{1.5cm}|p{1.5cm}|}
		\hline
		\multicolumn{1}{|c|}{\multirow{2}{*}{\textbf{\begin{tabular}[c]{@{}c@{}}Assessed variable\end{tabular}}}} & \multicolumn{2}{c|}{\textbf{Years of experience}} \\ \cline{2-3} 
		\multicolumn{1}{|c|}{} & \multicolumn{1}{c|}{$\rho$} & \multicolumn{1}{c|}{$p$} \\ \hline
		Overall quality & \multicolumn{1}{c|}{$-0.050$} & \multicolumn{1}{c|}{$.125$} \\ \hline
		Image quality & \multicolumn{1}{c|}{$-0.047$} & \multicolumn{1}{c|}{$.146$} \\ \hline
		Sound quality & \multicolumn{1}{c|}{$-0.053$} & \multicolumn{1}{c|}{$.105$} \\ \hline
		Focus & \multicolumn{1}{c|}{$-0.036$} & \multicolumn{1}{c|}{$.263$} \\ \hline
		Plot & \multicolumn{1}{c|}{$0.011$} & \multicolumn{1}{c|}{$.742$} \\ \hline
		Prior knowledge & \multicolumn{1}{c|}{$0.010$} & \multicolumn{1}{c|}{$.146$} \\ \hline
		Essence & \multicolumn{1}{c|}{$0.019$} & \multicolumn{1}{c|}{$.551$} \\ \hline
		Clutter & \multicolumn{1}{c|}{$-0.032$} & \multicolumn{1}{c|}{$.324$}\\ \hline
		Clarity & \multicolumn{1}{c|}{$-0.008$} & \multicolumn{1}{c|}{$.796$} \\ \hline
		Completeness & \multicolumn{1}{c|}{$-0.061$} & \multicolumn{1}{c|}{$.060$} \\ \hline
		Pleasure & \multicolumn{1}{c|}{$-0.054$} & \multicolumn{1}{c|}{$.094$} \\ \hline
		Intention & \multicolumn{1}{c|}{$-0.047$} & \multicolumn{1}{c|}{$.145$} \\ \hline
		Sense of responsibility & \multicolumn{1}{c|}{$0.037$} & \multicolumn{1}{c|}{$.259$} \\ \hline
		Support & \multicolumn{1}{c|}{$-0.007$} & \multicolumn{1}{c|}{$.841$} \\ \hline
		Stability & \multicolumn{1}{c|}{\cellcolor{gray!25}$0.067$} & \multicolumn{1}{c|}{\cellcolor{gray!25}$.038$} \\ \hline
	\end{tabular}
\end{table}

\subsection{Analysis: Binary Logistic Regression}
We used binary logistic regression to analyze the collected data since we wanted to investigate the relationship between the $15$ individual quality characteristics of vision videos (independent variables) and the overall quality of a vision video perceived by the subjects (dependent variable). The subsequent report of the logistic regression analysis follows the guidelines and recommendations by Peng et al. \cite{Peng.2002}. In particular, these guidelines and recommendations provide a concise reporting format containing all necessary tables, figures, and explanations required to evaluate the results of a logistic regression analysis.

\subsubsection{Assumptions}
Four assumptions need to be fulfilled to perform binary logistic regression \cite{Backhaus.2016, UZH}:

\begin{itemize}[leftmargin=0mm, topsep=0.5mm]
	\setlength\itemsep{-.25mm}
	\item[] $(1)$ The dependent variable is dichotomous.\\
	The dependent variable \textit{overall quality} is dichotomous with the two groups \textit{bad} coded as $0$ and \textit{good} coded as $1$.
	\item[] $(2)$ The independent variables are metric or categorical.\\
	 While \textit{video length} is measured in seconds (metric), the other $14$ characteristics are assessed on Likert scales (categorical).
	\item[] $(3)$ In the case of two or more metric independent variables, no multicollinearity is allowed to be present.\\
	This case is not fulfilled since \textit{video length} is the only metric independent variable.
	\item[] $(4)$ Both groups of the dichotomous dependent variable contain at least $25$ elements.\\
	This assumption is fulfilled: Group \textit{bad} contains $281$ elements and group \textit{good} contains $671$ elements (see section \ref{sec:data-set-description}).
\end{itemize}

\begin{table*}[!t]
	\renewcommand{\arraystretch}{1.1}
	\centering
	\caption{Details of the logistic regression model (The light gray cells show that the respective quality characteristic is a statistically significant predictor.)}
	\label{tbl:t1}
	\begin{tabular}{lllllllll}
		\hline
		\multicolumn{1}{|l|}{\multirow{2}{*}{\textbf{Predictor}}} & \multicolumn{1}{c}{\multirow{2}{*}{\textbf{$\beta$}}} & \multicolumn{1}{c}{\textbf{Standard}} & \multicolumn{1}{c}{\textbf{Wald's}} & \multicolumn{1}{c}{\multirow{2}{*}{\textbf{$df$}}} & \multicolumn{1}{c}{\multirow{2}{*}{\textbf{$p$}}} & \multicolumn{1}{c}{\multirow{2}{*}{\textbf{$Exp(\beta)$}}} & \multicolumn{2}{c|}{\textbf{$95\%$ Confidence interval}} \\ 
		\multicolumn{1}{|l|}{} & \multicolumn{1}{c}{} & \multicolumn{1}{c}{\textbf{error}} & \multicolumn{1}{c}{$\chi^{2}$} & \multicolumn{1}{c}{} & \multicolumn{1}{c}{} & \multicolumn{1}{c}{} & \multicolumn{1}{c}{\textbf{Lower value}} & \multicolumn{1}{c|}{\textbf{Upper value}} \\ \hline
		
		\multicolumn{1}{|l|}{Video length} & \multicolumn{1}{c}{$.020$} & \multicolumn{1}{c}{$.004$} & \multicolumn{1}{c}{\cellcolor{gray!25}$22.863$} & \multicolumn{1}{c}{\cellcolor{gray!25}$1$} & \multicolumn{1}{c}{\cellcolor{gray!25}$.000$} & \multicolumn{1}{c}{$1.020$} & \multicolumn{1}{c}{$1.012$} & \multicolumn{1}{c|}{$1.029$} \\ \hline
		
		\multicolumn{1}{|l|}{Focus (F)} & \multicolumn{1}{c}{} & \multicolumn{1}{c}{} & \multicolumn{1}{c}{\cellcolor{gray!25}$16.282$} & \multicolumn{1}{c}{\cellcolor{gray!25}$4$} & \multicolumn{1}{c}{\cellcolor{gray!25}$.003$} & \multicolumn{1}{c}{} & \multicolumn{1}{c}{} & \multicolumn{1}{c|}{} \\
		\multicolumn{1}{|l|}{\MyIndent F: Non-compact} & \multicolumn{1}{c}{$-.683$} & \multicolumn{1}{c}{$.276$} & \multicolumn{1}{c}{$6.109$} & \multicolumn{1}{c}{$1$} & \multicolumn{1}{c}{$.013$} & \multicolumn{1}{c}{$.505$} & \multicolumn{1}{c}{$.294$} & \multicolumn{1}{c|}{$.868$} \\
		\multicolumn{1}{|l|}{\MyIndent F: Neutral} & \multicolumn{1}{c}{$-.214$} & \multicolumn{1}{c}{$.205$} & \multicolumn{1}{c}{$1.086$} & \multicolumn{1}{c}{$1$} & \multicolumn{1}{c}{$.297$} & \multicolumn{1}{c}{$.807$} & \multicolumn{1}{c}{$.540$} & \multicolumn{1}{c|}{$1.207$} \\
		\multicolumn{1}{|l|}{\MyIndent F: Compact} & \multicolumn{1}{c}{$.469$} & \multicolumn{1}{c}{$.198$} & \multicolumn{1}{c}{$5.627$} & \multicolumn{1}{c}{$1$} & \multicolumn{1}{c}{$.018$} & \multicolumn{1}{c}{$1.599$} & \multicolumn{1}{c}{$1.085$} & \multicolumn{1}{c|}{$2.357$} \\
		\multicolumn{1}{|l|}{\MyIndent F: Very compact} & \multicolumn{1}{c}{$.515$} & \multicolumn{1}{c}{$.272$} & \multicolumn{1}{c}{$3.585$} & \multicolumn{1}{c}{$1$} & \multicolumn{1}{c}{$.058$} & \multicolumn{1}{c}{$1.673$} & \multicolumn{1}{c}{$.982$} & \multicolumn{1}{c|}{$2.852$} \\ \hline
		
		\multicolumn{1}{|l|}{Prior knowledge (PK)} & \multicolumn{1}{c}{} & \multicolumn{1}{c}{} & \multicolumn{1}{c}{\cellcolor{gray!25}$38.838$} & \multicolumn{1}{c}{\cellcolor{gray!25}$4$} & \multicolumn{1}{c}{\cellcolor{gray!25}$.000$} & \multicolumn{1}{c}{} & \multicolumn{1}{c}{} & \multicolumn{1}{c|}{} \\
		\multicolumn{1}{|l|}{\MyIndent PK: Unnecessary} & \multicolumn{1}{c}{$1.022$} & \multicolumn{1}{c}{$.213$} & \multicolumn{1}{c}{$22.926$} & \multicolumn{1}{c}{$1$} & \multicolumn{1}{c}{$.000$} & \multicolumn{1}{c}{$2.779$} & \multicolumn{1}{c}{$1.829$} & \multicolumn{1}{c|}{$4.223$} \\
		\multicolumn{1}{|l|}{\MyIndent PK: Neutral} & \multicolumn{1}{c}{$.203$} & \multicolumn{1}{c}{$.168$} & \multicolumn{1}{c}{$1.450$} & \multicolumn{1}{c}{$1$} & \multicolumn{1}{c}{$.229$} & \multicolumn{1}{c}{$1.225$} & \multicolumn{1}{c}{$.881$} & \multicolumn{1}{c|}{$1.703$} \\
		\multicolumn{1}{|l|}{\MyIndent PK: Necessary} & \multicolumn{1}{c}{$-.216$} & \multicolumn{1}{c}{$.187$} & \multicolumn{1}{c}{$1.338$} & \multicolumn{1}{c}{$1$} & \multicolumn{1}{c}{$.247$} & \multicolumn{1}{c}{$.806$} & \multicolumn{1}{c}{$.559$} & \multicolumn{1}{c|}{$1.162$} \\
		\multicolumn{1}{|l|}{\MyIndent PK: Very necessary} & \multicolumn{1}{c}{$-1.104$} & \multicolumn{1}{c}{$.227$} & \multicolumn{1}{c}{$23.714$} & \multicolumn{1}{c}{$1$} & \multicolumn{1}{c}{$.000$} & \multicolumn{1}{c}{$.331$} & \multicolumn{1}{c}{$.212$} & \multicolumn{1}{c|}{$.517$} \\ \hline
		
		\multicolumn{1}{|l|}{Clarity (C)} & \multicolumn{1}{c}{} & \multicolumn{1}{c}{} & \multicolumn{1}{c}{\cellcolor{gray!25}$36.644$} & \multicolumn{1}{c}{\cellcolor{gray!25}$4$} & \multicolumn{1}{c}{\cellcolor{gray!25}$.000$} & \multicolumn{1}{c}{} & \multicolumn{1}{c}{} & \multicolumn{1}{c|}{} \\
		\multicolumn{1}{|l|}{\MyIndent C: Unintelligible} & \multicolumn{1}{c}{$-.446$} & \multicolumn{1}{c}{$.265$} & \multicolumn{1}{c}{$2.829$} & \multicolumn{1}{c}{$1$} & \multicolumn{1}{c}{$.093$} & \multicolumn{1}{c}{$.640$} & \multicolumn{1}{c}{$.381$} & \multicolumn{1}{c|}{$1.076$} \\
		\multicolumn{1}{|l|}{\MyIndent C: Neutral} & \multicolumn{1}{c}{$.113$} & \multicolumn{1}{c}{$.230$} & \multicolumn{1}{c}{$.239$} & \multicolumn{1}{c}{$1$} & \multicolumn{1}{c}{$.625$} & \multicolumn{1}{c}{$1.119$} & \multicolumn{1}{c}{$.713$} & \multicolumn{1}{c|}{$1.757$} \\
		\multicolumn{1}{|l|}{\MyIndent C: Intelligible} & \multicolumn{1}{c}{$.714$} & \multicolumn{1}{c}{$.223$} & \multicolumn{1}{c}{$10.247$} & \multicolumn{1}{c}{$1$} & \multicolumn{1}{c}{$.001$} & \multicolumn{1}{c}{$2.043$} & \multicolumn{1}{c}{$1.319$} & \multicolumn{1}{c|}{$3.164$} \\
		\multicolumn{1}{|l|}{\MyIndent C: Very intelligible} & \multicolumn{1}{c}{$1.343$} & \multicolumn{1}{c}{$.259$} & \multicolumn{1}{c}{$26.876$} & \multicolumn{1}{c}{$1$} & \multicolumn{1}{c}{$.000$} & \multicolumn{1}{c}{$3.831$} & \multicolumn{1}{c}{$2.306$} & \multicolumn{1}{c|}{$6.367$} \\ \hline
		
		\multicolumn{1}{|l|}{Pleasure (PS)} & \multicolumn{1}{c}{} & \multicolumn{1}{c}{} & \multicolumn{1}{c}{\cellcolor{gray!25}$72.749$} & \multicolumn{1}{c}{\cellcolor{gray!25}$4$} & \multicolumn{1}{c}{\cellcolor{gray!25}$.000$} & \multicolumn{1}{c}{} & \multicolumn{1}{c}{} & \multicolumn{1}{c|}{} \\
		\multicolumn{1}{|l|}{\MyIndent PS: Unenjoyable} & \multicolumn{1}{c}{$-1.408$} & \multicolumn{1}{c}{$.285$} & \multicolumn{1}{c}{$24.431$} & \multicolumn{1}{c}{$1$} & \multicolumn{1}{c}{$.000$} & \multicolumn{1}{c}{$.245$} & \multicolumn{1}{c}{$.140$} & \multicolumn{1}{c|}{$.428$} \\
		\multicolumn{1}{|l|}{\MyIndent PS: Neutral} & \multicolumn{1}{c}{$-.147$} & \multicolumn{1}{c}{$.219$} & \multicolumn{1}{c}{$.453$} & \multicolumn{1}{c}{$1$} & \multicolumn{1}{c}{$.501$} & \multicolumn{1}{c}{$.863$} & \multicolumn{1}{c}{$.562$} & \multicolumn{1}{c|}{$1.325$} \\
		\multicolumn{1}{|l|}{\MyIndent PS: Enjoyable} & \multicolumn{1}{c}{$.914$} & \multicolumn{1}{c}{$.216$} & \multicolumn{1}{c}{$17.828$} & \multicolumn{1}{c}{$1$} & \multicolumn{1}{c}{$.000$} & \multicolumn{1}{c}{$2.493$} & \multicolumn{1}{c}{$1.631$} & \multicolumn{1}{c|}{$3.810$} \\
		\multicolumn{1}{|l|}{\MyIndent PS: Very enjoyable} & \multicolumn{1}{c}{$1.718$} & \multicolumn{1}{c}{$.302$} & \multicolumn{1}{c}{$32.248$} & \multicolumn{1}{c}{$1$} & \multicolumn{1}{c}{$.000$} & \multicolumn{1}{c}{$5.572$} & \multicolumn{1}{c}{$3.080$} & \multicolumn{1}{c|}{$10.081$} \\ \hline
		
		\multicolumn{1}{|l|}{Stability (ST)} & \multicolumn{1}{c}{} & \multicolumn{1}{c}{} & \multicolumn{1}{c}{\cellcolor{gray!25}$39.391$} & \multicolumn{1}{c}{\cellcolor{gray!25}$4$} & \multicolumn{1}{c}{\cellcolor{gray!25}$.000$} & \multicolumn{1}{c}{} & \multicolumn{1}{c}{} & \multicolumn{1}{c|}{} \\
		\multicolumn{1}{|l|}{\MyIndent ST: Unstable} & \multicolumn{1}{c}{$-1.006$} & \multicolumn{1}{c}{$.200$} & \multicolumn{1}{c}{$25.231$} & \multicolumn{1}{c}{$1$} & \multicolumn{1}{c}{$.000$} & \multicolumn{1}{c}{$.366$} & \multicolumn{1}{c}{$.247$} & \multicolumn{1}{c|}{$.542$} \\
		\multicolumn{1}{|l|}{\MyIndent ST: Neutral} & \multicolumn{1}{c}{$.181$} & \multicolumn{1}{c}{$.214$} & \multicolumn{1}{c}{$.717$} & \multicolumn{1}{c}{$1$} & \multicolumn{1}{c}{$.397$} & \multicolumn{1}{c}{$1.199$} & \multicolumn{1}{c}{$.788$} & \multicolumn{1}{c|}{$1.823$} \\
		\multicolumn{1}{|l|}{\MyIndent ST: Stable} & \multicolumn{1}{c}{$.270$} & \multicolumn{1}{c}{$.255$} & \multicolumn{1}{c}{$1.118$} & \multicolumn{1}{c}{$1$} & \multicolumn{1}{c}{$.290$} & \multicolumn{1}{c}{$1.309$} & \multicolumn{1}{c}{$.795$} & \multicolumn{1}{c|}{$2.158$} \\
		\multicolumn{1}{|l|}{\MyIndent ST: Very stable} & \multicolumn{1}{c}{$1.233$} & \multicolumn{1}{c}{$.536$} & \multicolumn{1}{c}{$5.287$} & \multicolumn{1}{c}{$1$} & \multicolumn{1}{c}{$.021$} & \multicolumn{1}{c}{$3.432$} & \multicolumn{1}{c}{$1.200$} & \multicolumn{1}{c|}{$9.816$} \\ \hline
		
		\multicolumn{1}{|l|}{Constant} & \multicolumn{1}{c}{$-.048$} & \multicolumn{1}{c}{$.253$} & \multicolumn{1}{c}{$.035$} & \multicolumn{1}{c}{$1$} & \multicolumn{1}{c}{$.851$} & \multicolumn{1}{c}{$.954$} & \multicolumn{1}{c}{} & \multicolumn{1}{c|}{} \\ \hline
	\end{tabular} 
\end{table*}

\begin{figure*}[!t]
	\captionsetup{justification=centering}
	\begin{subfigure}{.3\textwidth}
		\centering
		\includegraphics[width=\textwidth]{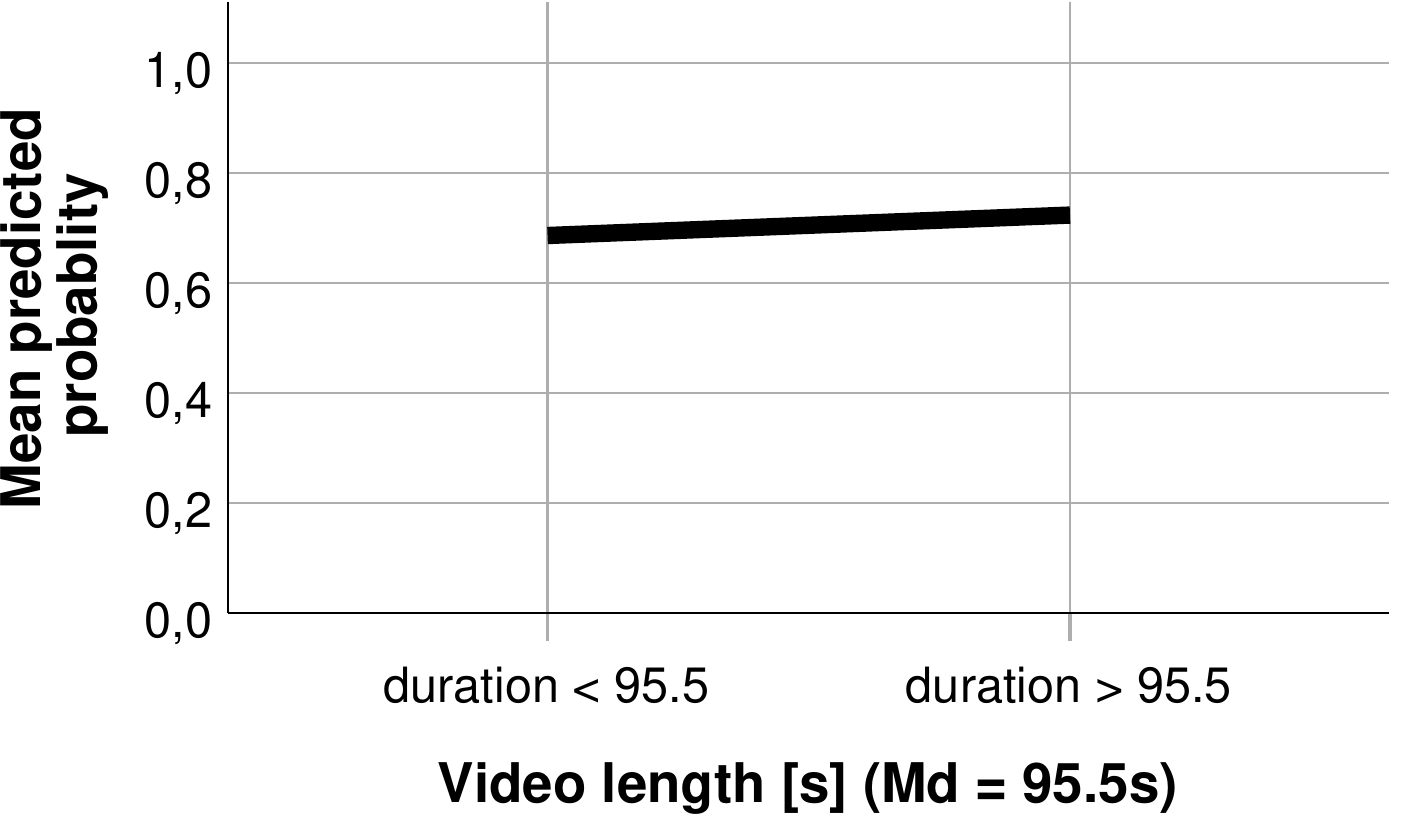}
		\caption{Mean predicted probability vs. video length}
		\label{fig10:sfig1}
	\end{subfigure}
	\begin{subfigure}{.3\textwidth}
		\centering
		\includegraphics[width=\textwidth]{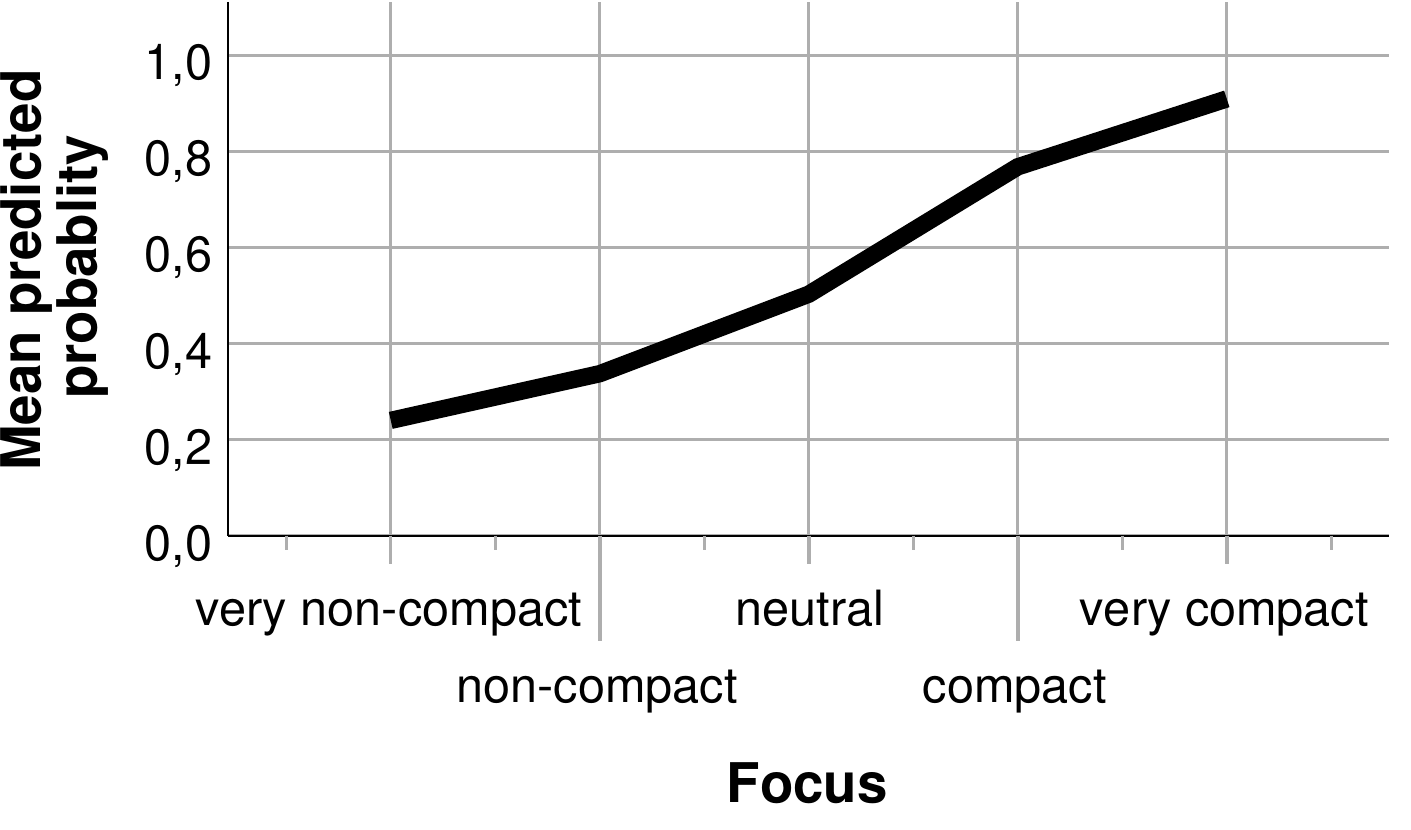}
		\caption{Mean predicted probability vs. focus}
		\label{fig10:sfig2}
	\end{subfigure}
	\begin{subfigure}{.3\textwidth}
		\centering
		\includegraphics[width=\textwidth]{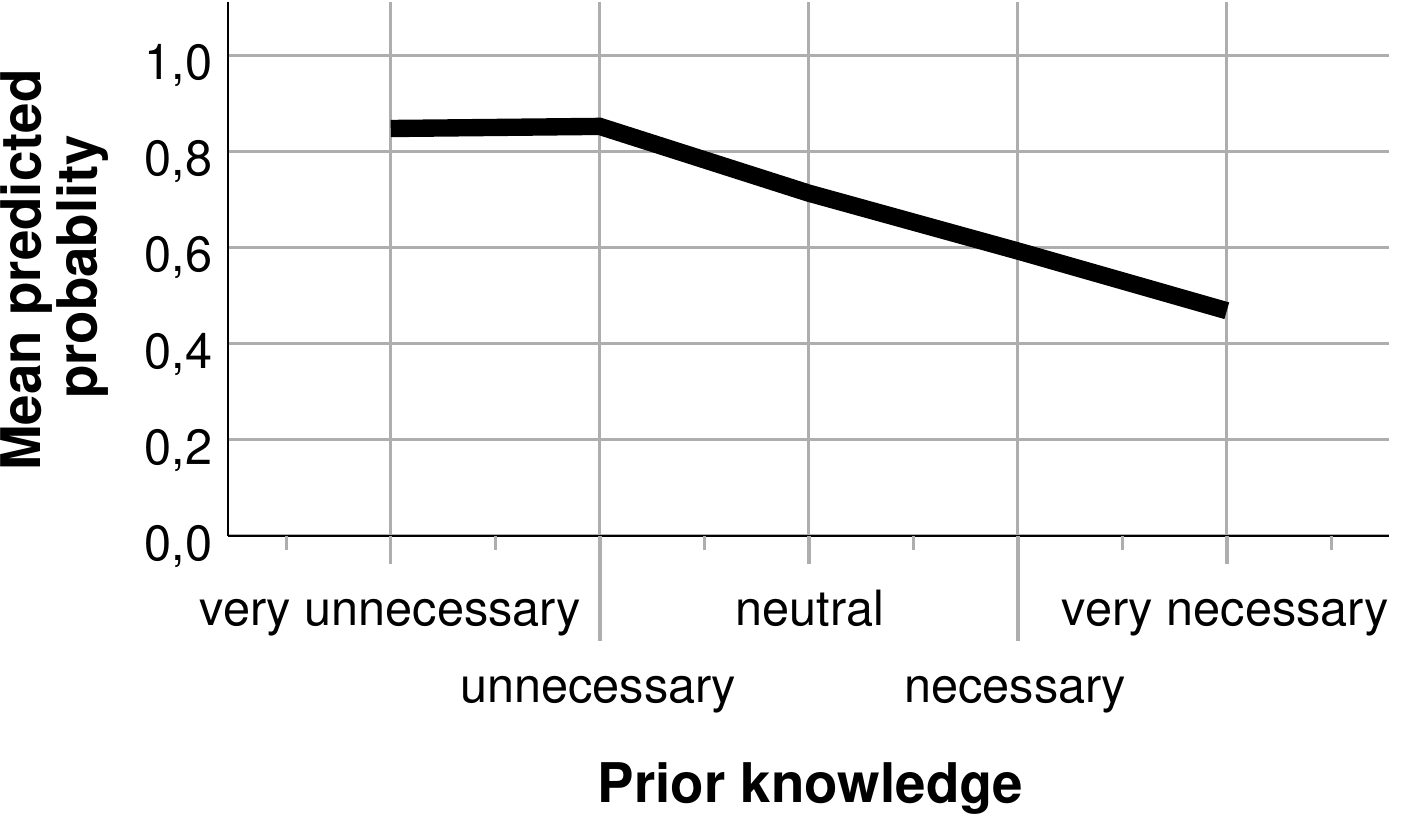}
		\caption{Mean predicted probability vs. prior knowledge}
		\label{fig10:sfig3}
	\end{subfigure}
	\begin{subfigure}{.3\textwidth}
		\centering
		\includegraphics[width=\textwidth]{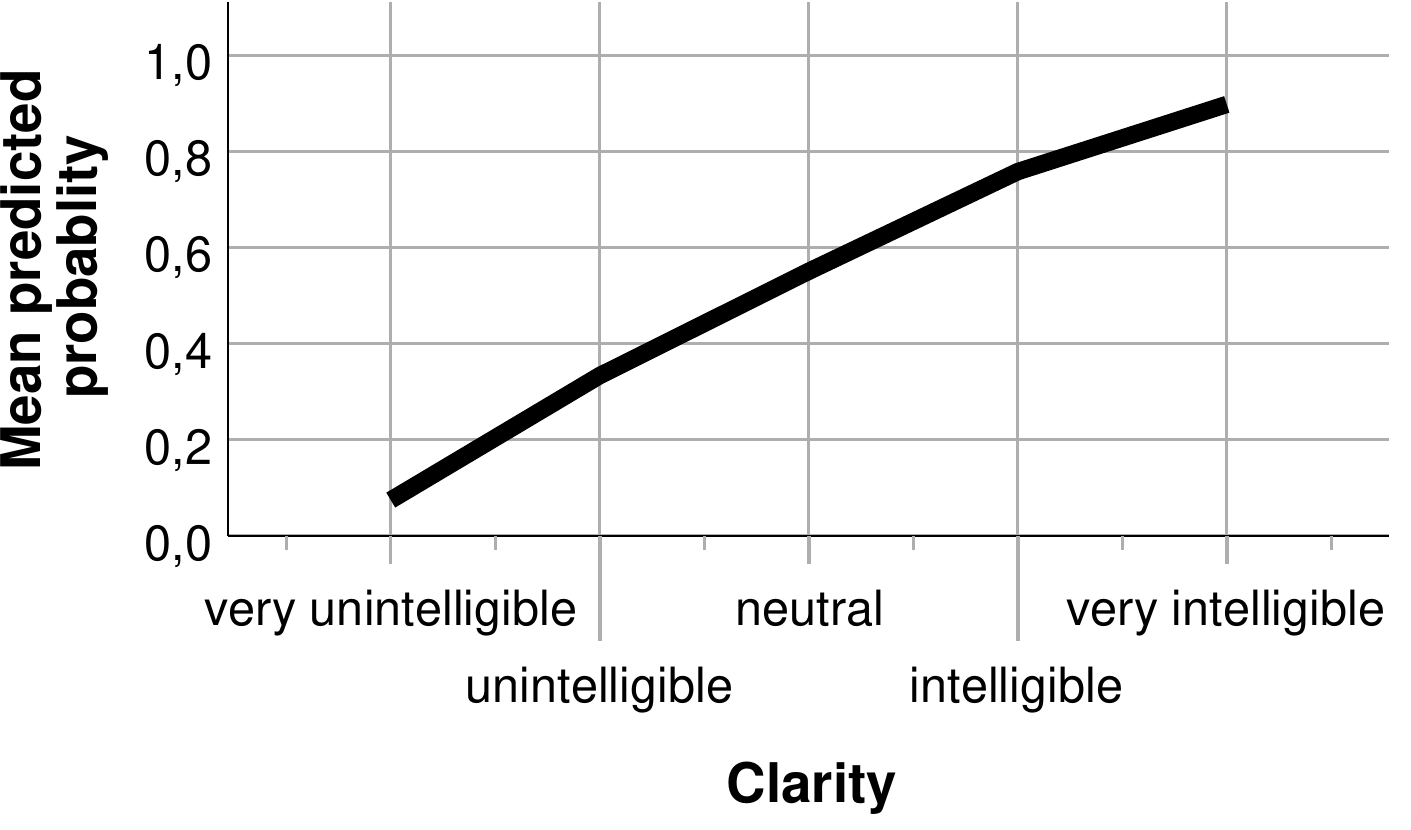}
		\caption{Mean predicted probability vs. clarity}
		\label{fig10:sfig4}
	\end{subfigure}
	\begin{subfigure}{.3\textwidth}
		\centering
		\includegraphics[width=\textwidth]{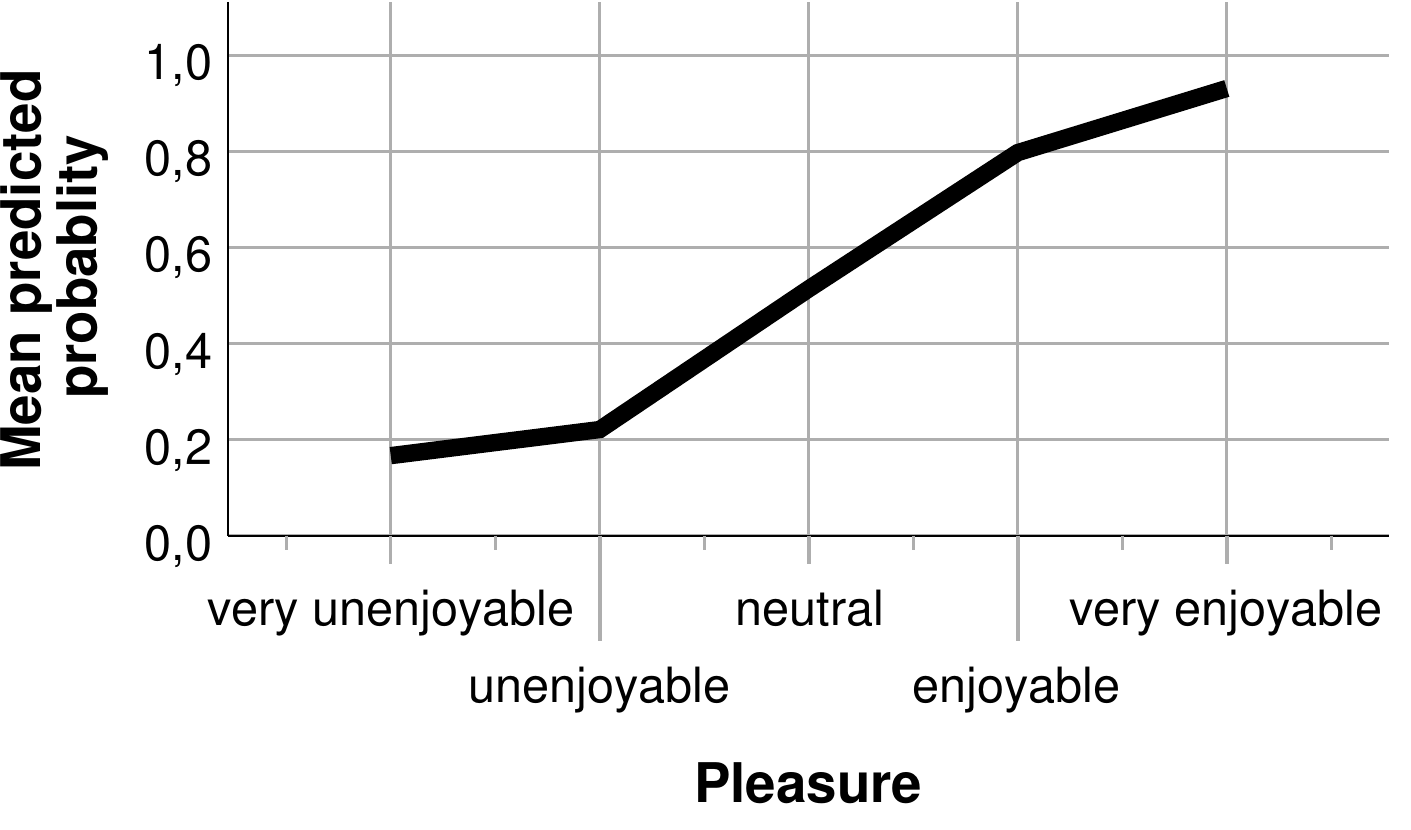}
		\caption{Mean predicted probability vs. pleasure}
		\label{fig10:sfig5}
	\end{subfigure}
	\begin{subfigure}{.3\textwidth}
		\centering
		\includegraphics[width=\textwidth]{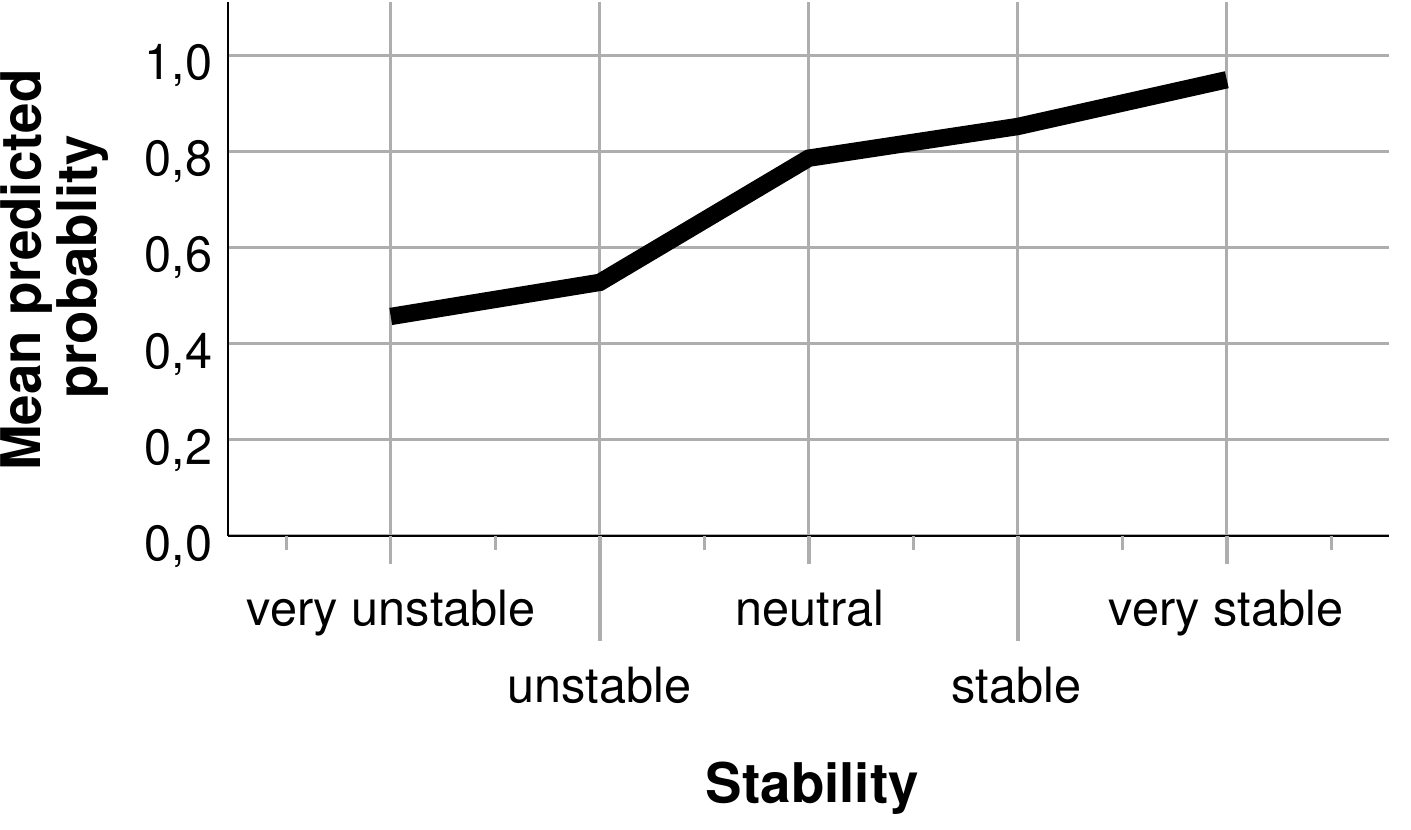}
		\caption{Mean predicted probability vs. stability}
		\label{fig10:sfig6}
	\end{subfigure}
	\centering
	\caption{Relationships between the mean predicted probability that the subjects perceive the overall quality of a vision video as good and each significant predictor}
	\label{fig:fig10}
\end{figure*}

\subsubsection{Results: Logistic Regression Model}
We fitted a logistic model to the data by using binary logistic regression to test our global and specific null hypotheses ($gH_{0}$ and $Hi_{0}$, $i \in \{1, \dots, 15\}$). The binary logistic regression was carried out by the binary logistic regression procedure in \textit{IBM SPSS Statistics}\footnote{\url{https://www.ibm.com/de-de/products/spss-statistics}} (Version 25). After two iterations, we obtained a logistic regression model only including quality characteristics of vision videos as predictors with statistical significance. 
In the first iteration, we performed binary logistic regression with all $15$ quality characteristics. As a result, we found that the nine characteristics \textit{image quality}, \textit{sound quality}, \textit{plot}, \textit{essence}, \textit{clutter}, \textit{completeness}, \textit{intention}, \textit{sense of responsibility}, and \textit{support} do not affect the likelihood that the subjects perceive the overall quality of a vision video as good. Thus, we cannot reject the specific null hypotheses $Hi_{0}$, $i \in \{1, 2, 5, 8, 9, 10, 12, 13, 14\}$. Therefore, we removed these nine characteristics from the $15$ characteristics in the second iteration. The resulting logistic regression model contained all six remaining quality characteristics as significant predictors for the overall quality of a vision video. These six characteristics are \textit{video length}, \textit{focus}, \textit{prior knowledge}, \textit{clarity}, \textit{pleasure}, and \textit{stability}. \tablename{ \ref{tbl:t1}} reports the details of the final logistic regression model. In the following, we report the identified relationships between the predicted probability of the logistic regression model and each predictor. \figurename{ \ref{fig:fig10}} illustrates these relationships for each predictor.\\
 
\noindent 
Based on the logistic model, \textit{the likelihood that the subjects perceive the overall quality of a vision video as good from a developer's point of view is} \dots

\begin{itemize}[leftmargin=0mm, topsep=0.5mm]
	\setlength\itemsep{-.25mm}
	\item[] \textbf{Finding 1}: \dots \textit{positively} related to the video length of a vision video $(p = .000, Exp(\beta) = 1.020 > 1$).\\
	\textit{Interpretation}: We can reject the specific null hypothesis $H3_{0}$. The closer the \textit{video length} is to the given maximum duration of $3$ minutes\footnote{We have to remark that the used vision videos had a maximum duration of $2$:$35$ minutes. It is possible that a much longer duration may have a negative impact on the overall quality of a vision video.}, i.e., the longer the duration of a vision video, the higher the likelihood that the subjects perceive the overall quality of a vision video as good (see \figurename{ \ref{fig10:sfig1}}).
	
	\item[] \textbf{Finding 2}: \dots ($1$) \textit{negatively} related to a non-compact $(p = .013, Exp(\beta) = .505 < 1$) and ($2$) \textit{positively} related to a compact representation of a vision $(p = .018, Exp(\beta) = 1.599 > 1$).\\
	\textit{Interpretation}: We can reject the specific null hypothesis $H4_{0}$.\\
	($1$) means: The lower the value for \textit{focus}, i.e, the less compact a vision video represents a vision, the lower the likelihood that the subjects perceive the overall quality of a vision video as good (see \figurename{ \ref{fig10:sfig2}}).\\
	($2$) means: The higher the value for \textit{focus}, i.e., the more compact a vision video represents a vision, the higher the likelihood that the subjects perceive the overall quality of a vision video as good (see \figurename{ \ref{fig10:sfig2}}).
	
	\item[] \textbf{Finding 3}: \dots ($1$) \textit{positively} related to unnecessary $(p = .000, Exp(\beta) = 2.779 > 1$) and ($2$) \textit{negatively} related to very necessary prior knowledge $(p = .000, Exp(\beta) = .331 < 1$).\\
	\textit{Interpretation}: We can reject the specific null hypothesis $H6_{0}$.\\
	($1$) means: The lower the value for \textit{prior knowledge}, i.e., the less prior knowledge is necessary to understand a vision video, the higher the likelihood that the subjects perceive the overall quality of a vision video as good (see \figurename{ \ref{fig10:sfig3}}).\\
	($2$) means: The higher the value for \textit{prior knowledge}, i.e., the more prior knowledge is necessary to understand a vision video, the lower the likelihood that the subjects perceive the overall quality of a vision video as good (see \figurename{ \ref{fig10:sfig3}}).
	
	\item[] \textbf{Finding 4}: \dots ($1$) \textit{positively} related to intelligible $(p = .001, Exp(\beta) = 2.043 > 1$) and ($2$) \textit{positively} related to very intelligible aspired goals of a vision $(p = .000, Exp(\beta) = 3.831 > 1$).\\
	\textit{Interpretation}: We can reject the specific null hypothesis $H7_{0}$.\\
	($1$) and ($2$) mean: The higher the value for \textit{clarity}, i.e., the more intelligible the aspired goals of a vision represented in a vision video, the higher the likelihood that the subjects perceive the overall quality of a vision video as good (see \figurename{ \ref{fig10:sfig4}}).
	
	\item[] \textbf{Finding 5}: \dots ($1$) \textit{negatively} related to an unenjoyable $(p = .000, Exp(\beta) = .245 < 1$), ($2$) \textit{positively} related to an enjoyable $(p = .001, Exp(\beta) = 2.043 > 1$), and ($3$) \textit{positively} related to a very enjoyable vision video $(p = .000, Exp(\beta) = 5.572 > 1$).\\
	\textit{Interpretation}: We can reject the specific null hypothesis $H11_{0}$.\\
	($1$) means: The lower the value for \textit{pleasure}, i.e., the less enjoyable a vision video is to watch, the lower the likelihood that the subjects perceive the overall quality of a vision video as good (see \figurename{ \ref{fig10:sfig5}}).\\
	($2$) and ($3$) mean: The higher the value for \textit{pleasure}, i.e., the more enjoyable a vision video is to watch, the higher the likelihood that the subjects perceive the overall quality of a vision video as good (see \figurename{ \ref{fig10:sfig5}}).
	
	\item[] \textbf{Finding 6}: \dots ($1$) \textit{negatively} related to an unstable $(p = .000, Exp(\beta) = .366 < 1$) and ($2$) \textit{positively} related to a very stable vision $(p = .021, Exp(\beta) = 3.432 > 1$).\\	
	\textit{Interpretation}: We can reject the specific null hypothesis $H15_{0}$.\\
	($1$) means: The lower the value for \textit{stability}, i.e., the less stable a vision presented in a vision video, the lower the likelihood that the subjects perceive the overall quality of a vision video as good (see \figurename{ \ref{fig10:sfig6}}).\\
	($2$) means: The higher the value for \textit{stability}, i.e., the more stable a vision presented in a vision video, the higher the likelihood that the subjects perceive the overall quality of a vision video as good (see \figurename{ \ref{fig10:sfig6}}).
\end{itemize}

\subsection{Evaluation of the Logistic Regression Model}
We need to evaluate the obtained logistic regression model regarding its effectiveness and soundness to draw a clear conclusion about the global null hypothesis $gH_{0}$. Below, we present the four required evaluations of a logistic regression model according to the reporting format by Peng et al. \cite{Peng.2002}.

\subsubsection{Overall Model Evaluation}
\label{sec:overall-model-evaluation}
We compared the obtained model with the intercept-only model, so-called null model. The null model serves as the baseline since it contains no predictors. Consequently, the null model would predict all observations to belong to the largest outcome category. The obtained logistic model provides a better fit to the data if it shows a significant improvement compared to the null model. We investigated whether there is such an improvement by using the \textit{Likelihood Ratio} test (see \tablename{ \ref{tbl:t2}}). The test indicated that the obtained logistic model is significantly better than the null model $(\chi^{2}(21) = 450.85, p = .000)$.

\begin{table}[htbp]
	\captionsetup{justification=centering}
	\renewcommand{\arraystretch}{1.1}
	\centering
	\caption{Significance and quality of the logistic regression model}
	\label{tbl:t2}
	\begin{tabular}{llllll}
		\hline
		\multicolumn{3}{|l}{\textbf{Model}} & \multicolumn{1}{c}{$ \chi^{2} $} & \multicolumn{1}{c}{$df$} & \multicolumn{1}{c|}{$ p $} \\ \hline
		\multicolumn{3}{|l}{Overall model evaluation} & \multicolumn{1}{c}{} & \multicolumn{1}{c}{} & \multicolumn{1}{c|}{} \\
		\multicolumn{3}{|l}{\MyIndent Likelihood ratio test} & \multicolumn{1}{c}{$450.851$} & \multicolumn{1}{c}{$21$} & \multicolumn{1}{c|}{$.000$} \\ \hline
		\multicolumn{3}{|l}{Goodness-of-fits test} & \multicolumn{1}{c}{} & \multicolumn{1}{c}{} & \multicolumn{1}{c|}{} \\
		\multicolumn{3}{|l}{\MyIndent Hosmer-Lemeshow test} & \multicolumn{1}{c}{$13.169$} & \multicolumn{1}{c}{$8$} & \multicolumn{1}{c|}{$.106$} \\ \hline
		\multicolumn{6}{|l|}{\multirow{2}{*}{\begin{tabular}[l]{@{}l@{}}	\textit{Note}. Nagelkerke's $R^{2} = 0.54$, effect size $\eta^{2} = 0.54$ \end{tabular}}} \\ 
		\multicolumn{6}{|l|}{} \\ \hline
	\end{tabular} 
\end{table}

\subsubsection{Statistical Tests of Individual Predictors}
\label{sec:statistical-tests-of-individual-predictors}
For each predictor, the logistic regression analysis performed a \textit{z}-test to investigate the statistical significance of the individual predictors. According to \tablename{ \ref{tbl:t1}} (light gray cells), all six characteristics \textit{video length}, \textit{focus}, \textit{prior knowledge}, \textit{clarity}, \textit{pleasure}, and \textit{stability} are statistically significant predictors of the overall quality of a vision video.

\subsubsection{Goodness-of-fit Statistics}
\label{sec:goodness-of-fit-statistics}
The goodness-of-fit statistics include the inferential \textit{Hosmer-Lemeshow} (H-L) test and the descriptive measure $R^{2}$ defined by Nagelkerke \cite{Nagelkerke.1991}. These statistics investigate the fit of the logistic model against the actual outcomes, i.e., whether the perceived overall quality of a vision video is good. For the H-L test, the $p$-value should be greater than $.05$ since the null hypothesis of this test assumes the model fits the data. The $R^{2}$ measure explains how much variation of the dependent variable is account for by the predictors in the model. $R^{2}$ ranges from $0$ to $1$. The closer the $R^{2}$ value is to $1$, the more variability in the actual data set can be explained by the model. The $R^{2}$ value can be converted to the effect size $\eta^{2}$ to assess the practical relevance of the findings. According to \tablename{ \ref{tbl:t2}, the \textit{Hosmer-Lemeshow} test showed that the obtained model fits the data well $(\chi^{2}(8) = 13.17, p = .106)$. The $R^{2}$ value is $0.54$. Thus, all six predictors in the model explained $54.0\%$ of the variability of the perceived overall quality of a vision video. The effect size value $(\eta^{2} = 0.54 > 0.20)$ indicates a large practical relevance of the model according to Cohen \cite{Cohen.2013}.

\subsubsection{Validations of Predicted Probabilities}
\label{sec:validations-of-predicted-probabilities}
A logistic regression model can be used to predict the probability that the dependent variable belongs to the desired group of its two possible groups. Based on the predicted probabilities, the model classifies each entry of the analyzed data set into one of the two groups. In general, the desired group is associated with a higher probability and the other group with a lower probability. However, the exact distinction between high and low depends on the individual data set. In our case, using the six significant predictors, the model predicts the probability that subjects perceive the overall quality of a vision video as good. The resulting classification of the model is compared with the actual classification of the analyzed data set to determine the accuracy of the logistic regression model. This accuracy can be expressed as a measure of association and a classification table.

An established measure of association is the \textit{c}-statistic. The \textit{c}-statistic represents the proportion of pairs with different observed groups for which the model correctly predicts a higher probability for observations with the desired group (perceived overall quality is \textit{good}) than for observations without the desired group (perceived overall quality is \textit{bad}). This measure ranges from $0.5$ to $1$. While $0.5$ means that the model is not better than assigning observations randomly to the groups, $1$ means that the model always assigns higher probabilities to observations with the desired group than to observations without the desired group. For the obtained model, the \textit{c}-statistic is $0.89$ (see \tablename{ \ref{tbl:t3}}). This means that for $89.0\%$ of all possible pairs of vision video assessments -- one with an overall quality rated as good and one with an overall quality rated as bad -- the model correctly assigned the higher probability to those assessments which rated a vision video as good.

\begin{table}[htbp]
	\captionsetup{justification=centering}
	\renewcommand{\arraystretch}{1.1}
	\centering
	\caption{Observed and predicted frequencies for overall vision video quality by logistic regression with cutoff value of 0.705}
	\label{tbl:t3}
	\begin{tabular}{|ll|cc|c|}
		\hline
		\multicolumn{2}{|c|}{\multirow{3}{*}{\textbf{Observed}}} & \multicolumn{2}{c|}{\textbf{Predicted}} & \multicolumn{1}{c|}{\multirow{3}{*}{\textbf{$\%$ Correct}}} \\ \cline{3-4}
		\multicolumn{2}{|c|}{} & \multicolumn{2}{c|}{\textbf{Overall quality}} & \\
		\multicolumn{2}{|c|}{} & Bad & \multicolumn{1}{c|}{Good} & \\ \hline
		\multirow{2}{*}{\begin{tabular}[c]{@{}c@{}}\textbf{Overall}\\ \textbf{quality}\end{tabular}} & Bad & $ 231 $ & $ 50 $ & $ 82.2 $ (Specificity)\\
		& Good & $ 127 $ & $ 544 $ & $ 81.1 $ (Sensitivity) \\ \hline
		\multicolumn{4}{|l}{\begin{tabular}[l]{@{}c@{}}\textbf{Overall correct prediction rate}\end{tabular}} & \multicolumn{1}{l|}{$ 81.4 $} \\ \hline
		\multicolumn{5}{|l|}{\multirow{1}{*}{\begin{tabular}[l]{@{}l@{}}	\textit{Note}. \textit{c}-statistic $= 0.89$\end{tabular}}}\\ \hline
	\end{tabular}
\end{table}

Besides the measure of association, a classification table illustrates the validity of predicted probabilities (see \tablename{ \ref{tbl:t3}}). The classification table shows the practical results of applying the logistic regression model on the actual data set. For each entry in the data set, the model calculates the predicted probability and classifies the entry into one of the two possible groups of the dependent variable. The classification depends on a defined cutoff value. This cutoff value typically corresponds to the proportion of entries in the actual data set that rated the dependent variable according to the desired group. In our case, the cutoff value is $0.705$ corresponding to the proportion of entries which rated the overall quality as good. Thus, the model classified an entry as \textit{good} if the predicted probability was greater than the cutoff value and otherwise as \textit{bad}. A classification table presents three important proportions: \textit{specificity}, \textit{sensitivity}, and \textit{overall correct prediction rate}. The \textit{specificity} measures the proportion of correctly classified observations without the desired group, i.e., perceived overall quality is bad. The \textit{sensitivity} measures the proportion of correctly classified observations with the desired group, i.e., perceived overall quality is good. The \textit{overall correct prediction rate} measures the proportion of all correctly classified observations.

The \textit{specificity} $(82.2\%)$ and the \textit{sensitivity} $(81.1\%)$ show that the prediction of entries which rated the overall quality as bad was slightly more accurate than the prediction of entries which rated the overall quality as good. However, in both cases, we have a high accuracy of correct predictions. This observation is supported by the \textit{overall correct prediction rate} which is $81.4\%$. This is an improvement of $10.9$ percentage points compared to the null model which has an overall correct prediction rate of $70.5\%$. Based on the \textit{c}-statistic and the results of the classification table, we can conclude that the obtained logistic regression model has high accuracy in predicting whether the subjects perceive a vision video as good.

\subsection{Summary: Binary Logistic Regression}
The binary logistic regression resulted in a significant model $(\chi^{2}(27) = 476.01, p = .000)$ that contains six characteristics (\textit{video length}, \textit{focus}, \textit{prior knowledge}, \textit{clarity}, \textit{pleasure}, and \textit{stability}) all of which are significant predictors of the perceived overall quality of a vision video. The other nine characteristics of the quality model for vision videos (\textit{image quality}, \textit{sound quality}, \textit{plot}, \textit{essence}, \textit{clutter}, \textit{completeness}, \textit{intention}, \textit{sense of responsibility}, and \textit{support}) are not significant and thus do not affect the likelihood that the subjects perceive the overall quality of a vision video as good. All predictors of the obtained model explain $54.0\%$ of the variability of the perceived overall quality of a vision video. Thereby, the effect size $\eta^{2} = 0.54$ indicates a large practical relevance of the obtained model. The model correctly classifies $81.1\%$ of the entries in the actual data set where the overall quality was assessed as good and $82.2\%$ of entries where the overall quality was assessed as bad, resulting in an overall correct prediction rate of $81.4\%$. Based on these results, we conclude that the obtained logistic regression model is effective and sound. These findings allow us to reject the null hypothesis $gH_{0}$ and accept the alternative hypothesis $gH_{1}$.
According to the logistic regression results, six quality characteristics (concretely three vision and three video characteristics) covering all three dimensions of our proposed quality model for vision videos affect the likelihood that the subjects perceive the overall quality of a vision video as good from a developer's point of view. Therefore, we conclude that the proposed quality model for vision videos is of fundamental relevance (see section \ref{sec:hypotheses}). We reached this conclusion with multiple evidence:

\begin{itemize}[leftmargin=7mm,topsep=0mm]
	\setlength\itemsep{-1.0mm}
	\item[($1$)] The logistic regression model is significantly better than the null model (see section \ref{sec:overall-model-evaluation}).
	\item[($2$)] All six quality characteristics included in the logistic regression model are significant predictors of the perceived overall quality of a vision video (see section \ref{sec:statistical-tests-of-individual-predictors}).
	\item[($3$)] The logistic regression model fits the data well and has a large practical relevance (see section \ref{sec:goodness-of-fit-statistics}).
	\item[($4$)] The logistic regression model has high accuracy in predicting whether a vision video is perceived as good or not (see section \ref{sec:validations-of-predicted-probabilities}).
\end{itemize}

\noindent
As an answer to our research question, we can summarize:

\begin{mdframed}
	\begin{itemize}[leftmargin=-2.5mm]
		\item[] \textbf{Answer:}
		Six out of the $15$ individual characteristics of the proposed quality model are statistically significantly related to the overall quality of vision videos. In particular, the better each of the six characteristics is fulfilled the higher is the likelihood that the overall quality of a vision video was perceived as good.
	\end{itemize}
\end{mdframed}

\section{Threats to Validity}
\label{sec:threats-to-validity}
We report threats to validity divided by the two phases of our research approach: \textit{literature reviews} and \textit{evaluation}. We considered threats to construct, external, internal, and conclusion validity according to Wohlin et al. \cite{Wohlin.2012}. 

\subsection{Literature Reviews: Threats to Validity}

\subsubsection{Construct Validity}
We performed two literature reviews which are probably one of the most simplified methods to fast and easily investigate literature compared to more complex methods such as a systematic literature review. However, this type of literature review is still a systematic approach. Although this simplified method has its weaknesses, its systematic reflection is nevertheless a useful and proven means for developing a valid body of knowledge \cite{Glinz.2015}.
Due to this simplified investigation of literature, we do not claim to present comprehensive systematic literature reviews on video production guidelines as well as software project vision. The goal of these initial literature reviews was not to achieve a comprehensive overview of all existing literature. Instead, we wanted to propose a first quality model for vision videos grounded and reflected on a valid body of knowledge. While we examined six generic video production guidelines for the video characteristics (see section \ref{sec:results_video}), we examined ten books and nine journal articles for the vision characteristics (see section \ref{sec:results_vision}). We verified the deduced characteristics of both literature reviews by evaluating the reliability of two raters using \textit{Cohen's kappa}. In both cases, the calculated \textit{Cohen's kappa} value shows an almost perfect raters' agreement. We are therefore confident that the identified categories and sub-categories in the analyzed literature represent major themes in terms of vision and video characteristics. However, the entire elaboration of the quality characteristics of vision videos is based only on the literature reviews. This use of one single method causes a mono-method bias. There is a lack of extraction of information about the quality characteristics of vision videos from practitioners producing vision videos. For example, other researchers working on the use of vision videos as well as the students, who produced the videos which we used in the evaluation, might be potential sources. Their experience in video production might help to verify the already known quality characteristics as well as identify additional characteristics not yet considered in the literature. This investigation requires the use of further methods such as questionnaires, focus groups, or brainstorming sessions. As future work, we plan this kind of expert evaluation. In particular, we want to conduct a delphi study. This type of study is a specialized form of survey using a questionnaire to achieve consensus between researchers and practitioners on a particular topic \cite{Delbecq.1975}.

\subsubsection{External Validity}
The selected method of literature review restricts the generalizability of our findings. We cannot guarantee that the proposed quality characteristics of videos and visions are complete since the literature reviews carried out are not systematic literature reviews and therefore not all relevant literature should be identified. However, to the best of our knowledge, this article provides the first quality model for assessing the rather ill-defined concept of (vision) \textit{video quality} which is based on mutually supportive references. Although the identified characteristics are only an initial proposal, the resulting quality model provides a viable basis for future refinements and extensions.

\subsubsection{Internal Validity}
The results of the literature reviews are based on six generic guidelines for video production and ten books as well as nine journal articles on software project vision. The selection of literature is probably the most crucial threat to internal validity since any bias in the selection affects the accuracy and quality of the final results. We mitigated this threat to validity by following specified search processes with defined exclusion and inclusion criteria. As part of the search process, the third authors reviewed the work the first authors to ensure the relevance and suitability of the literature used. In addition, two raters classified all extracted passages of both literature studies on their own by assigning the respective deduced characteristics. The calculated \textit{Cohen's kappa} values show an almost perfect raters' agreement. For these reasons, we are confident that we selected suitable literature that deals with video and vision characteristics. However, the number of analyzed guidelines, books, and journal articles is rather small and thus restricts the validity of the deduced characteristics. For the characteristics of videos, we consciously decided to focus on generic guidelines to deduce common characteristics of videos since we did not find any publication defining characteristics of videos. We excluded guidance focusing on specific kinds of videos, such as tutorials or demonstrations, to reduce any potential bias which might have been caused by a specific application context. For the vision characteristics, we identified several publications describing what constitutes a vision. We analyzed a larger number of publications on software project vision since we decided that a characteristic of a vision should be supported by at least three different sources to increase the reliability of the findings.

\subsubsection{Conclusion Validity}
The reproducibility of both literature reviews is limited due to their respective search process. We found the generic video production guidelines by performing a web search using the web search engines Google Scholar and Google. The literature on software project vision results from a manual search in an internal library as well as suggestions of the third author of this article regarding additional journal articles and books whose relevance for the literature review is based on his expertise. Therefore, both search processes can be hardly reproduced. However, we must emphasize that both literature reviews were not designed to identify all relevant literature in a reproducible manner as would be the case in a systematic literature review. A systematic literature review requires a more comprehensive and complex search process. We decided to perform the simplified literature reviews since we were interested in a valid body of knowledge as a starting point to develop a grounded and reflected quality model for vision videos. The results of both literature reviews are in turn based on the subjective interpretation of the coded data from the guidelines and publications by the authors of this article. We cannot completely exclude the misinterpretation of the coded data. However, we mitigated this threat to validity with a clear strategy. Two authors coded all data independently from each other and all authors cross-checked, discussed, and jointly agreed on the deduced characteristics after multiple iterations. We verified the identified characteristics of both literature reviews by asking two raters to code each extracted passages based on the respective deduced categories. The calculated reliability of the raters' classification ($\kappa_{video} = 0.81$ and $\kappa_{vision} = 0.84$) indicated an almost perfect agreement in both cases. Thus, we are confident that the deduced characteristics result from the analyzed guidelines, books, and journal articles.

\subsection{Evaluation: Threats to Validity}

\subsubsection{Construct Validity}
Although we used eight different vision videos of real projects with real customers, we had a mono-operation bias. All videos were created in the context of the \textit{Software Project} course at Leibniz Universität Hannover. Thus, the videos did not convey a comprehensive overview of the complexity in practice. Nevertheless, the videos were from real projects. Therefore, we expected a sufficient realistic complexity for a first evaluation. 
Apart from the duration of the videos, all data was collected with the single use of an assessment form which caused a mono-method bias. The use of one single subjective method only allows restricted explanations of our findings. However, we focused on the use of subjective judgments since asking humans for their opinion is considered as the ultimate standard and right way to assess video quality \cite{Seshadrinathan.2010}.
Conducting the experiment with all subjects at the same time caused an interaction of different treatments. Although we instructed the subjects to assess each video for itself without any comparison with the other ones, we cannot exclude that later assessments of videos were influenced by previous ones. We had to conduct the experiment with all subjects at the same time since there was no other option to handle the assessment of all eight vision video by all $139$ subjects during the course.
The given task of assessing the overall quality and individual quality characteristics of vision videos implied to analyze the relationships between both aspects. Thus, we had an interaction of testing and treatment. We could not exclude that the subjects tried to guess the potential outcome what might have affected their ratings. We did not expect any notable impact on the overall results due to the assessment of $14$ characteristics by over $100$ subjects. This threat to validity might have been mitigated by two separate groups. While the first group only assess the overall quality, the second group assess the individual characteristics. However, this type of evaluation has several disadvantages. First, two separate groups require spatial separation. Otherwise, the subjects notice that they evaluate different aspects. Second, the subjects who only rate the overall quality would be much faster than the other ones which in turn may affect the other group who assess the $14$ individual characteristics. Unfortunately, we had only the permission to carry out the evaluation in one $60$-minute appointment of the \textit{Software Project} course. Thus, we did not have the time to perform an individual session with each of the two groups since watching and assessing all eight vision videos already required the entire $60$ minutes (see section \ref{sec:setting-and-procedure}). Furthermore, we would have needed a second lecture hall which accommodates $70$ people (half of the subjects). This lecture hall would also need to have the same projector and sound system since the display devices have a major impact on the perceived video quality \cite{Winkler.2008}. Such a second lecture hall was not available since the evaluation took place during the lecture period. For these reasons, the chosen design was the only possible one.

\subsubsection{External Validity}
The selected subjects and the eight vision videos of real projects produced a good level of realism. The undergraduate students were close to their graduation, actively developed software at the time of the experiment, and had on average $2.4$ years of experience as developers. Thus, they were suitable to assess the vision videos from a developer's point of view. However, in contrast to professionals, the subjects formed a more homogeneous group which restricted the generalizability of the results. Furthermore, not only the point of view of a developer is important since other stakeholders and members of a development team also belong to the target audience of vision videos. The presented findings are restricted to the point of view of a developer and do not need to hold for other roles, such as stakeholders or video producers.
The experimental setting caused an interaction of setting and treatment. Watching a vision video in a lecture hall with over $100$ people is not a typical setting compared with an industrial environment. We had to accept this threat to validity to conduct the experiment. The given task of sharing the vision of the particular project had also no pragmatic value for the subjects.
As the next step in technology transfer from academia to industry, future evaluations need to be conducted in real projects with different roles belonging to the target audience of vision videos.

\subsubsection{Internal Validity}
Maturation is one crucial threat to internal validity. The whole experiment lasted $60$ minutes which might affect the subjects negatively by getting tired or bored. We had to accept this threat to validity since the constraints of the \textit{Software Project} course did not allow any other setting to conduct the experiment. However, we only lost $20$ out of $139$ subjects due to incomplete assessments forms. Thus, we assume that the duration of the experiment was still acceptable for most of the subjects.
The way of testing presented a further threat to validity since there might be a learning effect. The subjects repeated the same assessment for eight different vision videos one after the other. Thus, the later assessments might be affected by the previous ones.
The used assessment form for instrumentation is another potential threat to validity. In the case of bad design and wording, the evaluation might be affected negatively. We refined and tested the assessment form in a pilot study with $18$ subjects to improve the instrumentation. In addition, we explained the assessment form and quality characteristics to all subjects, asking the subjects to ask questions at any time if they needed clarification. 
We restricted the selection of subjects on computer science students who were active participants in the \textit{Software Project} course. In this way, we ensured that the subjects actively develop software and thus have the point of view of a developer. All subjects participated voluntarily in our evaluation. There was no financial reward and the participation did not influence the success of passing the course. Thus, there was little incentive to participate without being self-motivated. Nevertheless, even self-motivated subjects are a threat to validity since they might be more motivated and suited for the evaluation than the entire population.

\subsubsection{Conclusion Validity}
The validity of any scientific evaluation highly depends on the reliability of measures. Besides the video length measured in seconds, all other characteristics were assessed subjectively using an assessment form. A good wording, instrumentation, and instrumentation layout were crucial for the validity of the findings. We performed a pre-test of the assessment form with a small group of $18$ subjects to verify and improve the wording and instrumentation. Nevertheless, most of the data was based on subjective judgments which reduced the reliability and reproducibility. We consciously decided on this kind of assessment since the subjective judgment of video quality is the ultimate and right way to assess video quality \cite{Seshadrinathan.2010}. In contrast to typical video quality assessment experiments with $15$ to $30$ subjects, we had $139$ subjects to increase the reliability of our measurements.
By conducting the experiment in a lecture hall with all $139$ subjects at the same time, we ensured the reliability of treatment implementation. Thus, all subjects watched the eight videos under the same conditions which mitigated the risk that the implementation was not similar for all subjects.
The subjects formed a more homogeneous group than professionals from industry. This counteracted the threat of erroneous conclusion. A more homogeneous group mitigated the risk that the variation due to the subjects' random heterogeneity was larger than due to the investigated treatments. In turn, the subjects' homogeneity restricted external validity since a group of professionals is seldom homogeneous due to their different backgrounds.

\section{Discussion}
\label{sec:discussion}
Although videos are a frequently proposed solution to communicate and document a vision (or parts of it), there is a lack of guidance on how to produce a good video. This lack of guidance impedes the applicability of every given approach that focuses on the use of videos as a documentation option. Whether a video is good or not depends on its perceived quality by the target audience. However, \textit{video quality} is a rather ill-defined concept due to numerous influencing factors.\\

Below, we divided this section into three parts to discuss the following three aspects individually: ($1$) the proposed quality model for vision videos (see section \ref{sec:the-quality-model-for-vision-videos}), ($2$) the evaluation carried out in academia (see section \ref{sec:evaluation-in-academia}), and ($3$) our resulting ongoing future work (see section \ref{sec:future-work}).

\subsection{The Quality Model for Vision Videos}
\label{sec:the-quality-model-for-vision-videos}
In this article, we propose a quality model for vision videos in analogy to the ISO/IEC FDIS $25010$:$2010$ \cite{ISO25010.2010} for system and software quality models. Thus, we want to provide a first structured overview of the concept of vision video quality. For this purpose, we break down vision video quality into individual quality characteristics with familiar labels and concise definitions. 
We had the objectives to present a quality model for vision videos that can be used to (a) evaluate given vision videos and (b) guide video production by software professionals.

For goal (a), we provide a hierarchical decomposition of vision video quality. This representation provides a convenient breakdown of the overall quality of vision videos into its quality characteristics. According to literature, all identified quality characteristics constitute a good vision video and thus need to be assessed to evaluate its overall quality.

For goal (b), we propose a mapping of the individual quality characteristics to the single steps of the video production and use process. This perspective provides orientation and guidance by showing which quality characteristics are crucial for a particular step. Thus, the representation can be used as a checklist to ensure the comprehensive treatment of all dimensions of vision video quality during the entire process.

We developed the quality model by combining the results of two literature reviews on the characteristics and intentions of vision and video. We decided to proceed by analyzing existing knowledge in the form of generic guidelines on video production since all authors of this article lack detailed knowledge of video production. We have to note that we did not find any publication providing a comprehensive definition of video characteristics. In contrast, we found several publications on software project vision whose joint analysis resulted in a detailed description of what constitutes a vision. Thus, our findings are based on consistent and mutually supportive references. We strengthened the confidence in our findings by calculating the inter-rater reliability for the deduced characteristics of both literature reviews.

Despite all these efforts to objectively derive the quality model for vision videos, a potential researcher bias remains. The resulting quality model is based on our subjective interpretation of the considered guidelines and publications. As a consequence, the quality model is a theoretical description of potentially relevant characteristics of a vision video. Hence, we do not claim that the proposed quality model is complete. Instead, this model is a first starting point to better understand what constitutes a good vision video for its respective purposes. We provide a clearer definition of video quality in the context of representing software project visions by video.
As future work, we plan to conduct an expert evaluation of the proposed quality model for vision videos. Thus, we can match our findings with the opinions, expectations, and experiences of experts in RE and video production to validate, refine, and extend the model.

\subsection{Evaluation in Academia}
\label{sec:evaluation-in-academia}
Such an expert evaluation requires that the quality model as a candidate solution is sound. For this purpose, we conducted an evaluation with $139$ undergraduate students in the \textit{Software Project} course at the Leibniz Universität Hannover. The results of this evaluation indicate that there are clear correlations between individual quality characteristics and the likelihood that the perceived overall quality of vision videos is good. In particular, the findings validate obvious and logical relationships. It is not surprising that a vision video is more likely to be perceived as good if it is enjoyable to watch the video (\textit{pleasure}). Furthermore, it is obvious that the less \textit{prior knowledge} is necessary to understand a vision video, the higher the likelihood that a video is perceived as good since the video can be understood more easily. For the vision characteristics \textit{focus}, \textit{clarity}, and \textit{stability} the conclusions are similar. The more compact, clearer, and more stable a vision video is, the more the essence and aspired goals of the vision are consistently presented.
The relationship between a longer \textit{video length} and the likelihood that the overall quality is perceived as good seems surprising at first. According to Broll et al. \cite{Broll.2007}, a $3$-minute video appears as short but should still be presented in shorter clips to avoid that the viewers become mentally inactive. This recommendation corresponds to the findings of Guo et al. \cite{Guo.2014} which yielded that Massive Open Online Course (MOOC) videos are viewed at most $6$ minutes, regardless of total video length. For longer videos, viewers watch less than half of a video. In consideration of these findings, the identified relationship between video length and overall quality seems plausible. The used vision videos had a mean duration of $1$:$43$ minutes and a minimum and maximum duration of $1$:$09$ minutes and $2$:$35$ minutes. Thus, all used vision videos were shorter than the recommended upper limits. We assume that the videos with a duration closer to the given maximum of 3 minutes are perceived as better since they may provide more information and time to process and understand this information than shorter videos. This assumption is supported by Owens and Millerson who stated: \enquote{If the shots [and thus the video] are too brief, they may flick past the viewer’s eyes without entering the brain} \cite[p. 143]{Owens.2011}.
However, we have to remark that depending on the content of a vision video even a longer duration may have a negative impact on its perceived overall quality. According to Owens and Millerson \cite{Owens.2011}, if longer videos contain too many topics, they cannot address these contents adequately. In contrast, if longer videos cover too few topics, they appear to be slow and labored. Both cases negatively affect the target audience and their perception of the overall video quality either since they do not understand the video or get bored.

Although we found no relationships between the likelihood that the overall quality of a vision video is perceived as good and the nine characteristics \textit{image quality}, \textit{sound quality}, \textit{plot}, \textit{essence}, \textit{clutter}, \textit{completeness}, \textit{intention}, \textit{sense of responsibility}, and \textit{support}, we cannot exclude that there are relationships. 

In contrast, we assume that there are specific reasons why we could not find these relationships. First of all, the teams produced the used vision videos with similar simple equipment, i.e., smartphones. Thus, all assessed vision videos have comparable \textit{image quality} and \textit{sound quality} resulting in barely noticeable differences. One could argue that professional equipment might have caused a better quality and thus clearer differences. However, on the one hand, our research focus is on the simplicity regarding the equipment used as well as knowledge and skills required to simplify the production and use of vision videos in RE for software professionals. On the other hand, different researchers \cite{Brill.2010, Broll.2007, Schneider.2019} already showed that vision videos with a lower image and sound quality due to the use of simple equipment such as smartphones, tablets, and digital camcorders are completely sufficient for the purposes in RE. Owens and Millerson \cite{Owens.2011} support these views by emphasizing that it is important to know how to communicate visually no matter how professional the equipment is. Nevertheless, we presume that a vision video with a very poor image and sound quality affects its perceived overall quality negatively. Furthermore, we assume that the experimental design and the chosen point of view of developers have led to difficulties to adequately assess the other seven characteristics. All videos were only watched once. Therefore, we suppose that the subjects might not be able to directly recognize and assess the structured presentation (\textit{plot}), the important core elements (\textit{essence}), potentially disrupting and distracting elements (\textit{clutter}) as well as whether the vision is complete or not (\textit{completeness}). The subjects might also be confused about why we ask them whether the video is suitable for the given task (\textit{intention}) since the experiment implied that this aspect is fulfilled. The given task also had no pragmatic value for the subjects why there was no need for them to support the vision (\textit{support}). In contrast to our subjects who only watched the video, a video producer might be more concerned whether the legal regulations are fulfilled (\textit{sense of responsibility}) since this role is responsible and legally liable.

Nevertheless, the key finding of this evaluation is not individual relationships. Instead, the more important insight is that we found significant correlations between individual quality characteristics and the likelihood that the subjects perceive the overall quality of a vision video as good. Based on the results of the logistic regression, we rejected the global null hypothesis. This rejection, in turn, substantiates the fundamental relevance of the proposed quality model for vision videos. Therefore, we are confident that we developed a valid and sound solution that is sufficient to serve as a viable basis for future refinements and extensions. 

\subsection{Future Work}
\label{sec:future-work}
As previously mentioned, one part of our future work is an expert evaluation to verify the quality characteristics and possibly identify additional characteristics. In addition to this evaluation, we plan to work on two further topics: (a) analysis of interactions and (b) operationalization of the quality model.

The topic (a) focuses on the analysis of interactions among the quality characteristics and their impact on the overall vision video quality. The logistic regression performed only indicates what proportion of the change in the overall vision video quality can be explained by the change in the individual quality characteristics included in the regression model. This analysis does not take into account whether the individual characteristics influence each other and thus the overall quality. For example, a vision video that contains more relevant core elements (higher \textit{essence}) may be less compacted (lower \textit{focus}) and longer (higher \textit{video length}) since more content needs to be presented. Based on our data set, an analysis of interactions among the quality characteristics could be done with the multivariate analysis of variance (MANOVA). However, this analysis requires an interval-scaled dependent variable, i.e., the overall vision video quality needs to be metric. Although several objective video quality metrics are interval-scaled (see \ref{sec:video-quality-assessment}), none of them is widely applicable and universally recognized. Consequently, we must first either determine which of the existing objective metrics best reflects the subjectively perceived vision video quality or develop a new interval-scaled metric. When we have such an interval-scaled metric, we can do an analysis of interactions with our data set.

The topic (b) focus on two investigations to enhance the practical utility of the quality model by revealing how individual quality characteristics can be useful in practice.

First, we plan to create an overall set of recommendations on vision video production based on the extracted passages from the literature reviews. Each recommendation is linked to one or more quality characteristics as well as one or more steps of the production and use process. Thus, we can provide two different perspectives on the overall set of recommendations. On the one hand, this set can show all recommendations for a specific quality characteristic. On the other hand, a user can find all recommendations which are important in a specific process step. Both perspectives can help software professional with their video production. They provide guidance on what to consider regarding either a particular characteristic or process step.

Second, we are currently developing a software tool which supports continuously collecting assessment data for generic quality characteristics during the entire video playback \cite{Karras.2019c_2}. Established subjective video quality assessment methods ultimately determine video quality based on the average of all subjects' assessments. However, these assessments consist only of one single value that the respective subject assigns to the entire video. Thus, these established methods do not lend themselves to detailed analyzes of videos to determine how specific implementations of individual quality characteristics exactly affect the viewers' quality assessment. In contrast to the established methods, our software tool is intended to support such a detailed analysis of videos to provide fine-grained insights into the interrelationships of the specific implementation of a quality characteristic and its impact on the viewers' quality assessment. Based on these insights, we expect to conclude how quality characteristics should be implemented in a video so that they lead to an overall video quality that is perceived as good. These conclusions may help to specify the recommendations more precisely by providing substantiated rationales. These rationales, in turn, enhance the practical utility of the recommendations and thus of the quality model.

\section{Conclusion}
\label{sec:conclusion}
A clearly defined and shared vision serves as a basis for emerging a dialog between all stakeholders to define the scope of the future system. Although a vision is supposed to support communication, it is mainly documented in a textual representation which insufficiently supports a rich knowledge transfer. Therefore, different researchers proposed the use of video representing a vision or parts of it as a documentation option for communication, so-called vision videos. Despite several years of research on the use of vision videos in RE, the required video production is often considered a secondary task. There is a lack of knowledge of what constitutes a good video for representing a vision. In general, video quality is a complex and still insufficiently defined concept due to the wide variety of influencing factors. 

Inspired by the ISO/IEC FDIS $25010$:$2010$ \cite{ISO25010.2010} for system and software quality models, we propose a first quality model for vision videos that can be used to evaluate given vision videos and guide the video production by software professionals. The insights of two literature reviews on the characteristics and intentions of video and vision resulted in a hierarchical decomposition of vision video quality. We cover all three quality dimensions (\textit{representation}, \textit{content}, and \textit{impact}) comprising $15$ characteristics at the lowest level of the quality model.
Besides the hierarchical decomposition, we present a mapping of the individual quality characteristic to the steps of the video production and use process. This representation is intended to serve as a checklist for ensuring the comprehensive treatment of vision video quality by providing orientation and guidance during the entire process.
In an evaluation, we investigated whether the $15$ quality characteristics are related to the overall quality of vision videos. According to the findings, there are significant correlations between individual quality characteristics and the likelihood that the subjects perceive the overall quality of a vision video as good. These relationships substantiate the fundamental relevance of the proposed quality model for vision videos.

In particular, our work provides a clearer definition of the hitherto ill-defined concept of \textit{video quality} in the context of representing a software project vision. The proposed quality model is intended to engage the attention of software professionals on characteristics of vision video quality. We offer a basis for estimating the consequent effort and activities needed to produce a good video at moderate costs and with sufficient quality. The benefit of the model is its support of software professionals to identify and specify the quality characteristics that they believe are relevant for their particular vision video. When we know which quality characteristics are relevant for a particular vision video, we can specify requirements, criteria for their satisfaction and corresponding measures to guide the video production and evaluate the resulting video.

\section*{Acknowledgment}
This work was supported by the Deutsche Forschungsgemeinschaft (DFG) under Grant No.:~289386339, (2017 -- 2019).

\appendix

\section{Overview of Vision Videos in RE}
The \tablename{ \ref{tbl:t4}} is presented on the next page.

\label{sec:overview-of-vision-videos-in-re}
\begin{table*}[htbp]
	\renewcommand{\arraystretch}{1.1}
	\centering
	\caption{Overview of related work on vision videos in RE}
	\label{tbl:t4}
	\begin{tabular}{|c|l|c|l|l|l|c|}
		\hline
		\textbf{Paper} & \multicolumn{1}{c|}{\textbf{\begin{tabular}[c]{@{}c@{}}Supported\\ RE activities\end{tabular}}} & \multicolumn{1}{c|}{\textbf{\begin{tabular}[c]{@{}c@{}}Parts of \\ a vision\end{tabular}}} & \multicolumn{1}{c|}{\textbf{Video content}} & \multicolumn{1}{c|}{\textbf{Audience}} & \multicolumn{1}{c|}{\textbf{Producer}} & \textbf{Guidance} \\ \hline
		
		\cite{Creighton.} & \begin{tabular}[c]{@{}l@{}}Problem definition\\ Goal definition\\ Elicitation\\ Validation\end{tabular} & \begin{tabular}[c]{@{}c@{}}Problem \& \\ Solution\end{tabular} & \begin{tabular}[c]{@{}l@{}}Work practice\\ Visionary scenario\end{tabular} & \begin{tabular}[c]{@{}l@{}}Customer\\ User\\ RE Analyst\end{tabular} & Video producer & No \\ \hline 
		
		\cite{Brill.2010} & \begin{tabular}[c]{@{}l@{}}Elicitation\\ Validation\end{tabular} & Solution & Visionary scenario & \begin{tabular}[c]{@{}l@{}}Customer\\ RE Analyst\end{tabular} & RE Analyst & No \\ \hline 
		
		\cite{Pham.} & \begin{tabular}[c]{@{}l@{}}Elicitation\\ Validation\\ Documentation\end{tabular} & Solution & \begin{tabular}[c]{@{}l@{}}Project vision\\ Visionary scenario\end{tabular} & Stakeholder & RE Analyst & No \\ \hline 
		
		\cite{Karras.2017a} & \begin{tabular}[c]{@{}l@{}}Elicitation\\ Documentation\end{tabular} & Solution & \begin{tabular}[c]{@{}l@{}}Prototype\\ Visionary scenario\end{tabular} & \begin{tabular}[c]{@{}l@{}}User\\ Developer\end{tabular} & RE Analyst & No \\ \hline
		
		\cite{Xu.2012} & \begin{tabular}[c]{@{}l@{}}Problem definition\\ Elicitation\\ Validation\end{tabular} & \begin{tabular}[c]{@{}c@{}}Problem \& \\ Solution\end{tabular} & \begin{tabular}[c]{@{}l@{}}Work practice\\ Visionary scenario\\ Implementation\end{tabular} & Stakeholder & Team member & Yes \\ \hline
		
		\cite{Xu.2013} & \begin{tabular}[c]{@{}l@{}}Problem definition\\ Elicitation\\ Validation\end{tabular} & \begin{tabular}[c]{@{}c@{}}Problem \& \\ Solution\end{tabular} & \begin{tabular}[c]{@{}l@{}}Work practice\\ Visionary scenario\\ Implementation\end{tabular} & \begin{tabular}[c]{@{}l@{}}Stakeholder\\ Team member\end{tabular} & Team member & Yes \\ \hline
		
		\cite{Brouse.1992} & \begin{tabular}[c]{@{}l@{}}Problem definition\\ Goal definition\\ Elicitation\\ Documentation\end{tabular} & Problem & Environment & Decision maker & RE Analyst & No \\ \hline 
		
		\cite{BrunCottan.1995} & Elicitation & Problem & Work practice & \begin{tabular}[c]{@{}l@{}}Developer\\ User\end{tabular} & Team member & No\\ \hline
		
		\cite{Haumer.1998} & Goal definition & Problem & Work practice & \begin{tabular}[c]{@{}l@{}}Team member\\ Stakeholder\end{tabular} & RE Analyst & No \\ \hline 
		
		\cite{Mackay.1999} & \begin{tabular}[c]{@{}l@{}}Problem definition\\ Goal definition\\ Elicitation\\ Validation\end{tabular} & \begin{tabular}[c]{@{}c@{}}Problem \& \\ Solution\end{tabular} & \begin{tabular}[c]{@{}l@{}}Work practice\\ Visionary scenario\end{tabular} & \begin{tabular}[c]{@{}l@{}}User\\ Manager\\ Designer\end{tabular} & Designer & Yes \\ \hline
		
		\cite{Zachos.2004} & \begin{tabular}[c]{@{}l@{}}Elicitation\\ Documentation\end{tabular} & Problem & Environment & Stakeholder & RE Analyst & No \\ \hline
		
		\cite{Jirotka.2006} & Elicitation & Problem & Work practice & RE Analyst & RE Analyst & Yes \\ \hline 
		
		\cite{Rabiser.2006} & \begin{tabular}[c]{@{}l@{}}Elicitation\\ Documentation\end{tabular} & Problem & Work practice & \begin{tabular}[c]{@{}l@{}}RE Analyst\\ User\end{tabular} & \begin{tabular}[c]{@{}l@{}}RE Analyst\\ User\end{tabular} & No \\ \hline 
		
		\cite{Broll.2007} & Elicitation & \begin{tabular}[c]{@{}c@{}}Problem \& \\ Solution\end{tabular} & \begin{tabular}[c]{@{}l@{}}Project vision\\ Visionary scenario\end{tabular} & Stakeholder & Video producer & Yes \\ \hline
		
		\cite{Shan.2008} & \begin{tabular}[c]{@{}l@{}}Goal definiton\\ Elicitation\end{tabular} & Problem & Environment & Domain expert & No information & No \\ \hline
		
		\cite{Bruegge.b} & \begin{tabular}[c]{@{}l@{}}Elicitation\\ Validation\end{tabular} & \begin{tabular}[c]{@{}c@{}}Problem \& \\ Solution\end{tabular} & \begin{tabular}[c]{@{}l@{}}Project vision\\ Visionary scenario\end{tabular} & \begin{tabular}[c]{@{}l@{}}Customer\\ Supplier\end{tabular} & Video producer & No \\ \hline
		
		\cite{Bruegge.2008} & \begin{tabular}[c]{@{}l@{}}Elicitation\\ Validation\end{tabular} & Solution & Visionary scenario & \begin{tabular}[c]{@{}l@{}}Customer\\ Developer\end{tabular} & Video producer & No \\ \hline 
	
		\cite{Schneider.2011} & Elicitation & Problem & \begin{tabular}[c]{@{}l@{}}Environment\\ Work practice\end{tabular} & RE Analyst & User & No \\ \hline
	
		\cite{Stangl.} & \begin{tabular}[c]{@{}l@{}}Elicitation\\ Validation\end{tabular} & Solution & \begin{tabular}[c]{@{}l@{}}Prototype\\ Visionary scenario\end{tabular} & Stakeholder & RE Analyst & No \\ \hline 		
		
		\cite{Karras.2017} & Documentation & \begin{tabular}[c]{@{}c@{}}Problem \& \\ Solution\end{tabular} & Project vision & Developer & Team member & No \\ \hline
	\end{tabular}
\end{table*}

\newcommand\tabrotate[2][5em]{\rotatebox[origin=c]{90}{\begin{varwidth}{#1}\centering#2\end{varwidth}}}

\section{Data Set}
\label{sec:data-set}

\begin{center}
	\centering
	\renewcommand{\arraystretch}{1.1}
	\tablecaption{Detailed description of the data set -- Part 1}
	\tablefirsthead{}
	\tablehead{}
	\tabletail{\hline}
	\tablelasttail{}
	
	\begin{supertabular}{|c|l|c|c|c|}
		\hline 
		\multicolumn{2}{|l|}{\multirow{2}{*}{}} & \multicolumn{2}{c|}{\textbf{Overall quality}} & \multirow{2}{*}{\textbf{In total}} \\ \cline{3-4}
		\multicolumn{2}{|l|}{} & \textbf{Bad} & \textbf{Good} & \\ \hline
		\multicolumn{2}{|l|}{Total sample $(N)$} & 281 & 671 & 952 \\ \hline \hline
		\multicolumn{2}{|l|}{\multirow{2}{*}{\textbf{Vision video characteristic}}} & \multicolumn{2}{c|}{\textbf{Overall quality}} & \multirow{2}{*}{\textbf{In total}} \\ \cline{3-4}
		\multicolumn{2}{|l|}{} & \textbf{Bad} & \textbf{Good} & \\ \hline\hline
		\multirow{5}{*}{\tabrotate[\widthof{Image quality}]{Image quality}} & Very bad & 9 & 2 & 11 \\ \cline{2-5} 
		& Bad & 38 & 18 & 56 \\ \cline{2-5} 
		& Neutral & 68 & 73 & 141 \\ \cline{2-5} 
		& Good & 95 & 229 & 324 \\ \cline{2-5} 
		& Very good & 71 & 349 & 420 \\ \hline \hline
		\multirow{5}{*}{\tabrotate[\widthof{Sound quality}]{Sound quality}} & Very bad & 21 & 25 & 46 \\ \cline{2-5} 
		& Bad & 51 & 88 & 132 \\ \cline{2-5} 
		& Neutral & 70 & 97 & 167 \\ \cline{2-5} 
		& Good & 83 & 201 & 284 \\ \cline{2-5} 
		& Very good & 56 & 260 & 316 \\ \hline \hline
		\multirow{2}{*}{\tabrotate[\widthof{length}]{Video length}} & Mean [s]& 106.03 & 102.26 & 103.38 \\ \cline{2-5} 
		& Std. Deviation [s]& 26.26 & 24.98 & 25.40 \\ \hline
	\end{supertabular}
	\label{tbl:t5}
\end{center}

\pagebreak
\begin{center}
	\centering
	\renewcommand{\arraystretch}{1.1}
	\tablecaption{Detailed description of the data set -- Part 2}
	\tablefirsthead{}
	\tablehead{}
	\tabletail{\hline}
	\tablelasttail{}
	\begin{supertabular}{|c|l|c|c|c|}
		\hline
		\multicolumn{2}{|l|}{\multirow{2}{*}{\textbf{Vision video characteristic}}} & \multicolumn{2}{c|}{\textbf{Overall quality}} & \multirow{2}{*}{\textbf{In total}} \\ \cline{3-4}
		\multicolumn{2}{|l|}{} & \textbf{Bad} & \textbf{Good} & \\ \hline \hline
			\multirow{5}{*}{\tabrotate[\widthof{Focus}]{Focus}} & Very non-compact & 19 & 6 & 25 \\ \cline{2-5} 
		& Non-compact & 57 & 29 & 86 \\ \cline{2-5} 
		& Neutral & 94 & 95 & 189 \\ \cline{2-5} 
		& Compact & 85 & 281 & 366 \\ \cline{2-5} 
		& Very compact & 26 & 260 & 286 \\ \hline \hline
		\multirow{5}{*}{\tabrotate[\widthof{Plot}]{Plot}} & Very bad & 12 & 3 & 15 \\ \cline{2-5} 
		& Bad & 37 & 26 & 63 \\ \cline{2-5} 
		& Neutral & 79 & 74 & 153 \\ \cline{2-5} 
		& Good & 119 & 299 & 418 \\ \cline{2-5} 
		& Very good & 34 & 269 & 303 \\ \hline \hline
		\multirow{5}{*}{\tabrotate[\widthof{knowledge}]{Prior\\ knowledge}} & Very unnecessary & 18 & 100 & 118 \\ \cline{2-5} 
		& Unnecessary & 32 & 184 & 216 \\ \cline{2-5} 
		& Neutral & 87 & 216 & 303 \\ \cline{2-5} 
		& Necessary & 77 & 112 & 189 \\ \cline{2-5} 
		& Very necessary & 67 & 59 & 126 \\ \hline \hline
		\multirow{5}{*}{\tabrotate[\widthof{Clarity}]{Clarity}} & Very unintelligible & 25 & 2 & 27 \\ \cline{2-5} 
		& Unintelligible & 68 & 34 & 102 \\ \cline{2-5} 
		& Neutral & 76 & 93 & 169 \\ \cline{2-5} 
		& Intelligible & 78 & 244 & 322 \\ \cline{2-5} 
		& Very intelligible & 34 & 298 & 332 \\ \hline \hline
		\multirow{5}{*}{\tabrotate[\widthof{Essence}]{Essence}} & Very little & 23 & 12 & 35 \\ \cline{2-5} 
		& Little & 58 & 41 & 99 \\ \cline{2-5} 
		& Neutral & 100 & 170 & 270 \\ \cline{2-5} 
		& Much & 76 & 263 & 339 \\ \cline{2-5} 
		& Very much & 24 & 185 & 209 \\ \hline \hline
		\multirow{5}{*}{\tabrotate[\widthof{Clutter}]{Clutter}} & Very little & 69 & 140 & 209 \\ \cline{2-5} 
		& Little & 92 & 136 & 228 \\ \cline{2-5} 
		& Neutral & 89 & 233 & 322 \\ \cline{2-5} 
		& Much & 27 & 109 & 136 \\ \cline{2-5} 
		& Very much & 4 & 53 & 57 \\ \hline \hline
		\multirow{5}{*}{\tabrotate[\widthof{Completeness}]{Completeness}} & Very incomplete & 23 & 14 & 37 \\ \cline{2-5} 
		& Incomplete & 53 & 73 & 126 \\ \cline{2-5} 
		& Neutral & 80 & 96 & 176 \\ \cline{2-5} 
		& Complete & 85 & 230 & 315 \\ \cline{2-5} 
		& Very complete & 40 & 258 & 298 \\ \hline \hline
		\multirow{5}{*}{\tabrotate[\widthof{Pleasure}]{Pleasure}} & Very unenjoyable & 15 & 3 & 18 \\ \cline{2-5} 
		& Unenjoyable & 67 & 19 & 86 \\ \cline{2-5} 
		& Neutral & 106 & 112 & 218 \\ \cline{2-5} 
		& Enjoyable & 75 & 295 & 370 \\ \cline{2-5} 
		& Very enjoyable & 18 & 242 & 260 \\ \hline
	\end{supertabular}
	\label{tbl:t6}
\end{center}

\pagebreak
\begin{center}
	\centering
	\renewcommand{\arraystretch}{1.1}
	\tablecaption{Detailed description of the data set -- Part 3}
	\tablefirsthead{}
	\tablehead{}
	\tabletail{\hline}
	\tablelasttail{}
	\begin{supertabular}{|c|l|c|c|c|}
		\hline
		\multicolumn{2}{|l|}{\multirow{2}{*}{\textbf{Vision video characteristic}}} & \multicolumn{2}{c|}{\textbf{Overall quality}} & \multirow{2}{*}{\textbf{In total}} \\ \cline{3-4}
		\multicolumn{2}{|l|}{} & \textbf{Bad} & \textbf{Good} & \\ \hline \hline
			\multirow{5}{*}{\tabrotate[\widthof{Intention}]{Intention}} & Very unsuitable & 10 & 4 & 14 \\ \cline{2-5} 
		& Unsuitable & 38 & 11 & 49 \\ \cline{2-5} 
		& Neutral & 107 & 89 & 196 \\ \cline{2-5} 
		& Suitable & 95 & 326 & 421 \\ \cline{2-5} 
		& Very suitable & 31 & 241 & 272 \\ \hline \hline
		\multirow{5}{*}{\tabrotate[\widthof{Responsibility}]{Sense of responsibility}} & Very non-compliant & 77 & 150 & 227 \\ \cline{2-5} 
		& Non-compliant & 102 & 170 & 272 \\ \cline{2-5} 
		& Neutral & 85 & 252 & 337 \\ \cline{2-5} 
		& Compliant & 11 & 67 & 78 \\ \cline{2-5} 
		& Very compliant & 6 & 32 & 38 \\ \hline \hline
		\multirow{5}{*}{\tabrotate[\widthof{Support}]{Support}} & Totally disagree & 5 & 3 & 8 \\ \cline{2-5} 
		& Disagree & 30 & 17 & 47 \\ \cline{2-5} 
		& Neutral & 76 & 100 & 176 \\ \cline{2-5} 
		& Agree & 121 & 263 & 384 \\ \cline{2-5} 
		& Totally agree & 49 & 288 & 337 \\ \hline \hline
		\multirow{5}{*}{\tabrotate[\widthof{Stability}]{Stability}} & Very unstable & 56 & 47 & 103 \\ \cline{2-5} 
		& Unstable & 130 & 145 & 275 \\ \cline{2-5} 
		& Neutral & 63 & 231 & 294 \\ \cline{2-5} 
		& Stable & 27 & 155 & 182 \\ \cline{2-5} 
		& Very stable & 5 & 93 & 98 \\ \hline
	\end{supertabular}
	\label{tbl:t7}
\end{center}

\bibliographystyle{elsarticle-num} 
\bibliography{references}

\end{document}